\documentclass[11pt, oneside]{article}   	
\usepackage{jheppub}
\usepackage{amsmath}
\usepackage{amssymb}
\usepackage{amsfonts}
\usepackage{amsthm}
\usepackage{amsbsy}
\usepackage{array}
\usepackage{mathtools}
\usepackage[usenames,dvipsnames]{xcolor}
\usepackage{subcaption}
\usepackage{dsdshorthand}
\usepackage{graphicx}
\usepackage[vcentermath]{youngtab}
\usepackage{tikz}
\usetikzlibrary{snakes}
\usetikzlibrary{decorations}
\usetikzlibrary{trees}
\usetikzlibrary{decorations.pathmorphing}
\usetikzlibrary{decorations.markings}
\usetikzlibrary{external}
\usetikzlibrary{intersections}
\usetikzlibrary{shapes,arrows}
\usetikzlibrary{arrows.meta}
\usetikzlibrary{calc}
\usetikzlibrary{shapes.misc}
\usetikzlibrary{decorations.text}
\usetikzlibrary{backgrounds}
\usetikzlibrary{fadings}
\usetikzlibrary{tikzmark,calc,arrows,shapes,decorations.pathreplacing}
\tikzset{
        cross/.style={cross out, draw=black, minimum size=2*(#1-\pgflinewidth), inner sep=0pt, outer sep=0pt},
	branchCut/.style={postaction={decorate},
		snake=zigzag,
		decoration = {snake=zigzag,segment length = 2mm, amplitude = 2mm}	
    }}

\newcommand{\bea}{\setlength\arraycolsep{2pt} \begin{eqnarray}}
\newcommand{\eea}{\end{eqnarray}}

\def\fft#1#2{{\frac{#1}{#2}}}

\newcommand{\baa}{\begin{align}}
\newcommand{\eaa}{\end{align}}

\def\mlight{m_\ell}
\def\pmin{m_{\rm IR}}
\def\lsim{\mathrel{\hbox{\rlap{\lower.55ex \hbox{$\sim$}} \kern-.3em \raise.4ex \hbox{$<$}}}}
\def\gsim{\mathrel{\hbox{\rlap{\lower.55ex \hbox{$\sim$}} \kern-.3em \raise.4ex \hbox{$>$}}}}
\def\hatg{\hat{g}}
\def\Mpl{M_{\rm pl}}

\makeatletter
\def\@fpheader{\ }
\makeatother

\title{Causality constraints on corrections to Einstein gravity}
\author{Simon Caron-Huot$^1$, Yue-Zhou Li$^1$, Julio Parra-Martinez$^2$, David Simmons-Duffin$^2$}
\affiliation{
${}^1$Department of Physics, McGill University, 3600 Rue University, Montr\'eal, H3A 2T8, QC Canada \\
${}^2$Walter Burke Institute for Theoretical Physics, Caltech, Pasadena, California 91125, USA \\
}
\emailAdd{schuot@physics.mcgill.ca}
\emailAdd{liyuezhou@physics.mcgill.ca}
\emailAdd{jparram@caltech.edu}
\emailAdd{dsd@caltech.edu}
\date{}
\abstract{We study constraints from
causality and unitarity on $2\to2$ graviton scattering in four-dimensional weakly-coupled effective field theories. Together, causality and unitarity imply dispersion relations that connect low-energy observables to high-energy data.
Using such dispersion relations, we derive two-sided bounds on gravitational Wilson coefficients in terms of the mass $M$ of new higher-spin states.
Our bounds imply that gravitational interactions must shut off uniformly in the limit $G \to 0$,
and prove the scaling with $M$ expected from dimensional analysis (up to an infrared logarithm).
We speculate that causality, together with the non-observation of gravitationally-coupled higher spin states at colliders,
severely restricts modifications to Einstein gravity that could be probed by experiments in the near future.}

\preprint{CALT-TH 2021-003}

\begin{document}

\maketitle
\pagenumbering{roman}
\setcounter{page}{2}
\newpage
\pagenumbering{arabic}
\setcounter{page}{1}

\section{Introduction}

Einstein's theory of general relativity (GR) has been extraordinarily successful since its inception over a century ago. 
Nevertheless, modifications to GR are often discussed in relation to various puzzles,
such as the nature of dark energy and dark matter, see \cite{Clifton:2011jh,Bull:2015stt,Ishak:2018his} for reviews.
An important class of modifications add higher-derivative terms to the equations of motion,
which lead to physical effects which grow at short distances.
Such corrections, handily classified in \cite{Bueno:2016ypa},
arise naturally in string theory, and presumably in any UV-complete theory of quantum gravity.

In this paper, we consider $2\to 2$ scattering of gravitons and ask a simple question:
Assuming that graviton scattering respects causality at {\it all\/} energies, 
by how much can the low-energy amplitude differ from the predictions of general relativity?
A well-known result by \cite{Weinberg:1964ew,Weinberg:1965rz} shows any Lorentz-invariant
theory of massless spin-2 particles must reproduce general relativity at large distances.
Our goal will be to bound the corrections to this limit, assuming that relativistic causality (as we understand it) holds.

Our answer will depend on the spectrum of the theory.  Let us describe our setup and assumptions.
At low energies, we assume there exists a massless graviton,
together with a finite number of fields of spin $\leq 2$, which can be described 
by an effective field theory (EFT).
Schematically, the low-energy effective action encodes modifications to Einstein gravity:
\begin{equation}
 S = \frac{1}{16\pi G} \int d^4x \sqrt{g}
 \left( R + g_{R^{(3)}}{\rm Riem}^3+  g_{R^{(4)}} {\rm Riem}^4+ \ldots \right) +S_{\rm matter}\,,\label{eq: schematically S}
\end{equation}
where ${\rm Riem}^n$ denotes (possibly non-unique) contractions of products of $n$ Riemann tensors.

An important idea is that the sizes of the Wilson coefficients $g_{R^{(3)}},g_{R^{(4)}},\dots$ are constrained by causality,
that is, the notion that signals cannot travel faster than light.
For example in~\cite{Camanho:2014apa} it was observed
by Camanho-Edelstein-Maldacena-Zhiboedov (CEMZ) that in the presence of $g_{R^{(3)}}$,
the two polarization modes of the graviton would move at different velocities in certain backgrounds,
and inevitably one of them moves faster than ``light''.
By considering a setup with large enough black holes and mirrors, this effect could lead to closed timelike curves,
with ensuing grandparent-type paradoxes; the conclusion is that a classical theory with $g_{R^{(3)}}\neq 0$ is inconsistent.
Ref.~\cite{Camanho:2014apa} further pointed out that paradoxes can be avoided at the quantum level
if the graviton couples to higher-spin states. Denoting the mass of the lightest higher-spin state (spin 4 or higher) by $M$,
this led to a parametric bound: $|g_{R^{(3)}}|\lsim \frac{1}{M^4}$.

Our main goal in this paper will be to quantitatively bound higher-derivative corrections
in \eqref{eq: schematically S} in terms of the mass $M$ of higher-spin states. This mass $M$ provides a UV-cutoff scale for the low-energy EFT in (\ref{eq: schematically S}).
We will assume a large hierarchy between the Plank scale and this cutoff
\begin{equation}
  M^2 \ll \Mpl^2\,,
\end{equation}
or, equivalently, $GM^2 \ll 1$, so that gravity is weakly-interacting below the cutoff.
Our methods will test causality of graviton scattering with arbitrary center-of-mass energies,
although physically the most important region for us will be near the cutoff $M$.

The notion of causality is subtle in gravitational EFTs because there is no globally well-defined lightcone in nontrivial backgrounds. As discussed recently in \cite{Gao:2000ga,Chen:2021bvg}, 
one should contrast ``asymptotic causality," which exploits a fixed causal structure at large distances,
with ``infrared causality" which compares local time delays between species \cite{Chen:2021bvg}.
We will use asymptotic causality, but crucially, imposed at all energy scales and not only within the EFT regime.
This leads to sharp mathematical statements involving crossing symmetry, analyticity, and Regge boundedness of scattering amplitudes \cite{Adams:2006sv}.
Many works have examined how these conditions constrain EFTs and their UV completions, see e.g.\ \cite{deRham:2017avq,Caron-Huot:2020cmc,Caron-Huot:2021rmr,Tolley:2020gtv,Arkani-Hamed:2020blm,Chiang:2021ziz,Bellazzini:2020cot,Sinha:2020win,Alberte:2020bdz,Li:2021lpe,Haldar:2021rri,Raman:2021pkf,Henriksson:2021ymi,Zahed:2021fkp,Alberte:2019xfh, Alberte:2021dnj,Chowdhury:2021ynh,Bellazzini:2021oaj,Zhang:2021eeo,Wang:2020jxr, Trott:2020ebl,deRham:2017zjm,deRham:2018qqo,Wang:2020xlt} and references therein. In particular, these conditions give rise to dispersion relations that relate high and low energies. In some cases, dispersion relations can be interpreted as expressing the commutativity of coincident shockwaves \cite{Kologlu:2019bco}. Initially, the use of dispersion relations in gravitational EFTs was hindered by divergences related to the forward limit of the low-energy graviton amplitude. These technical issues were recently partially overcome in \cite{Caron-Huot:2021rmr} by studying dispersion relations in impact parameter space;
hence our renewed interest in this problem.

There is a natural motivation to study scattering events which have center-of-mass energies above the ``cutoff'' $M$.
In any scenario where a higher-derivative correction to GR might be observed in an astrophysical
or other large-distance process, the suppression scale would have to be very low,
$M\ll 1\ {\rm TeV}=(2\times 10^{-19}{\rm m})^{-1}$:
energies above such a cutoff are routinely probed at colliders.
However, collider experiments have not yet reported any higher-spin particles of the type suggested above.
Is it at all possible to modify GR in a way that simultaneously:
(1) satisfies collider constraints, (2) is relevant at large scales, and (3) respects causality as we understand it?


In this paper we take a modest step toward answering this question,
by quantitatively relating higher-derivative corrections to the mass $M$ of higher-spin states, using methods from \cite{Caron-Huot:2021rmr}.
Our main results will be that dimensionless ratios, of the schematic form $g_{R^{(3)}}M^4$ or $g_{R^{(4)}}M^6$,
are bounded by order-unity constants, times an infrared logarithmic divergence $\log (M/m_{\rm IR})$.  
Alternatively, given a measurement of such couplings, we constrain the mass $M$ of new states
and their couplings to gravitons.
The task of bounding their couplings to Standard Model fields is left to future work. 

The operational definition of ``gravity'' in this paper is a force which grows linearly with energy at high speeds,
corresponding in particle physics language to exchange of a spin-2 particle.
We stress that static long-range forces,
which could come from direct interactions between matter
and light spin-0 or spin-1 particles (also sometimes called fifth forces), are unconstrained by our arguments.
 
The infrared logarithmic divergence in our bounds is related to the divergence of the eikonal phase in four dimensions. In the context of scalar scattering in AdS/CFT, this logarithmic divergence gets regulated by the AdS curvature scale, yielding rigorous, finite bounds proportional to $\log M R_\mathrm{AdS}$ \cite{Caron-Huot:2021enk}. We expect the same mechanism to apply to graviton scattering as well. Thus, our bounds can be interpreted as finite bounds on gravitational Wilson coefficients in AdS$_4$. 
The key feature of our bounds is the absence of power-law infrared divergences;
eventually, we hope that infrared logarithms can be removed by studying suitable IR finite observables.
In our view, an incredibly conservative assumption would be to replace $m_\mathrm{IR}$ with the Hubble scale, which for $M\sim 1\ \mathrm{TeV}$, multiplies our bounds by a factor of only $\log M/m_\mathrm{IR}\sim 100$. Note that this would still strongly rule out a modification of GR that simultaneously satisfies (1), (2), and (3) above.

This paper is organized as follows. In section \ref{sec:review sum rules}, we state our assumed axioms encoding causality of graviton scattering, review dispersive sum rules, and some of their known implications, notably positivity bounds from forward limits.
In section \ref{sec:example functional}, we review the impact parameter approach of \cite{Caron-Huot:2021rmr} and provide example positive functionals which prove upper bounds on gravitational EFT coefficients. We explain why the bounds are only weakly affected by possible  light matter fields. We then describe our numerical strategy to systematically search for optimal bounds, and how CEMZ-like bounds are automatically included.
In section \ref{sec:results}, we report the bounds obtained with this method, comment on their relations with known theories, and speculate about the possibility of modifications associated with a low scale $M$.
We summarize in section \ref{sec:conclusion}. In appendix \ref{app:Wigner}, we review partial waves for graviton scattering amplitudes. In appendix \ref{app:kk} we present graviton amplitudes from light exchanges of spin-0 and spin-2 particles.
Finally, we record our numerical set-ups in appendix \ref{app: numerics}.
\section{Review: Dispersive sum rules}\label{sec:review sum rules}

In this section, we describe our key physical assumptions and tools:
effective field theory at low energies (section \ref{sec:low}), unitarity (section \ref{sec:unitarity}),
causality (section \ref{sec2: Regge bound}), and Kramers-Kronig-type dispersion relations and some of their consequences (section \ref{sec2: dispersive sum rule}).
Most of this material is standard, except perhaps
for our discussion of Regge boundedness  in section \ref{sec2: Regge bound}, using compact-support wavefunctions.

Let us first briefly motivate our assumptions, before we state them technically.

By unitarity, we mean that initial and final states of scattering processes are elements of positive-definite Hilbert spaces,
whose norms are conserved by time evolution. This represents the idea that the probabilities of all possible events
are positive and add up to one.  The reason we assume this is obvious: we wouldn't know how to
interpret negative probabilities.

By causality, we refer to the notion that ``signals can't travel faster than light''.
We assume causality because of its tremendous explanatory power and past successes:
by forbidding instantaneous action at a distance,
it qualitatively explains why electromagnetic and gravitational waves \emph{must} exist, why
forces in nature are mediated by particles, why antiparticles exist, how their interactions are quantitatively
related \cite{Weinberg:1995mt}, and so much more. 
Abandoning causality without a good replacement principle seems akin to opening Pandora's box.\footnote{If one were not worried about instantaneous action at a distance, any many-body Hamiltonian such as
\be\nonumber
 H= \sum_i \frac{\vec{p}_i{}^2}{2m_i} +\sum_{i<j} V_{ij}
\ee
with potential $V_{ij} =\frac{-Gm_im_j}{|\vec{x}_i-\vec{x}_j|}$, or others, would trivially define a ``quantum theory of gravity''.
}

There is an interesting interplay between unitarity and causality, as displayed by quantum fields with wrong-sign kinetic terms,
sometimes called ``ghosts''.
In one quantization, positive-frequency modes propagate forward in time but have negative norms.
In an alternative quantization choice, norms are fine but positive-frequency modes propagate backward in time.
Lee and Wick famously proposed that the ensuing acausality could be made unobservably small if one treats
backward-moving modes as resonances which decay to normal forward-moving modes \cite{Lee:1969fy},
as is indeed seen at low orders in perturbation theory \cite{Grinstein:2008bg,Donoghue:2019fcb}.
A problem is that, as soon as interactions with normal matter are included,
negative-frequency modes make the vacuum unstable against resonant particle production.
This instability can be nicely discussed in connection with a classical theorem by Ostrogradsky \cite{Woodard:2015zca};
it seems incompatible with a long-lived Universe \cite{Cline:2003gs}.\footnote{As was pointed out with much deference to the original authors, the Lee-Wick prescription is ambiguous
and ``an additional prescription would be needed to completely define the theory'' \cite{Cutkosky:1969fq}.
In our view, the Lee-Wick idea fails to address the vacuum stability issue for the simple reason that
the timescales in the relevant vacuum diagram are shorter than the decay time of Lee-Wick quanta.}

In short, we assume unitarity and causality because \emph{we see no alternatives}.
It is possible that Nature does not conform to these principles as we understand them,
but  resulting bounds can be viewed as tests of these principles.


\subsection{Helicity amplitudes and low-energy EFT} \label{sec:low}

Four-dimensional gravitons possess two helicity states. A complete set of independent amplitudes for graviton-graviton scattering is
\begin{align}
{\cal M}(1^+ 2^- 3^- 4^+) &=   \langle 23\rangle^4 [14]^4\, f(s,u)\,,\\
 \mathcal{M}(1^{+}2^{+}3^{+}4^{-}) &= ([12][13]\langle 14\rangle)^4 \,g(s,u)\,,\\
 \mathcal{M}(1^{+}2^{+}3^{+}4^{+}) &= \fft{[12]^2 [34]^2}{\langle 12\rangle^2 \langle 34\rangle^2}\,
h(s,u)\,,
\label{eq:Mdef}
\end{align}
and permutations thereof.
At tree-level in Einstein's theory, only the maximal-helicity-violating (MHV) amplitude $f(s,u)$ is nonvanishing,
and a large fraction of our results will be derived by studying only this amplitude.
Here, we use spinor-helicity variables (see \cite{Elvang:2013cua}) and we have introduced the Mandelstam invariants
\begin{equation}
        s = - (p_1+ p_2)^2\,, \quad t = - (p_2+ p_3)^2\,, \quad u = - (p_1+ p_3)^2\,,
\end{equation}
with $s+t+u=0$.
The functions $f(s,u)$, $g(s,u)$, and $h(s,u)$ are analytic in the upper-half plane with ${\rm Im}\,s>0$, and crossing symmetric:
\bea
f(s,u) = f(u,s)\,,\quad g(s,u)= g(t,s)=g(u,t)\,,\quad h(s,u)=h(t,s)=h(u,t)\,. \label{crossing}
\eea
Other helicity amplitudes may be obtained by complex conjugation and Schwarz reflection,
for example the ${-}{-}{-}{+}$ amplitude is $\bar{g}(s,u)=(g(s^*,u^*))^*$. 

At low energies, we assume a spectrum comprising massless gravitons
together with possible light particles of spin $\leq 2$. 
These can be described by an effective field theory (EFT) 
of the generic form \eqref{eq: schematically S}, with higher derivative terms 
encoding modifications to Einstein gravity generated by physics above the EFT cutoff $M$.
We assume $M\ll \Mpl$ and thus neglect loops within the EFT.
States with spin greater than two are genuinely gravitational, and assumed to have mass above the cutoff, $m>M$.

By contrast, particles of spin two or less and mass $\mlight \ll M$ can be interpreted as additional states in the Standard Model of particle physics and its extensions, or states arising from Kaluza-Klein reduction of massless gravity in higher-dimensions.
Angular momentum conservation forbids the decay of a state of half-integer or odd spin to a pair of gravitons,
so only matter fields of spins 0 and 2 can affect graviton scattering at tree level.
We will refer to both as ``matter'', even though this nomenclature is slightly unconventional for spin 2 fields.

\begin{figure}[t]
\centering
\includegraphics[width=.8\linewidth]{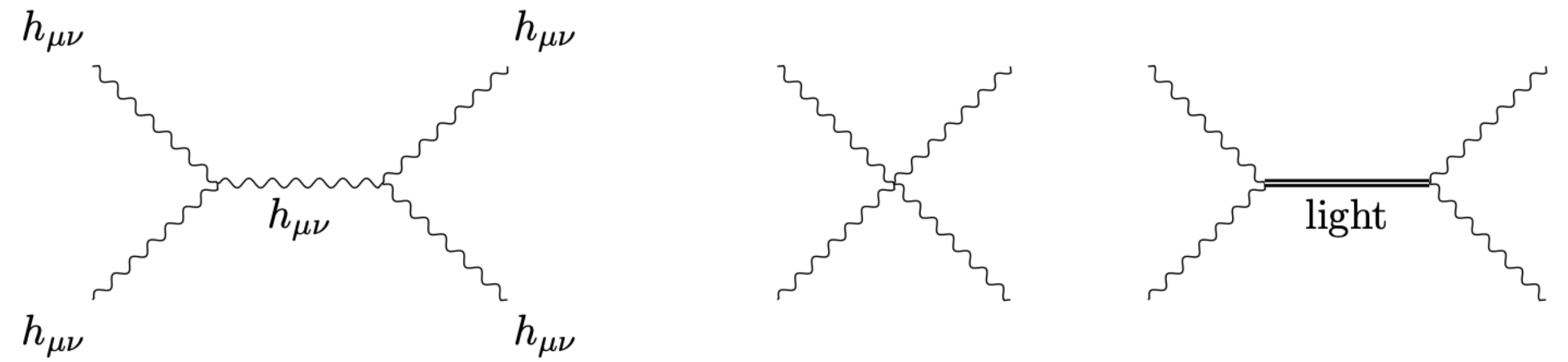} 
\caption{$2$-to-$2$ scattering amplitudes of gravitons within the low-energy effective theory.
We include at tree-level both the graviton exchange and (higher-derivative) contact diagrams,
as well as exchanges of possible light spin-0 and spin-2 particles.
Other spins are forbidden by angular momentum conservation.
}
\end{figure}

The best way to enumerate EFT couplings is to list how they modify graviton scattering amplitudes.
On-shell three-particle vertices are determined by Lorentz invariance up to overall parameters
\be
\mathcal{M}(1^+,2^+,3^-) = \sqrt{8\pi G}\fft{[12]^6}{[13]^2[23]^2}\,,\quad \mathcal{M}(1^+,2^+,3^+)=\fft{\hatg_3}{2} \sqrt{8\pi G} ([12][13][23])^2\,.
\ee
The tree-level four-particle amplitudes \eqref{eq:Mdef} may then be written in terms of exchange diagrams,
plus a sum of contact interactions, which are simply polynomials with the symmetry \eqref{crossing}:
\begin{subequations}
\label{eq:fghlow}
\begin{align}
  f_{\rm low}(s,u) =&
        \frac{8\pi G}{s t u}  +  \frac{2\pi G su}{t}|\hatg_3|^2  +  g_{4} + g_{5} t +   g_{6}\, t^2 -g_{6}' \, s u + \ldots+f_{\rm matter}(s,u)
        \nonumber\\&+O({\rm loops})\,,\label{eq:flow} \\
g_{\rm low}(s,u) =&\ \frac{4\pi G}{stu} \hatg_3+ \fft{1}{2}\hatg^{\prime\prime}_6+ \ldots+g_{\rm matter}(s,u)+O({\rm loops})\,,\\
 h_{\rm low}(s,u) =&\ 40\pi G \hatg_3 s t u + \frac12 \hatg_4(s^2+t^2+u^2)^2 + 2\hatg_5 s t u (s^2+t^2+u^2)
\nonumber\\ &+ \hatg_6(s^2+t^2+u^2)^3 + \hatg_6^{\prime} s^2t^2u^2 + 
\ldots+ h_{\rm matter}(s,u)+O({\rm loops})\,,    
\end{align}
\end{subequations}
where hatted couplings are complex (the real and imaginary part representing parity-even and parity-odd couplings, respectively).
The subscript ``low" emphasizes that this expansion is used only for $|s| < M^2$.  
The signs on the first line have been chosen so that our couplings relate simply to
those in \cite{Bern:2021ppb}.\footnote{The conversion is simply:
\be \{g_4,\ g_5,\ g_6,\ g_6'\}^{\rm here} = \{a_0,\ a_1,\ a_{2,0},\ a_{2,1}\}^{\rm there}.
\ee
In our notation the subscript always denotes half the number of derivatives in the contact interaction.}
The matter contributions $f_\mathrm{matter}(s,u)$, $g_\mathrm{matter}(s,u)$, and $h_\mathrm{matter}(s,u)$ are recorded in appendix~\ref{app:kk}.


It is straightforward to write down Lagrangians that give rise to the above amplitudes.
Before doing so, it is important to note that Lagrangian densities
are only defined modulo field redefinitions (which change contact interactions
by equation of motions) and total derivatives. In particular, any
higher-derivative term involving the Ricci tensor $R_{\mu\nu}$ or scalar $R$ is
removable, so only powers of the Riemann curvature $R_{\mu\nu\sigma\rho}$ must be
 kept.\footnote{It is well-known for example that $f(R)$ gravity is
        equivalent to standard Einstein gravity minimally coupled to a scalar
field with a specific potential.  From our perspective, $f(R)$ gravity thus
does not constitute a higher-derivative correction to Einstein's gravity. Instead, it is
a specific choice of matter sector.}
Furthermore, numerous identities relate various
contractions of Riemann tensors and derivatives. This is the reason why we do not include $R^2$: $R^2$-terms can be recast into the Gauss-Bonnet term, which is topological in $d=4$. In contrast, the amplitudes
\eqref{eq:fghlow} are unambiguous. 

With this being said, it is straightforward to list a minimal set of irreducible higher-dimension operators
and map them to the amplitudes \eqref{eq:fghlow} by computing the resulting tree-level amplitudes.  For example, the parity-even sector of cubic gravity contains $10$ different operators, but field redefinitions and various identities leave us with only one independent operators \cite{Bueno:2019ltp}.
Up to dimension eight, our effective action is
\bea
 S &=& \frac{1}{16\pi G} \int  d^4x \sqrt{-g} \Big[
 R
 -\fft{1}{3!} \left(\alpha_3 R^{(3)} + \tilde{\alpha}_3 \tilde{R}^{(3)}\right) 
         \cr && \hspace{2cm}
+ \frac14\left(\alpha_4 (R^{(2)})^2 +\alpha_4' (\tilde{R}^{(2)})^2 + 2 \tilde\alpha_4 R^{(2)}\tilde{R}^{(2)}\right)+\ldots \Big]
 + S_{\rm matter}\,, \label{eq: full S to R4}
\eea
where we defined
\begin{equation}
\begin{aligned}
&R^{(2)} = R_{\mu\nu\rho\sigma}R^{\mu\nu\rho\sigma}\,,\quad &&\tilde{R}^{(2)}=R_{\mu\nu\rho\sigma}\tilde{R}^{\mu\nu\rho\sigma}\,,\quad && \tilde{R}_{\mu\nu\rho\sigma} \equiv \tfrac12\epsilon_{\mu\nu}\,
^{\alpha\beta}R_{\alpha\beta\rho\sigma}\,,
\cr
&R^{(3)}=R_{\mu\nu}\,^{\rho\sigma}R_{\rho\sigma}\,^{\alpha\beta}
R_{\alpha\beta}\,^{\mu\nu}\,, \quad
&&\tilde{R}^{(3)}=R_{\mu\nu}\,^{\rho\sigma}R_{\rho\sigma}\,^{\alpha\beta}
\tilde{R}_{\alpha\beta}\,^{\mu\nu}\,. &&
\end{aligned}
\end{equation}
It is then straightforward to expand $g_{\mu\nu}=\eta_{\mu\nu}+\sqrt{32\pi G}h_{\mu\nu}$ and apply the standard Feynman techniques to evaluate scattering amplitudes and compare with eqs.~\eqref{eq:fghlow}:
\be
\hatg_3 = \alpha_3+i\tilde{\alpha}_3,\qquad
g_4=8\pi G(\alpha_4 +\alpha^\prime_4)\,,\qquad \hat{g}_4=8\pi G(\alpha_4-\alpha^\prime_4 + i \tilde{\alpha}_4)\,.\label{eq: g4hat Lagrangian}
\ee
Note that we absorbed a factor of $8\pi G$ in three-point couplings but not in four-point couplings. 
   
\subsection{High energies:  partial waves and unitarity} \label{sec:unitarity}

We will assume that graviton scattering remains sensible even at
center-of-mass energies that exceed the EFT cutoff $M$
(where the parametrization \eqref{eq:fghlow} no longer applies). 
Our minimal assumptions are that the amplitude remains causal (that is, analytic) and unitary, and that the spectrum is relativistic so that it can be organized in terms of mass, $m^2$, and spin, $J$. 

In other words, the amplitude admits a partial wave expansion of the form \cite{Hebbar:2020ukp}
\begin{equation}
        {\cal M}(1^{h_1} 2^{h_2} 3^{h_3} 4^{h_4}) =  16\pi \sum_J (2J+1)\, a_J^{\{h\}}(s)\, d^J_{h_{12},h_{34}}\!\left(1+\tfrac{2t}{s}\right) \,.\label{eq: partial wave expansion}
\end{equation}
Here, $d_{\alpha,\beta}^J(x)$ are the well-known Wigner-$D$ functions, which are explicitly written in appendix \ref{app:Wigner}, and $h_{ij} = h_i-h_j$.  The partial wave coefficients, $a_J^{\{h\}}(s)$, encode all dynamical information.
Without boost invariance, partial waves would be more complicated,
but study of causality constraints has been initiated in \cite{Pajer:2020wnj,Grall:2021xxm}.

Unitarity of the S-matrix $S=1+i{\cal M}$ imposes crucial positivity properties on the ``absorptive part'' of the amplitude,
through the familiar relation: $i({\cal M}^\dagger-{\cal M})={\cal M}^\dagger {\cal M}$.
The matrix structure will be important and is in contrast to the scalar case studied in \cite{deRham:2017avq,Caron-Huot:2020cmc,Caron-Huot:2021rmr,Tolley:2020gtv,Arkani-Hamed:2020blm,Chiang:2021ziz,Bellazzini:2020cot}.
In terms of partial waves,
\be
 i\left[\left( a_J^{-h_4,-h_3,-h_2,-h_1}(s)\right)^*-a_J^{h_1,h_2,h_3,h_4}(s) \right] =
 \sum_X \left(a_J^{-h_3,-h_4\to X}(s)\right)^*a_J^{h_1,h_2\to X}(s)\,,
\ee
where $X$ runs over intermediate states.  The right-hand-side is a positive semi-definite matrix.
To capture its positive properties we adopt an abbreviated notation from \cite{Du:2021byy}
and omit the $X$ sum, writing the right-hand-side simply as $2(c^{-h_4,-h_3})^*c^{h_1h_2}$.
Specializing to the MHV amplitude, we have
\be
  {\rm Im}\ a_J^{+-+-}(s) = |c_{J,s}^{+-}|^2,\qquad
  {\rm Im}\ a_J^{++--}(s) = |c_{J,s}^{++}|^2\,, \label{Im a from c}
\ee
where $c^{+-}_{J,s}$ is real. In particular, the quantities in (\ref{Im a from c}) are positive.\footnote{Even though $c_{J,s}^{+-}$ is real, we nonetheless write the absolute value sign $|c_{J,s}^{+-}|^2$ throughout to emphasize that its square is positive.} Note this would not be so for the permutation $a_J^{+--+}=(-1)^J a_J^{+-+-}$.
Partial waves admit two-sided bounds, which follow from applying 
the same argument to the unitary matrix $-S=1+i(2i-{\cal M})$. In particular, we have\footnote{In more detail: In the even spin sector, there are three incoming states for the helicities $|h_1h_2\>$: namely $|++\>$, $|--\>$, and $\frac 1 {\sqrt 2} (|+-\> + |-+\>)$. The corresponding diagonal elements of the $S$-matrix are $1+i a^{--++}$, $1+i a^{++--}$, and $1+2i a^{+-+-}$. In the odd-spin sector, the $S$-matrix is a $1\x 1$ matrix with element $1+2i a^{+-+-}$. By unitarity, each of these diagonal elements must have real part in $[-1,1]$, which leads to (\ref{a two-sided bound}).}
\be
\label{a two-sided bound}
 0 \leq \Im a^{+-+-}_{J}(s) \leq 1,\quad \textrm{and} \quad 0 \leq \Im a^{++--}_{J}(s) \leq 2\,.
\ee 
Positivity of the spectral density is key for establishing bounds on the low-energy EFT Wilson coefficients \cite{deRham:2017avq,Caron-Huot:2020cmc,Caron-Huot:2021rmr,Tolley:2020gtv,Arkani-Hamed:2020blm,Chiang:2021ziz}.

Explicitly, the MHV amplitude $f(s,u)$ has distinct discontinuities in the $s$- and $t$- channels:
\begin{subequations}
\begin{align}
 s=m^2>0:&&\quad {\rm Im}\, f(m^2,-p^2)&= \frac{16\pi}{m^8}\sum_{J\geq 4} (2J+1)|c^{+-}_{J,m^2}|^2 \, \tilde{d}_{4,4}^J\left(1-\tfrac{2p^2}{m^2}\right)\,,\\
t=m^2>0:&&\quad {\rm Im}\, f(-p^2,p^2-m^2) &= \frac{16\pi}{m^8}\sum_{\substack{J\geq 0 \\ \rm even}}( 2J+1)
|c^{++}_{J,m^2}|^2  
\tilde{d}_{0,0}^J\left(1-\tfrac{2p^2}{m^2}\right)\,, \label{f: tpole}
\end{align}
\label{eq:Imf}
\end{subequations}
where $\tilde{d}_{\alpha,\beta}^J$ are Wigner-$D$ functions with stripped helicity factors, see appendix \ref{app:Wigner} for more details.
The overall $m^{-8}$ originates from the prefactor in \eqref{eq:Mdef}.

For other helicity configurations, we have similar relations,
except that the ``imaginary part'' gets replaced by the discontinuity $\widetilde{\rm Im}\ a\equiv [a(s+i0)-a(s-i0)]/(2i)$,
and the right-hand-sides are now complex numbers:
\be
  \widetilde{\rm Im}\ a_J^{+++-}(s) = c_{J,s}^{++}c_{J,s}^{+-},\qquad
  \widetilde{\rm Im}\ a_J^{++++}(s) = (c_{J,s}^{++})^2\,. \label{Im a from c other}
\ee
The corresponding partial wave expansions are
\begin{align}
& \widetilde{\rm Im}\ g(m^2,-p^2)\Big|_{s=m^2} = \frac{16\pi}{m^{12}}\sum_{\substack{J\geq 4 \\ \rm even}}(2J+1)c_{J,s}^{++}c_{J,s}^{+-}\, \tilde{d}_{4,0}^J\left(1-\tfrac{2p^2}{m^2}\right)\,,\\
& \widetilde{\rm Im}\ h(m^2,-p^2)\Big|_{s=m^2} = 16\pi\sum_{\substack{J\geq 0 \\ \rm even}} (2J+1)(c_{J,s}^{++})^2 \, \tilde{d}_{0,0}^J\left(1-\tfrac{2p^2}{m^2}\right)\,.
\end{align}

\subsection{Regge boundedness and all that}
\label{sec2: Regge bound}


Low and high energies are related by Kramers-Kronig-type dispersion relations. It will be crucial that we can
predict beforehand which dispersion relations converge.

Typically one assumes a Froissart-Martin-like bound at fixed momentum transfer and large complex energies:
\begin{equation}
        \lim_{|s|\to \infty}{\cal M}/s^2 \to 0  \qquad \text{at fixed} \qquad t<0 \qquad\mbox{(\emph{not}
        what we'll assume)}\,.
        \label{eq:reggeM not}
\end{equation}
For example, in tree-level string theory, $\mathcal{M}\sim s^{2+\alpha't}< s^2$.
However the validity of this bound is not generally established in an abstract theory of quantum gravity.
Martin's original proof of the Froissart-Martin bound in axiomatic field theory \cite{Martin:1962rt}
does not apply to gravity, due to the absence of a mass gap. For holographic theories it has been argued that the behavior \eqref{eq:reggeM not} holds for physical kinematics as a consequence of the chaos bound \cite{Maldacena:2015waa,Chandorkar:2021viw}.

This difficulty is a physical one and not merely technical:
to bound amplitudes at large complex energies, one must generally combine analyticity with some
boundedness property on the real axis, as we do shortly.
The difficulty is that analyticity holds at fixed momentum, while boundedness
holds at fixed impact-parameter; these two spaces are related by a Fourier transform which is not easy to control.
Namely, it is not straightforward to estimate large-impact-parameter contributions
in the absence of a mass gap or of an explicit model of the dynamics.
Thankfully, large-impact-parameter physics however seems immaterial
for bounding EFT couplings at the scale $M$.
The intuition, stressed in \cite{Tolley:2020gtv,Caron-Huot:2020cmc},
is that  EFT parameters at the scale $M$ satisfy sum rules saturated by impact parameters $b\sim M^{-1}$.

Let us explain how we sidestep \eqref{eq:reggeM not}
by adapting a recent method from \cite{Caron-Huot:2021enk}, which showed that the conclusions from flat space
sum rules apply to quantum gravity in AdS (defined as a CFT with large but finite central charges and single-trace gap).  The method is simple: we integrate scattering amplitudes against wavepackets that have finite support in momentum space and decay rapidly at large impact parameters $b$.  Formally, for a wavefunction $\Psi(p)$, we define the smeared amplitude:
\be
 {\cal M}_\Psi(\tilde{s}) = \int_0^M dp \Psi(p) {\cal M}(\tilde{s}+\tfrac12p^2,-p^2)\,. \label{def smearing}
\ee
It is apparent that for $|\tilde{s}|>\frac12M^2$, all amplitudes on the right-hand-side
are in the physical region where the partial wave expansion \eqref{eq: partial wave expansion} applies.
(The offset of $s$ by $\frac12p^2$ is not essential but ensures that
$s\leftrightarrow u$ crossing symmetry is simply reflection of $\tilde{s}$.)
Furthermore, thanks to compactness of the integral, ${\cal M}_\Psi(\tilde{s})$ inherits
the analyticity properties of the original amplitude: our fundamental assumption is that 
a crossing path exists which connects the two points $\tilde{s}=\pm\frac12M^2$, and that the amplitude is analytic outside of that arc.

Fast decay in $b$ requires $\Psi(p)$ to be smooth and
to vanish rapidly enough at the endpoints; the precise condition is detailed below (see \eqref{eq: condition for wave function}).
The upshot is that if the decay sets in at some $b>b_{*}$,
then the spin sum in \eqref{eq: partial wave expansion} is effectively limited to $J\leq \sqrt{s}b_{*}$.
Since individual $a_J$ are bounded (see \eqref{a two-sided bound}), one trivially gets the bound
\be
 |{\cal M}_\Psi(s)| \leq s\times \mbox{constant}\qquad (|s|>\tfrac12M^2, \mbox{ real})\,.
\ee
We thus have an analytic function which is bounded on the real axis.
Unless this function grows exponentially at complex energies
(which would imply blatant time advances when Fourier transformed to the time domain, a behavior which
was not seen in theories of quantum gravity in AdS realized by unitary CFTs \cite{Caron-Huot:2021enk}),
it must be bounded in all complex directions by a version of the maximum principle called
Phragm\'en-Lindel\"of principle (see \cite{Maldacena:2015waa}): 
\be  \label{eq:reggeM}
 |{\cal M}_\Psi(s)| \leq |s|\times \mbox{constant} \qquad (\mbox{complex $s$ outside $s\sim M^2$ arc})\,.
\ee
The results presented in this paper rely only on the above properties of
smeared amplitudes ${\cal M}_\Psi$, and not on \eqref{eq:reggeM not}. (In fact, we will only use that $\lim_{|s|\to \oo}|\cM_\Psi(s)|/|s|^2=0$, which is easily implied by (\ref{eq:reggeM}).)
We believe that these are conservative assumptions directly traceable to causality and unitarity.

We stress that causality is stronger than ``particles cannot move faster than light''. Since particles are waves,
the notion that ``signals'' cannot travel faster than light is taken to mean that (asymptotic) measurements at space-like separated points $A$ and $B$ commute. This relates amplitudes for a particle moving from $A$ to $B$ to  an antiparticle moving the other way: causality is entwined with crossing symmetry \cite{Gell-Mann:1954ttj}.
While this physical picture is compelling, we should note that crossing symmetry and analyticity are nontrivial
to prove mathematically (for recent discussions see \cite{DeLacroix:2018arq,Mizera:2021fap}). 
The use of standard S-matrix axioms in the context of quantum gravity is supported by
the recent work \cite{Caron-Huot:2021enk}, which showed
that well-established CFT axioms imply that graviton scattering in AdS space satisfy dispersion relations.

The implications of \eqref{eq:reggeM} depend on the helicity of scattered particles.
Recalling that for physical kinematics $\langle ij \rangle = \pm [ij]^*$, and
\begin{equation}
        |\langle 12 \rangle| = |\langle 34 \rangle| = \sqrt{|s|}\,, \qquad |\langle 23 \rangle| =|\langle 14 \rangle| = \sqrt{|t|}\,, \qquad |\langle 13 \rangle| =|\langle 24 \rangle| = \sqrt{|u|}\,,
\end{equation}
we find that for the component amplitudes $f$, $g$ and $h$ from \eqref{eq:Mdef}, the condition \eqref{eq:reggeM} yields:
\bea
&& \lim_{|s|\to \infty}f(s,-p^2) \leq Cs^{-3}\,, \qquad \lim_{|s|\to \infty}f(s,p^2-s) \leq Cs\,, 
\cr &&\lim_{|s|\to \infty}g(s,-p^2) \leq Cs^{-3}\,, \qquad \lim_{|s|\to \infty}h(s,-p^2) \leq Cs\,, \quad
\mbox{(after smearing in $p$)}\,.
\label{eq:reggef}
\eea
The first line gives respectively the fixed-$u$ and fixed-$t$ Regge limits of the MHV amplitude.
Notice that certain limits enjoy improved behavior $\sim s^{-3}$:
when amplitudes are normalized so that contact interactions are polynomial (see eq.~\eqref{eq:fghlow}),
they vanish in some high-energy limits.
This phenomenon is known as superconvergence
and is the main reason why we will find stronger constraints on graviton contact interactions than for scalars.\footnote{In general, superconvergence occurs in scattering of particles of spins $J_1$ and $J_2$
whenever $J_1+J_2-1>J_0$, where $J_0$ is the Regge intercept of the theory, see e.g.\ \cite{Fu:2013cza,Kologlu:2019bco}. The bound \eqref{eq:reggeM} amounts to $J_0\leq 1$ but all we
ultimately use in this paper is $J_0<2$.}

Superconvergence is also related to the observation of \cite{Chowdhury:2019kaq} that a very limited
number of graviton contact interactions obey (or more precisely, saturate) the classical bound \eqref{eq:reggeM not}.
Although it is simpler to prove, the bound \eqref{eq:reggeM} is stronger and is not satisfied by any individual graviton contact interaction.

\subsection{Dispersive sum rules}\label{sec2: dispersive sum rule}

We are now ready to write dispersive sum rules for the amplitudes $f(s,u)$, $g(s,u)$, and $h(s,u)$.
We begin with the MHV amplitude $f$.
From the behavior \eqref{eq:reggef}, we get two types of constraints: from fixed-$u$ and fixed-$t$.
For fixed-$u$ we can separate $f$ into combinations that are even/odd under $s{\leftrightarrow}t$
and obtain the following basis of sum rules, for integer $k$:
\begin{subequations}
\label{defB}
\begin{align}
     \!   B^{(1)}_{k}(p^2) &= \oint_{\infty} \frac{ds}{4\pi i} \frac{2s-p^2}{[s(s-p^2)]^{\frac{k-2}{2}}} \, \Big[
        f(s,-p^2)+ f(p^2-s,-p^2)\Big] = 0  & (k\geq 2 \; \text{even})\,, \\
      \!  B^{(1)}_{k}(p^2) &= \oint_{\infty} \frac{ds}{4\pi i} \frac{1}{[s(s-p^2)]^{\frac{k-3}{2}}} \Big[-f(s,-p^2)+f(p^2-s,-p^2)\Big] = 0 & (k\geq 3 \; \text{odd}) \,,
\end{align}
\end{subequations}
where the integrals are along a large circle at infinity.
We additionally have three fixed-$t$ dispersion relations, which also integrate to zero for $k\geq 2$ even:
\begin{align}
\Big\{ B^{(2)}_{k},B^{(3)}_{k},B^{(4)}_{k}\Big\}(p^2) &=
\oint_{\infty} \frac{ds}{4\pi i}(2s-p^2) \left\{
\frac{f(s,p^2{-}s)}{[s(s-p^2)]^{\frac{k+2}{2}}},
\frac{g(s,p^2{-}s)}{[s(s-p^2)]^{\frac{k-2}{2}}},
\frac{h(s,p^2{-}s)}{[s(s-p^2)]^{\frac{k+2}{2}}}
\right\}\,.
\end{align}
To avoid confusion between different channels, we always write the fixed momentum transfer as $p$.
These sum rules become strictly valid after the $p$-dependence is integrated against appropriate wavepackets as
in eq.~\eqref{def smearing}.

The subscript $k$ indicates the Regge spin of a sum rule.
This concept is closely related, but distinct, from the ``number of subtractions'' or power of $1/s$ inserted to improve high-energy convergence.
For example, $B^{(1)}_2$ has fewer subtractions than $B^{(2)}_2$
(and is even ``anti-subtracted'' since it has no denominator!), yet they possess the same convergence properties.
The nomenclature is motivated by the fact that exchange of a single $t$-channel particle of spin $J$ yields an amplitude that grows like $\mathcal{M}\sim s^J$: we say that a sum rule has spin $k$ if it converges on exchanges with $J<k$ (and marginally diverges on spin $k$).

Regge spin is more important than subtraction-counting because the Regge growth \eqref{eq:reggeM} translates into
the simple convergence criterion $k> 1$.
This is the same criterion as convergence of the Froissart-Gribov formula which extracts partial waves of spin $J>1$,
or of the analogous Lorentzian inversion formula \cite{Caron-Huot:2017vep,Simmons-Duffin:2017nub,Kravchuk:2018htv} which extracts CFT data for spin $J>1$. 

\begin{figure}[t]
\centering 
\begin{tikzpicture}[decoration={markings, 
    mark= at position 0.52 with {\arrow{stealth}}}]
	\draw (-3,0) -- (3,0);
    \draw[thick, rotate=3, postaction={decorate}] (-3,0) arc (-180:-6:3) ;
    \draw[thick, rotate=3, postaction={decorate}] (3,0) arc (0:174:3) ;
    \draw[red, decorate, decoration={snake=zigzag,segment length=1.5mm, amplitude=0.5mm}]       (-3,0) -- (-1.1,0);
    \fill (-1.1,0) node[red, cross=2pt] {};
	\draw[red, decorate, decoration={snake=zigzag,segment length=1.5mm, amplitude=0.5mm}]       (1.5,0) -- (3,0);
    \fill (0,0) node[cross=2pt] {};
	\draw (0.4,0) node[yshift=0.4cm] {\scriptsize $-t$};
    \fill (0.4,0) node[cross=2pt] {};
    \fill (1.5,0) node[cross=2pt] {};
	\node (a) at (2.75,3) {$s$};
	\draw (a.north west) -- (a.south west) -- (a.south east);
	\draw (0,0) node[yshift=0.4cm] {\scriptsize $0$};
	\draw (1.5,0) node[xshift=0.1cm,yshift=0.4cm] {\scriptsize $M^2$};
	\draw (-1.5,0) node[yshift=-0.4cm,xshift=0.2cm] {\scriptsize $-M^2 -t$};
	\end{tikzpicture}
    \raisebox{86pt}{$\quad\longrightarrow\quad$}
\begin{tikzpicture}
	\draw (-3,0) -- (3,0);
    \draw[white, thick, rotate=-3, postaction={decorate}] (3,0) arc (0:-174:3) ;
    \draw[white, thick, rotate=3, postaction={decorate}] (3,0) arc (0:174:3) ;
    \draw[thick, rotate=7, postaction={decorate}, decoration={markings, 
        mark= at position 0.52 with {\arrow{stealth}}}] (-1.105,-0.05) arc (-180:-14:1.3) ;
    \draw[thick, rotate=7, postaction={decorate}, decoration={markings, 
        mark= at position 0.52 with {\arrow{stealth}}}] (1.5,0) arc (0:166:1.3) ;
	\draw[thick, thick, postaction={decorate}, decoration={markings, 
            mark= at position 0.5 with {\arrow{stealth}}}] (1.5,-0.2) --(3,-0.2) ;
	\draw[thick, thick, postaction={decorate}, decoration={markings, 
            mark= at position 0.5 with {\arrow{stealth}}}] (-3,-0.2) --(-1.1,-0.2) ;
    \draw[thick, postaction={decorate}, decoration={markings, 
            mark= at position 0.5 with {\arrow{stealth}}}] (3,0.2) --(1.5,0.2) ;
	\draw[thick, thick, postaction={decorate}, decoration={markings, 
            mark= at position 0.5 with {\arrow{stealth}}}] (-1.1,0.2) --(-3,0.2) ;
    \draw[red, decorate, decoration={snake=zigzag,segment length=1.5mm, amplitude=0.5mm}]       (-3,0) -- (-1.1,0);
    \fill (-1.1,0) node[cross=2pt] {};
	\draw[red, decorate, decoration={snake=zigzag,segment length=1.5mm, amplitude=0.5mm}]       (1.5,0) -- (3,0);
    \fill (0,0) node[cross=2pt] {};
	\draw (0.4,0) node[yshift=0.3cm] {\scriptsize $-t$};
    \fill (0.4,0) node[cross=2pt] {};
    \fill (1.5,0) node[cross=2pt] {};
	\draw (1.5,0) node[xshift=0.3cm,yshift=0.4cm] {\scriptsize $M^2$};
	\node (a) at (2.75,3) {$s$};
	\draw (a.north west) -- (a.south west) -- (a.south east);
	\draw (0,0) node[yshift=0.3cm] {\scriptsize $0$};
	\draw (-1.5,0) node[yshift=-0.4cm,xshift=-0.3cm] {\scriptsize $-M^2-t$};
	\end{tikzpicture}
\caption{Contour deformation which gives rise to sum rules eq.~\eqref{sum rule schematic}. The final contour relates low-energy EFT data along the arcs to heavy discontinuities along the branch cuts.}\label{fig: contour}
\end{figure}
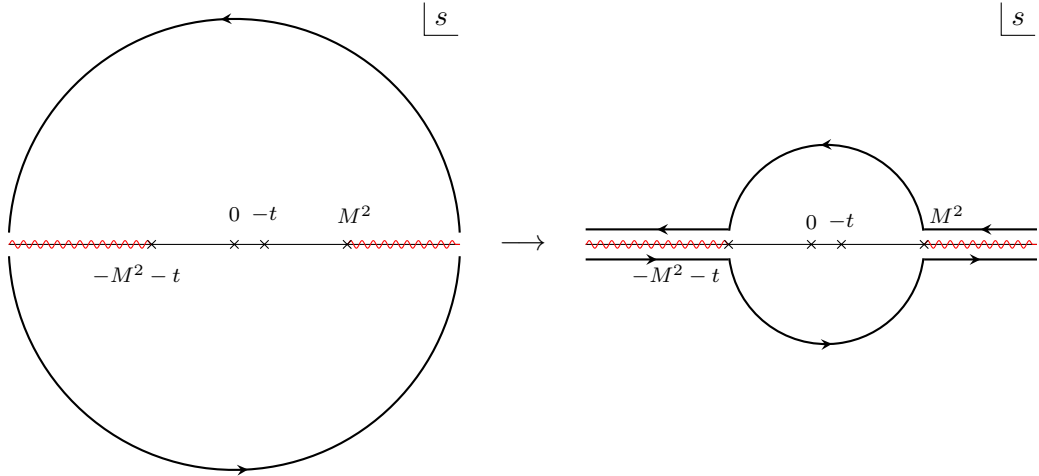

Sum rules are obtained by deforming the contour towards the real axis but avoiding the low-energy region:
the contour in figure \ref{fig: contour} relates low-energy data at the scale $M$ and heavy data above $M$:
\begin{align} \label{sum rule schematic}
 -B^{(i)}_k(p^2)\Big|_{\rm low} &= B^{(i)}_k(p^2)\Big|_{\rm high} \\
\oint_{s=M^2}^{u=M^2} \frac{ds}{4\pi i} \left(\cdots\right) &=
\int_{M^2}^\infty \frac{ds}{\pi} {\rm Im} \left(\cdots\right)\,.
\end{align}

Note that the $s$ and $u$ channel cuts contribute identically due to symmetry of eqs.~\eqref{defB},
so we included only the right cut.
(The contour on the left is in reality the union of upper and lower half-circles,
separated by the branch cut of the amplitude.)

Let us focus on the first sum rule for simplicity.
At tree-level we find only two residues, from $s=0$ and $u=0$, which contribute the same amount:
\begin{align}
-B^{(1)}_k(p^2)\Big|_{\rm low} &=
\underset{s=0}{\text{Res}} \left( \frac{2s-p^2}{[s(s-p^2)]^{\frac{k-2}{2}}} \, \big[f(s,-p^2)+ f(p^2-s,-p^2)\big]\right)
\quad \mbox{(tree-level)}\,.
\end{align}
Substituting in the low-energy amplitude \eqref{eq:flow}, only the exchange graphs contribute for $k=2,3$:
\begin{subequations}
 \label{B1 low}
 \begin{align}
-B^{(1)}_2(p^2)\Big|_{\rm low} &= \frac{16\pi G}{p^2} + 2\pi G |\hatg_3|^2p^6 +O(\mbox{matter and loops}),\\
-B^{(1)}_3(p^2)\Big|_{\rm low} &= -2\pi G|\hatg_3|^2 p^4 +O(\mbox{matter and loops}) \,.
\end{align}
\end{subequations}
The absence of contact term contributions is a hallmark of superconvergent sum rules.
Examples that probe contact interactions include:
\begin{subequations}
\begin{align}
-B^{(1)}_4(p^2)\Big|_{\rm low} &= 2g_4 + (4\pi G |\hatg_3|^2 +g_5)p^2 + (g_6+g_6')p^4 + \ldots\,,
\label{B4 low} \\
-B^{(1)}_5(p^2)\Big|_{\rm low} &= g_5 + (g_6-g_6')p^2 + \ldots\,, \\
-B^{(2)}_2(p^2)\Big|_{\rm low} &= 2\pi G|\hatg_3|^2 \frac{1}{p^2} +g_6' + \ldots\,.
\end{align}
\label{B lows}
\end{subequations}
A salient feature is that the same couplings appear in multiple sum rules: this reflects crossing symmetry.
Another feature is the appearance of the cubic self-coupling in $B_4^{(1)}$:
this is due to the rapid growth with $t$ of the $t$-channel exchange diagram with derivatives.
This rapid energy growth at zero impact parameter
will turn out to be a powerful mechanism to bound $\hatg_3$, as was proposed in section 7 of \cite{Chowdhury:2019kaq};
this mechanism is distinct from the spin-2 growth at large impact parameter that was exploited by CEMZ \cite{Camanho:2014apa}.

At high energies $s\geq M^2$, the amplitudes are beyond our knowledge. We can nevertheless evaluate the contribution to the dispersive sum rule, by inserting the partial wave decomposition
and using eq.~\eqref{eq:Imf} for even $k$:
\begin{align} \label{B1 heavy}
        B_k^{(1)}(p^2)\Big|_{\rm high} &=
          \left\langle (2m^2-p^2)
          \frac{ |c^{++}_{J,m^2}|^2\, \tilde{d}_{0,0}^J(x)+
            |c^{+-}_{J,m^2}|^2\,\tilde{d}_{4,4}^J(x)}{m^{k+4}(m^2-p^2)^{\frac{k-2}{2}}}\right\rangle \qquad\mbox{($k\geq 2$ even)}\,, 
\end{align}
where $x=1-\frac{2p^2}{m^2}$ and
for later convenience we define the heavy densities $\cal C$ and (dimensionless) averages\footnote{To avoid unnecessary clutter,
we set undefined coefficients to zero, {\it ie.} $c^{+-}=0$ when $J<4$ and $c^{++}=0$ when $J$ is odd.}
\begin{align}
      B_k(p^2)\Big|_{\rm high} =  \left\langle {\cal C}_{k,-p^2}[m^2,J]\right\rangle &= 16 \,\, \sum_{J} \,\, (2J+1) \int_{M^2}^\infty \frac{dm^2}{m^2} {\cal C}_{k,-p^2}[m^2,J] \,.
      \label{eq: average}
\end{align}

The remaining four sum rules are:
\begin{subequations}
\begin{align}
        B_k^{(1)}(p^2)\Big|_{\rm high} &=
          \left\langle
          \frac{ |c^{++}_{J,m^2}|^2\, \tilde{d}_{0,0}^J(x)-
            |c^{+-}_{J,m^2}|^2\,\tilde{d}_{4,4}^J(x)}{m^{k+3}(m^2-p^2)^{\frac{k-3}{2}}}\right\rangle \qquad\mbox{($k\geq 3$ odd)},
\\
 \Big\{B_k^{(2)},B_k^{(3)},B_k^{(4)}\Big\}(p^2)\Big|_{\rm high} &=
          \Bigg\langle(2m^2-p^2) \\
           &\hspace{-40mm}\times
          \left\{
          \frac{(-1)^J |c^{+-}_{J,m^2}|^2\,\tilde{d}_{4,-4}^J(x)}{m^{k+8}(m^2-p^2)^{\frac{k+2}{2}}},
          \frac{c_{J,m^2}^{++}c_{J,m^2}^{+-}\,\tilde{d}_{4,0}^J(x)}{m^{k+8}(m^2-p^2)^{\frac{k-2}{2}}},
           \frac{ (c^{++}_{J,m^2})^2\,\tilde{d}_{0,0}^J(x)}{m^{k}(m^2-p^2)^{\frac{k+2}{2}}}
           \right\}\Bigg\rangle \quad\mbox{($k\geq 2$ even)\,.} \nonumber
\end{align}
\label{eq: more sum rules}
\end{subequations}

In summary, the constraints of analyticity and unitarity used in this paper are
embodied in the relation \eqref{sum rule schematic},
which connects EFT couplings (eqs.~\eqref{B1 low}-\eqref{B lows})
to heavy averages like \eqref{B1 heavy} that involve a positive measure.


\subsection{Review of simple bounds from forward limits} \label{sec2: simple bounds}

Let us explore the above sum rules.
Evidently, the spin-2 sum rules \eqref{B1 low} diverge in the forward limit $p\to 0$.
This is the famous ``graviton pole'' problem, whose resolution using smeared sum rules
will be described in the next section.
However, higher-spin sum rules have smooth limits which have been extensively studied.

For the spin-4 sum rule, the forward limit of \eqref{B4 low}  and \eqref{B1 heavy} yields simply \cite{Bellazzini:2015cra,Arkani-Hamed:2021ajd}:
\be
 g_4 = \left\langle \frac{|c^{++}_{J,m^2}|^2+|c^{+-}_{J,m^2}|^2}{m^8} \right\rangle \geq 0\,.
\label{g4 lower bound}
\ee
Similarly, taking the forward limit of the same-helicity amplitude ($B^{(4)}$ sum rule) we get
\be
\hatg_4 = \left\langle \frac{(c^{++}_{J,m^2})^2}{m^8}\right\rangle\,, \label{g4 sum rule}
\ee
where we recall that the complex couplings $\hat{g}_4$ and $c^{++}_{J,m^2}$ combine parity even and odd parts.
Comparison yields the obvious bound
\be
|\hatg_4| \leq g_4\,. \label{positivity 4}
\ee
In particular this implies the two positivity constraints \cite{Bellazzini:2015cra,Arkani-Hamed:2021ajd}:
$\alpha_4, \alpha_4^\prime\geq 0$, in the Lagrangian \eqref{eq: full S to R4} and \eqref{eq: g4hat Lagrangian}.

As stressed before, forward limits can be dangerous. 
Should we trust these bounds?

From the present perspective, one can argue that the answer is: yes, as long as the size of $g_4$ is much larger than the loop effects
which cause the danger at loop level.  Let us estimate this in detail.
The one-loop amplitude diverges in the forward limit $p\to 0$ like $\mathcal{M}\sim i\pi \frac{G^2s^3}{p^2}\log p^2$. This divergence can be avoided simply by replacing the limit with evaluation at a small scale
$p\sim p_*\ll M$.
(This will spoil positivity at large $b=\frac{2J}{m}$, but if we assume that this region is controlled by known low-energy-computable eikonal physics, positivity should not be crucial there.)
The bound \eqref{g4 lower bound} becomes schematically (up to logarithms)
\be g_4\geq -\frac{G^2}{M^2p_*^2} \qquad\mbox{including loops}\,.
\ee

On the other hand, as discussed in the next section, in the presence of spin-4 particles at the scale $M$
the expected size is $g_4\sim \frac{G}{M^6}\gg \frac{G^2}{M^4}$.
By choosing appropriately $p_*$, it is easy to make the loop corrections negligible in comparison:
\be \label{forward limit correctness}
\frac{G^2}{M^2p_*^2} \ll \frac{G}{M^6}
\quad\Rightarrow\quad \frac{M^2}{\Mpl}\ll p_* \ll M\,.
\ee
This can be satisfied since we assume that the EFT cutoff is parametrically below the Planck scale.
A similar argument was described in \cite{Arkani-Hamed:2021ajd}.

By the same token we can trust the Taylor expansion around the forward limit
of other sum rules with spin $k\geq 3$.
For $k=5$ we find 
\be
 g_5 = \left\langle \frac{|c^{++}_{J,m^2}|^2-|c^{+-}_{J,m^2}|^2}{m^{10}} \right\rangle\,,\label{eq: to do g5/g4 bound}
\ee
from which the two-sided bound $-g_4\leq g_5M^2\leq g_4$ readily follows (again up to loop corrections). 
From the forward limit of $B_6^{(1)}$, we have
\be
g_6=\left\langle \frac{|c^{++}_{J,m^2}|^2+|c^{+-}_{J,m^2}|^2}{m^{12}} \right\rangle \quad\Rightarrow\quad
0\leq g_6 \leq \frac{g_4}{M^4}\,.
\label{eq: g6}
\ee
We get additional sum rules of the same scaling dimensions by considering also forward limit derivatives
of $k=4$ and $k=5$ sum rules: 
\begin{subequations}\def\JJ{\mathcal{J}}
\begin{align}
g_6^\prime &= \left\langle \frac{1}{m^{12}}
\left(\JJ|c^{++}_{J,m^2}|^2-(\JJ-22)|c^{+-}_{J,m^2}|^2\right) \right\rangle,\\
0 &=
\left\langle \frac{1}{m^{12}}\left(
\JJ(\JJ-6) |c^{++}_{J,m^2}|^2 +
(\JJ^2- 50\JJ+596)|c^{+-}_{J,m^2}|^2\right) \right\rangle,
\end{align}
\label{eq: g6 more}
\end{subequations}
where $\mathcal{J}=J(J+1)$.
The novelty is the presence of a ``null constraint'', that is, a sum rule with vanishing low-energy contribution.
By taking linear combinations of \eqref{eq: g6} and \eqref{eq: g6 more} that maintain the positivity of RHS, these reproduce the known bounds derived in \cite{Bern:2021ppb}:\footnote{If we denote the three sum rules in the order they appear as $A,B,C$,
the relevant linear combinations are respectively: $6A-B+C\geq 0$ and $90A+11B+\frac12C\geq 0$.}
\bea
g_6\geq 0\,,\quad -\fft{90}{11} \leq \fft{g_6^\prime}{g_6} \leq 6\,.\label{eq: bound g6 and g6prime}
\eea
We will see in section \ref{D4R4} that the above bounds are not optimal;
we will get tighter ones by using more null constraints.
In particular we will show that indeed $g_6^\prime\geq 0$ (up to $1/\Mpl^2$ corrections from loops,
like other bounds in this paper).

A generic observation is that all dimensionless ratios of the form $g_k M^{2(k-4)}/g_4$ satisfy two-sided bounds,
consistent with dimensional analysis scaling.
We do not have a formal proof for all couplings, but we expect
this to hold in analogy with the scalar theory case studied in \cite{Caron-Huot:2021rmr}.
The novel feature for gravitons is that we will be able to upper-bound $g_4$ itself by a multiple of $G/M^6$,
as we now discuss.

\section{Bounds that relate gravity and higher derivatives}\label{sec:example functional}

\subsection{Impact parameter functionals}

As we reviewed in the previous section, the simplest positivity bound is $g_4\geq 0$, which is trivially established by evaluating $B_4^{(1)}$ at $p\rightarrow 0$. However, this says little about the size of $g_4$ since Newton's constant does not appear. On the other hand, one may expect that all graviton interactions shut down if $G=0$, a concrete example being the CEMZ bound \cite{Camanho:2014apa}. To probe Newton's constant with dispersion relations, we must inevitably use the spin-$2$ sum rules $B_2$ and deal with the $1/p^2$ pole.
This precludes doing Taylor expansions around the forward limit. 

Our strategy will follow \cite{Caron-Huot:2021rmr}: we tame the graviton pole by considering processes with finite impact parameter $b\sim1/M$.
In spacetime dimensions $D>4$ this removes all divergences. In $D=4$, as we are considering in this paper,
we will be left with infrared logarithms, still a major improvement over power-law divergences in our opinion.
Being mindful of the compact-support property (see section \ref{sec2: Regge bound}), we consider functionals of the generic form:
\bea
\mathcal{F}= \int_0^{M} dp \sum_i \psi_i(p) B_i(p^2)\,,\label{eq: impact parameter functionals}
\eea
where $B_i(p^2)$ generally denotes dispersive sum rules and $f_i(p)$ is compact functions in $p$.
The reader might worry that the integral in Eq.~\eqref{eq: impact parameter functionals} is done all the way up to the cutoff,
pushing the convergence of the EFT expansion to the limit. For the time being we will focus on super-convegent sum rules, for which this is not an issue. In section \ref{sec: improved sum rules} we will review how this problem can be avoided by considering \emph{improved} sum rules \cite{Caron-Huot:2020cmc,Caron-Huot:2021rmr}, which only receive contributions from a finite number of EFT coefficients.

In order to get a bound, the idea is to look for functionals whose action on any state above the cutoff is nonnegative:
\bea
    && \text{if}\,\quad \mathcal{F}[J,m^2] = \int_0^M dp \sum_i \psi_i(p) \mathcal{C}_{i,-p^2}[J,m^2] \geq 0\,,\quad \forall\, m\geq M\,, J\in \mathbb{Z}\,,
\cr && \text{then}\,\quad -\int_0^M dp \sum_i \psi_i(p) B_{i}(p^2)\Big|_{\rm low} \geq 0\,,\label{eq: impact parameter functional}
\eea
where $\mathcal{C}_{i,-p^2}[J,m^2]$ are the heavy densities defined in Eq.~\eqref{eq: average}.
In words, any nonnegative functional yields an inequality on low-energy observables,
thanks to the relation between low and high energies \eqref{sum rule schematic}.

In $D=4$ there will be a tension between finiteness and positivity:
finiteness on the graviton pole requires $\psi_i(p)$ to vanish faster than $p$ at the origin,
which is impossible for the Fourier transform of a positive function.
In practice, we begin by finding functionals which are rigorously positive at all impact parameters
but logarithmically diverge on the graviton pole. We then regulate by adding an infrared cutoff $\pmin\ll M$,
and accept that this causes negativity at large impact parameters.

Before explaining how we produce positive functionals, we first detail how we ascertain positivity.


\subsection{Example positive functionals involving gravity}

We now present explicit bounds that combine two ingredients:
\begin{itemize}
\item The spin-2,3 sum rules $B^{(1)}_{2}(p), B^{(1)}_{3}(p)$, smeared against suitable wavepackets $\psi(p)$
\item The forward limit of spin-4 sum rules and its first derivative: $B^{(1)}_{4}(0)$, $\partial_{p^2}B^{(1)}_{4}(0)$.
\end{itemize}
These use only $B^{(1)}$ \eqref{defB}: fixed-$u$ dispersion relation for the MHV amplitude.

In principle the spin-4 sum rules should also be smeared to make them rigorously valid, but as discussed in section \ref{sec2: Regge bound}
this is a technical modification, which we will ignore here.

Using these ingredients and the method of section \ref{sec2: dispersive sum rule}, we have constructed the following two functionals:
\begin{subequations}
\label{functionals g3g4}
\bea 
 \mathcal{F}_{g_3}&&=
 \int_{m_{\rm IR}}^M dp\ (1-p)^3 \big(p(65+155p+47p^2) B_2^{(1)}+2 (81-1927 p+4601 p^2)B_3^{(1)}\big) 
\cr && - 5 \partial_{p^2} B_4^{(1)}|_{p=0}\,, \\
 \mathcal{F}_{g_4}&&=  \int_{m_{\rm IR}}^M dp (1-p)^3  \big(p (99+282 p+49 p^2)B_2^{(1)}+(21-474 p+1216 p^2) B_3^{(1)}\big)
\cr && -\big(7+ \partial_{p^2}\big)B_4^{(1)}|_{p=0}\,,
\eea
\end{subequations}
in units where $M=1$.
Notice the lower cutoff $m_{\rm IR}$ makes the functionals infrared-safe.

We claim that:
\begin{itemize}
\item Without the cutoff $\pmin$, the functionals are positive for all states with $m>M$
\item With the cutoff $\pmin$, they only become negative at some large $b\sim \pmin^{-2/3}$
\item They imply the respective bounds:
\begin{subequations}
\bea
\label{g3 bound}
|\hatg_{3}|^2M^8 &\leq& 37.8 \log(M/\pmin)- 45.4-\mathcal{F}_{g_3}\big|_{\rm matter}\,,\\
\label{g4 bound}
g_4 M^6&\leq& 8\pi G\left(14.1 \log(M/\pmin)- 15.5-0.04|\hatg_3|^2\right)-\mathcal{F}_{g_4}\big|_{\rm matter}\,,
\eea
\label{eq: example bound}
\end{subequations}
where the matter contributions (coming from possible light scalars or Kaluza-Klein modes and detailed in eq.~\eqref{F KK} below)
is sign definite: $\mathcal{F}\big|_{\rm matter}\geq 0$.
\end{itemize}

The bound \eqref{g3 bound} is a sharp version of the CEMZ constraint \cite{Camanho:2014apa}:
$\hatg_{3}\lsim \frac{1}{M^4}$, up to logarithms. This shows that a cubic coupling of size
$\frac{1}{M^4}R^3$ cannot be turned on without having
a heavy state at the mass $M$ or lighter.
The bound \eqref{g4 bound} is similar  for quartic couplings.

The functionals \eqref{functionals g3g4} are not optimal: their main virtue is to be explicit enough to be analyzed in full detail
in this section. They establish our main conceptual result: higher dimensional couplings can be bounded in terms of Einstein 
gravity.  Optimal bounds are presented in the next section, see eq.~\eqref{optimal}.


How do we ascertain that a functional is positive on all heavy states?
A simple strategy is to plot the action $\mathcal{F}[m^2,J]$ as a function of $m$ for various discrete spins $J=0,2,4,5,6,\ldots$.
Note that the heavy contribution \eqref{B1 heavy} is the sum of two positive unknowns ($|c^{++}_{J,m^2}|^2$ and $|c^{+-}_{J,m^2}|^2$) and we must ascertain that the coefficient of each is positive:
we will refer to those as $\mathcal{F}^{++}$ and $\mathcal{F}^{+-}$ below.
Thus two plots must be made for each (even) value of the spin, and one plot for each odd spin.

Since we cannot plot infinitely many spins, it is
fruitful to exploit regularity of the functionals.  We find that when plotted as
a function of ``impact parameter'' $b=2J/m$, and $m$, the curves vary smoothly with spin and display simple asymptotic trends.
This allows us to draw contour plots where positivity is easy to ascertain,
and potentially dangerous regions can be easily identified for finer sampling.

\begin{figure}[ht]
\begin{subfigure}{.48\textwidth}
\centering
\includegraphics[height=0.31\textheight]{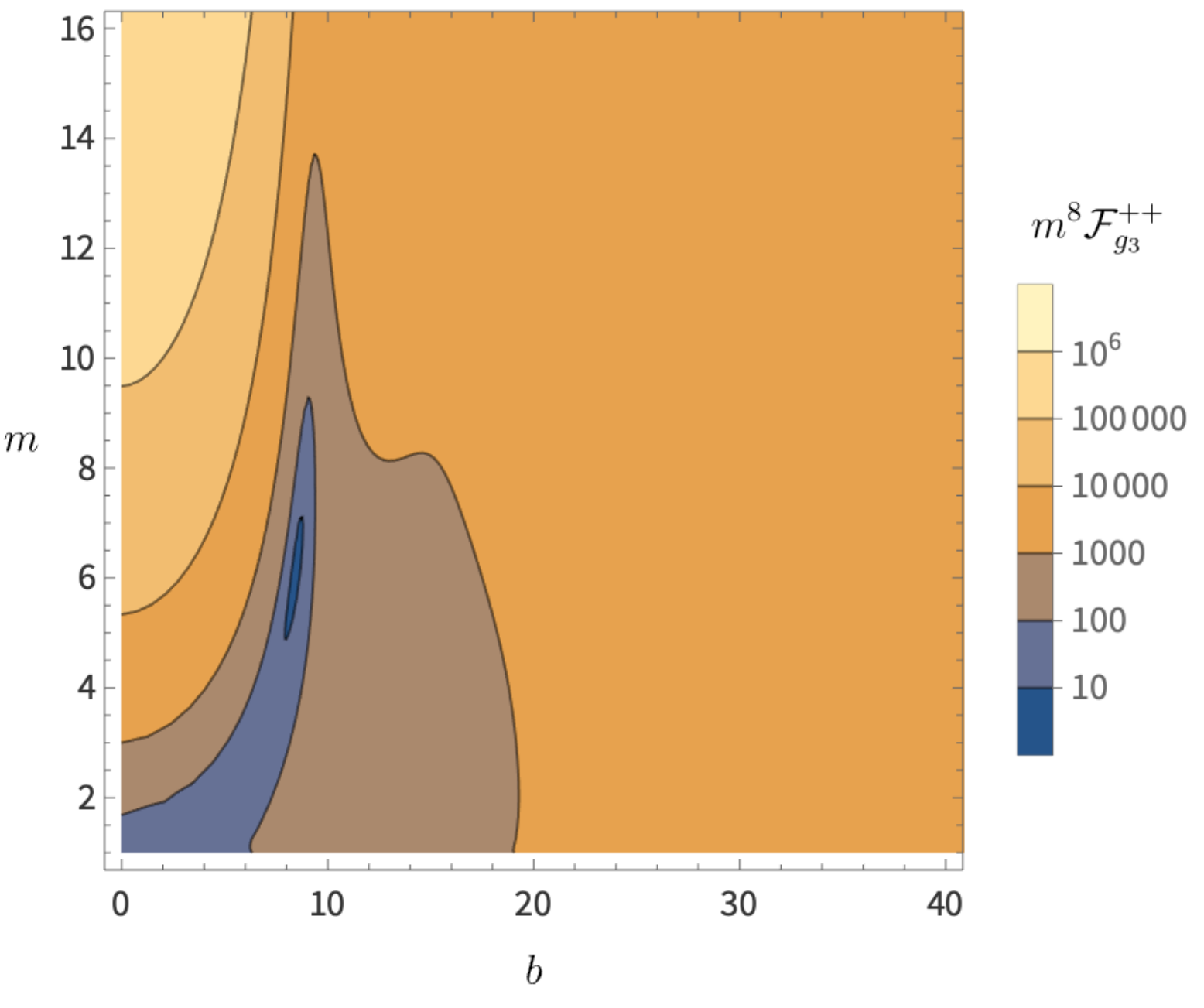} 
\caption{action of $\mathcal{F}_{g_3}$ on ${+}{+}$ states}
\label{subfig: pp}
\end{subfigure}
\hfill
\begin{subfigure}{.48\textwidth}
\centering
\includegraphics[height=0.31\textheight]{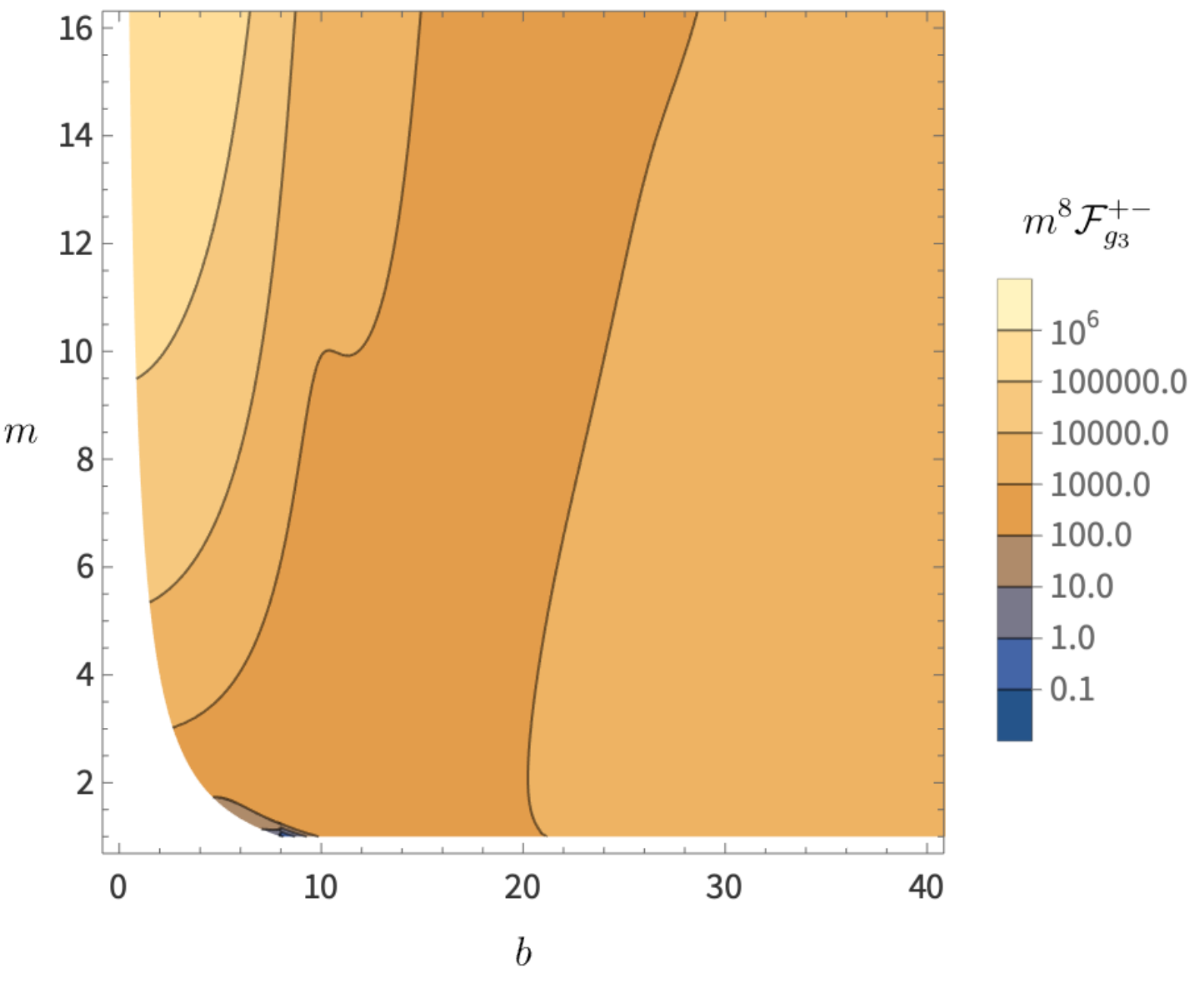}  
\caption{action of $\mathcal{F}_{g_3}$ on ${+}{-}$ states}
\label{subfig: pm}
\end{subfigure}
\caption{Contour plots which confirm non-negativity of the functional $\mathcal{F}_{g_3}$ giving
the upper bound \eqref{g3 bound}, in units where $M=1$.
We scaled the functional by $m^{10}$ to make scaling limits more manifest.
The sampling points are detailed in the text.
The lower-left corner is blank in the second figure due to the selection rule $J\geq4$.
}
\label{fig: functional_g3}
\end{figure}

\begin{figure}[ht]
\begin{subfigure}{.48\textwidth}
\centering
\includegraphics[height=0.31\textheight]{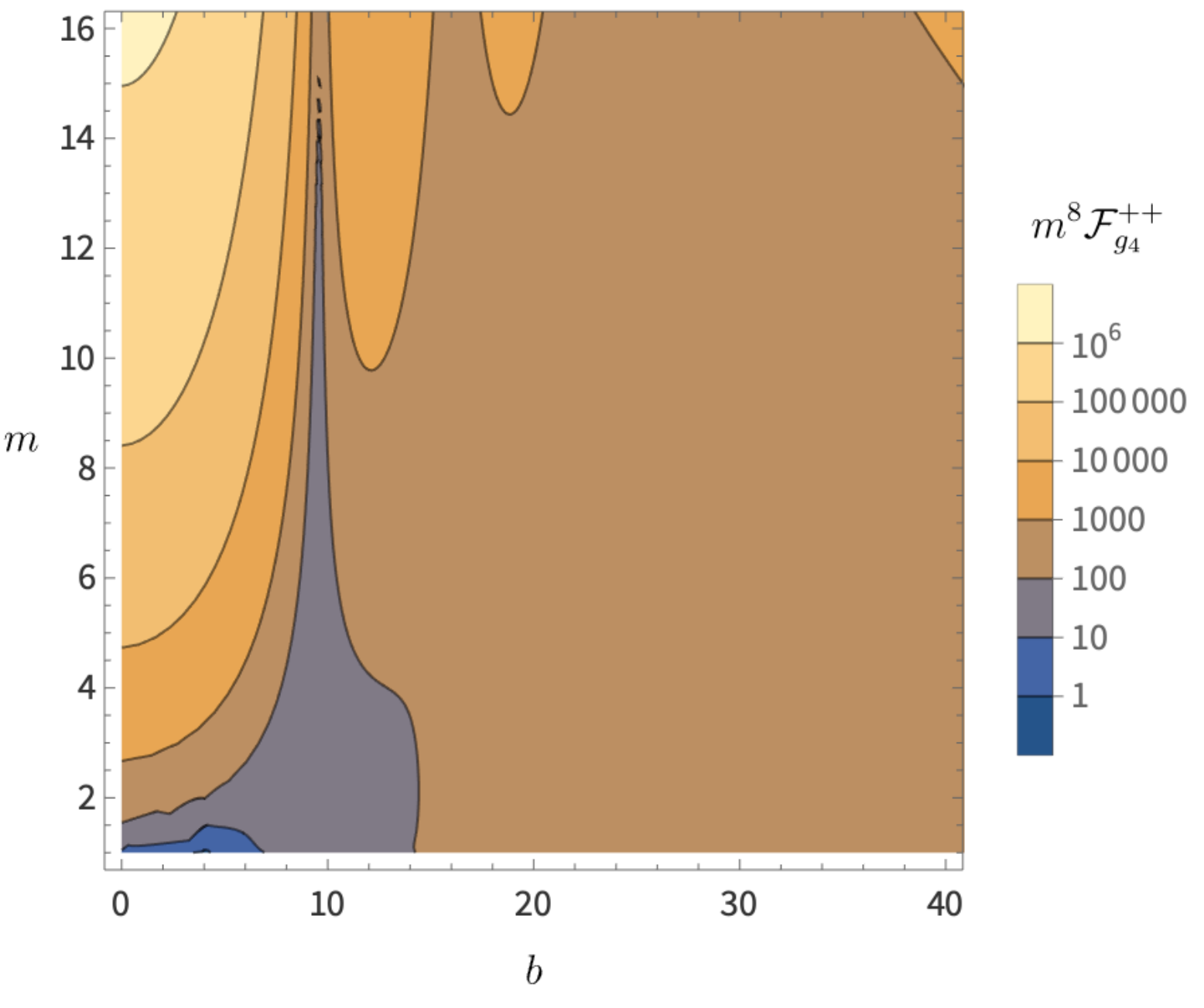} 
\caption{action of $\mathcal{F}_{g_4}$ on ${+}{+}$ states}
\label{subfig: pp2}
\end{subfigure}
\hfill
\begin{subfigure}{.48\textwidth}
\centering
\includegraphics[height=0.31\textheight]{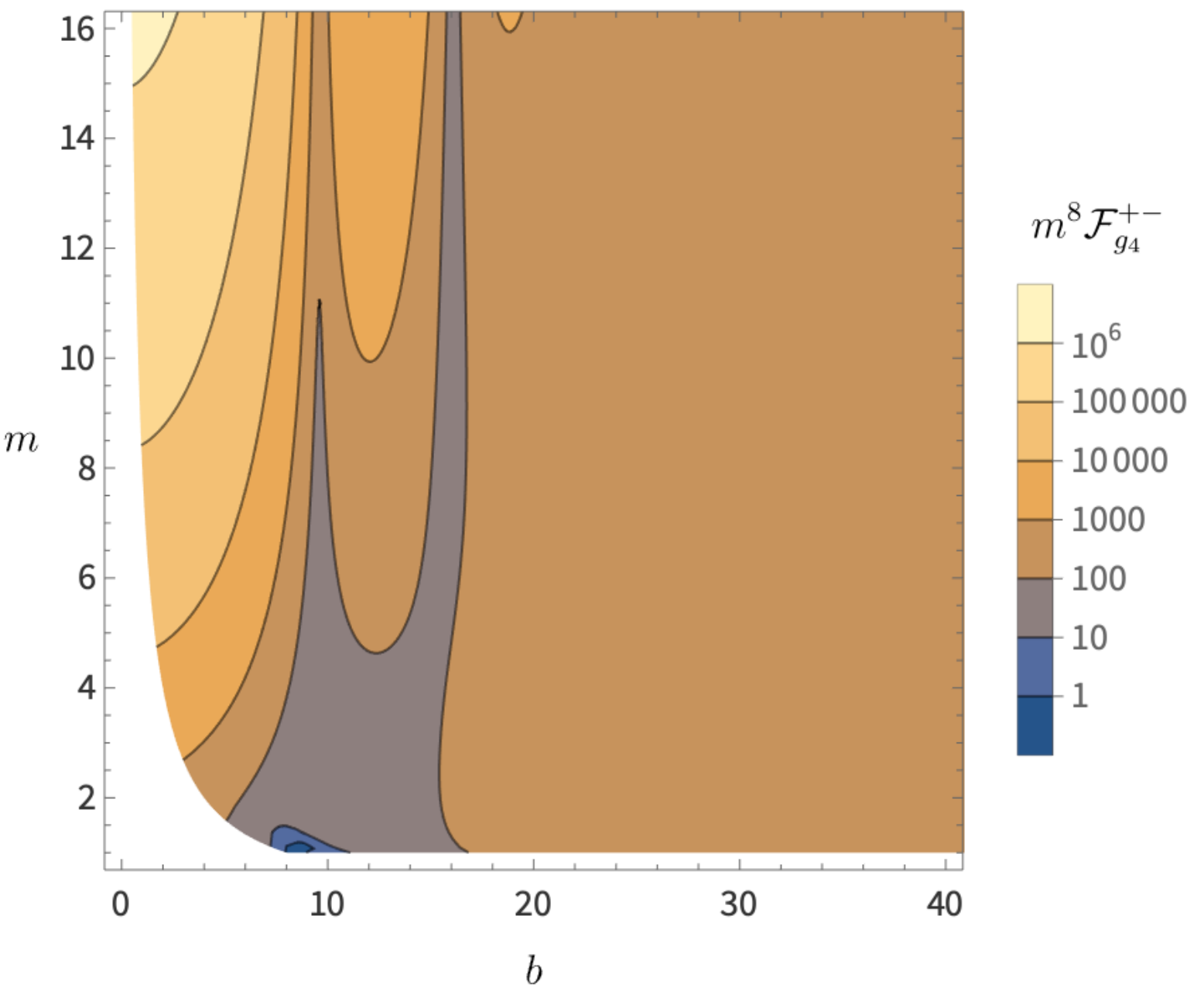}  
\caption{action of $\mathcal{F}_{g_4}$ on ${+}{-}$ states}
\label{subfig: pm2}
\end{subfigure}
\caption{Similar to figure \ref{fig: functional_g3}: Contour plots which confirm non-negativity of $\mathcal{F}_{g_4}$, establishing the bound \eqref{g4 bound}.
}
\label{fig: functional_g4}
\end{figure}

We display $\mathcal{F}_{g_3}$ in figure \ref{fig: functional_g3}.
Unless noted otherwise, all plots in this section are in units where $M=1$.
To make the plot, we considered data up to $J_{\rm max}=300$, more specifically, $J=0, 2,\ldots,300$ for $\mathcal{F}_{g_3}^{++}$ and $J=4,5,\ldots,300$ for $\mathcal{F}_{g_3}^{+-}$, accounting for spin selection rules.
For each spin, we sample the interval $m\in [1,16]$, with a larger density of points closer to the origin
(for example 80 points between 1 and 1.2 and 700 between 1.2 and 16).  
At the highest value $m=16$ we have safely reached the $m\to\infty$ limit, further discussed below,
and the value $b=40$ is well past any interesting structure.
Therefore figure \ref{fig: functional_g3} establishes positivity of $\mathcal{F}_{g_3}$. Similar plots are displayed for $\mathcal{F}_{g_4}$ in fig \ref{fig: functional_g4}, where we used the same sampling.

\begin{figure}[t]
\centering
\begin{subfigure}{.49\textwidth}
\centering
\includegraphics[height=0.19\textheight]{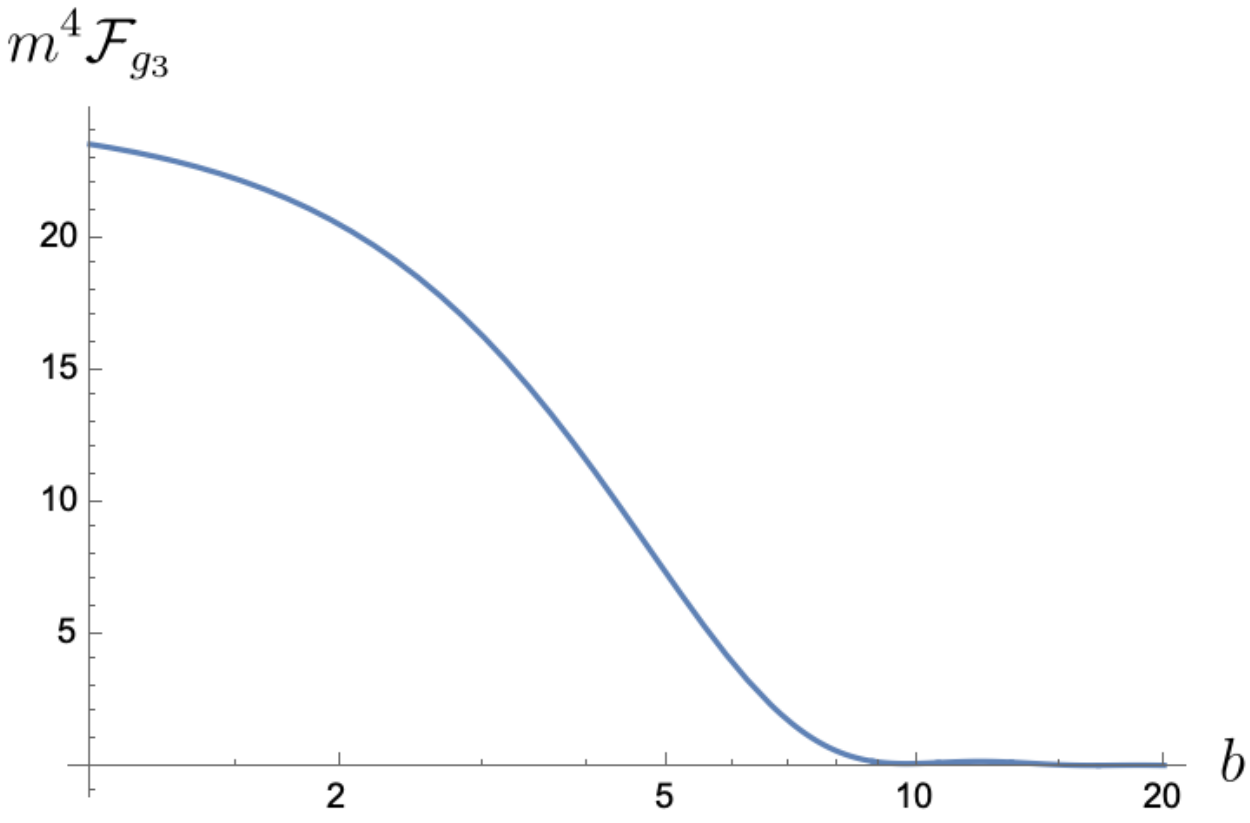} 
\caption{}
\end{subfigure}
\hfill
\begin{subfigure}{.49\textwidth}
\centering
\includegraphics[height=0.19\textheight]{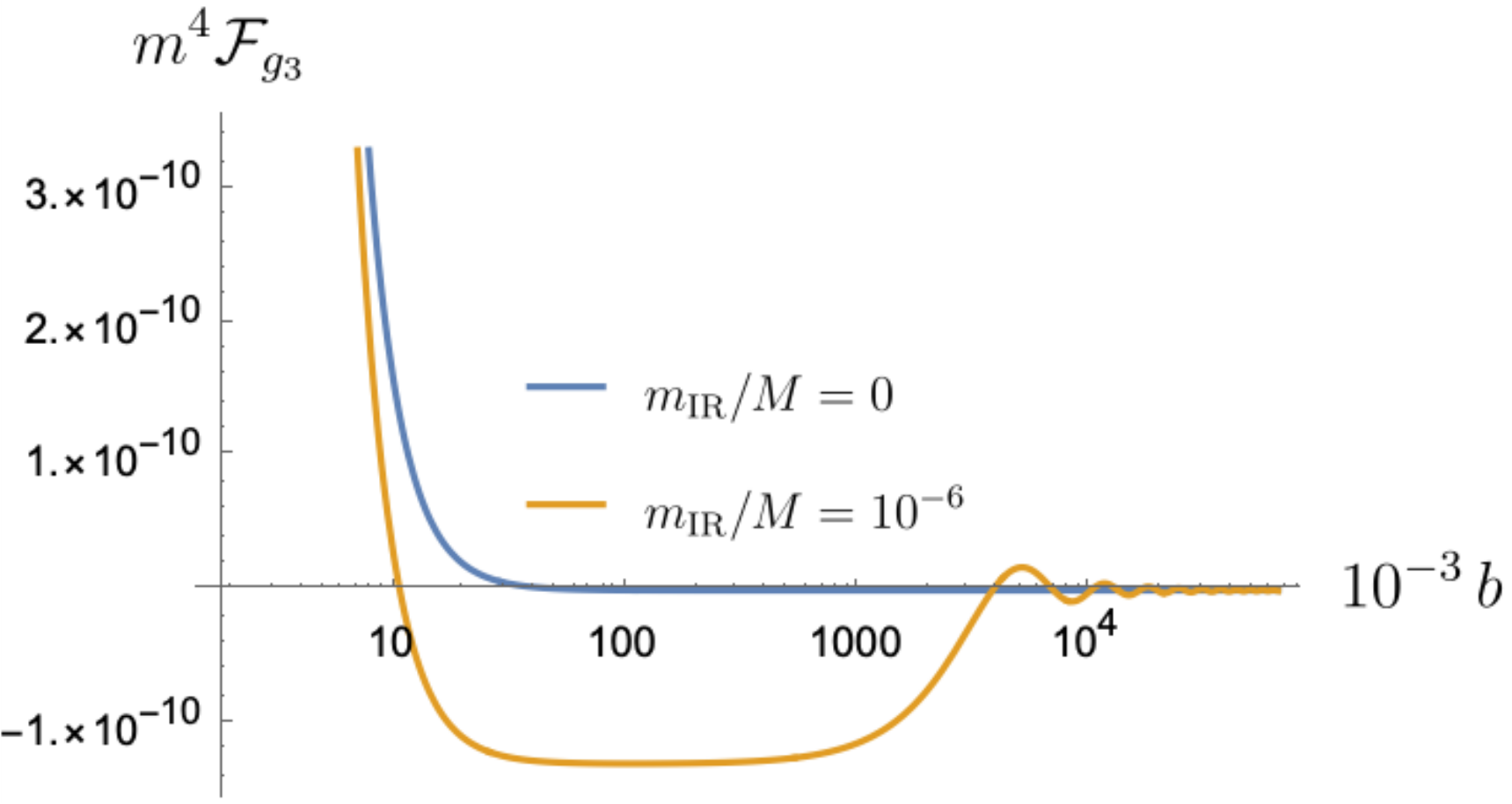} 
\caption{}
\end{subfigure}
\caption{Plots of functionals $\mathcal{F}_{g_3}$ at the scaling limit $J,m\rightarrow\infty$ where $b=2J/m$. (a) confirms the positivity at finite $b$ where the cutoff $\pmin$ makes no difference. (b) displays the large $b$ behavior  for both $\pmin=0$ (blue curve) and $\pmin/M=10^{-6}$ (orange curve).
There is a negative plateau in the range $\pmin^{-2/3} \lsim b\lsim \pmin^{-1}$ as detailed in the text.
}
\label{fig: funcs_scaling_lim_cut}
\end{figure}

In addition to sampling a finite range, we find it useful to verify
positivity in a $m\to\infty$ scaling limit. The limit is nontrivial if keeping impact parameter $b=2J/m$ is fixed,
and is dominate by the $B_2$ component of the sum rules. 
It is essentially the Fourier transform of its coefficient \cite{Caron-Huot:2021rmr}:
\be
\lim_{m\to\infty}\mathcal{F}[m^2,\tfrac{bm}{2}] = \frac{2}{m^4}\int_0^1 dp\, \psi(p) \times \left(\left(|c^{++}_{J,m^2}|^2 + |c^{+-}_{J,m^2}|^2\right)J _{0}(p b) + \ldots\right)\,,
\ee
where $\ldots$ represents higher orders in $1/m$. With $\pmin=0$, positivity is easy to ascertain at all $b$, as shown for $\mathcal{F}_{g_3}$ in figure \ref{fig: funcs_scaling_lim_cut}.
(In this limit, $\mathcal{F}^{++}$ and $\mathcal{F}^{+-}$ coincide, so there is only one curve.)
However, with $\pmin=0$ the action on low-energy gravity is infinite, and the resulting upper bound is vacuous.
We thus add a cutoff $\theta(p>\pmin)$, which creates negativity (we take $\mathcal{F}_{g_3}$ as example to illustrate):
\be
\lim_{b\to\infty} \lim_{m\to\infty}\mathcal{F}_{g_3}[m^2,\tfrac{bm}{2}] \approx \frac{160}{m^4b^3} -260 \frac{\pmin^2}{m^4} \fft{J(\pmin b)}{\pmin b}\,.
\ee
The second term overwhelms the first at $b_{\rm max}\sim \pmin^{-2/3}$ and creates negative plateau
up til $b\sim \pmin^{-1}$, where the Bessel function gets damped.
While the total area under the functional is necessarily 0 (because of vanishing at $p=0$),
the behavior past $\pmin^{-1}$ is not universal and could be modified by using a smoother cutoff.
This is depicted in figure \ref{fig: funcs_scaling_lim_cut}, where we contrast the functionals with $\pmin=0$ and $\pmin/M=10^{-6}$.

Note that the functionals $\cF_{g_3}$ and $\cF_{g_4}$ include derivatives of the spin-4 sum rule $B_4^{(1)}$ around the forward limit. For the purposes of bounding $|\hat g_3|$, one can also find functionals that are pure linear combinations of $B_2^{(1)}$ and $B_3^{(1)}$, with no forward limit component.\footnote{One example is 
\begin{align}
  &(1-p)^2(-154.020 p^7+2006.03 p^6-3443.55 p^5-2322.66 p^4+5294.87 p^3-2194.19 p^2+798.968 p)B_2^{(1)}\nn\\
  &+(3701.82 p^6-18251.2 p^5+32684.9 p^4-25581.3 p^3+7603.72 p^2-157.911 p+0.00171)B_3^{(1)} ,
\end{align}
which proves the bound $|\hat g_3|^2 \leq 1598 \log(M/m_\mathrm{IR})-3211$. This could certainly be improved by including more powers of $p$.} Such functionals may have useful technical applications, since they avoid the subtleties with forward limits discussed in section~\ref{sec2: simple bounds}.

\subsection{Light spin-0 and spin-2 matter fields don't lower the cutoff}

In addition to the graviton, we allow for the presence of possible spin-0 and spin-2 light states (for example, Kaluza-Klein modes),
whose amplitudes are given by eq.~\eqref{eq: KK MHV amp}.
They contribute to the low-energy part of sum rules (integral over the arc at $s\sim M^2$): 
\begin{subequations}
\begin{align}
        B_{2}^{(1)}(p^2)|_{\rm matter} =& -\sum_{\mlight<M} |g_0(\mlight)|^2 (p^2-2\mlight^2) \nn \\ 
                                       &- \sum_{\mlight<M}  \frac{|g_2(\mlight)|^2}{\mlight^4} (p^2-2\mlight^2)(\mlight^2-6 \mlight p^2+6 p^4)\,,\\
        B_{3}^{(1)}(p^2)|_{\rm matter}  =&
\sum_{\mlight<M} |g_0(\mlight)|^2+ \sum_{\mlight<M} \frac{|g_2(\mlight)|^2}{\mlight^4} (\mlight^2-6 \mlight p^2+6 p^4)\,,\\
        B_{4}^{(1)}(p^2)|_{\rm matter}  = & -\sum_{\mlight<M}  \frac{|g_2(\mlight)|^2}{\mlight^4} 12 \,p^2 \,.
\end{align}
\end{subequations}
The spin-2 contribution with $\mlight\to 0$ is proportional to that from the cubic coupling
$|\hatg_3|^2$, see \eqref{B1 low}.

\begin{figure}[t]
\centering
\includegraphics[width=.65\linewidth]{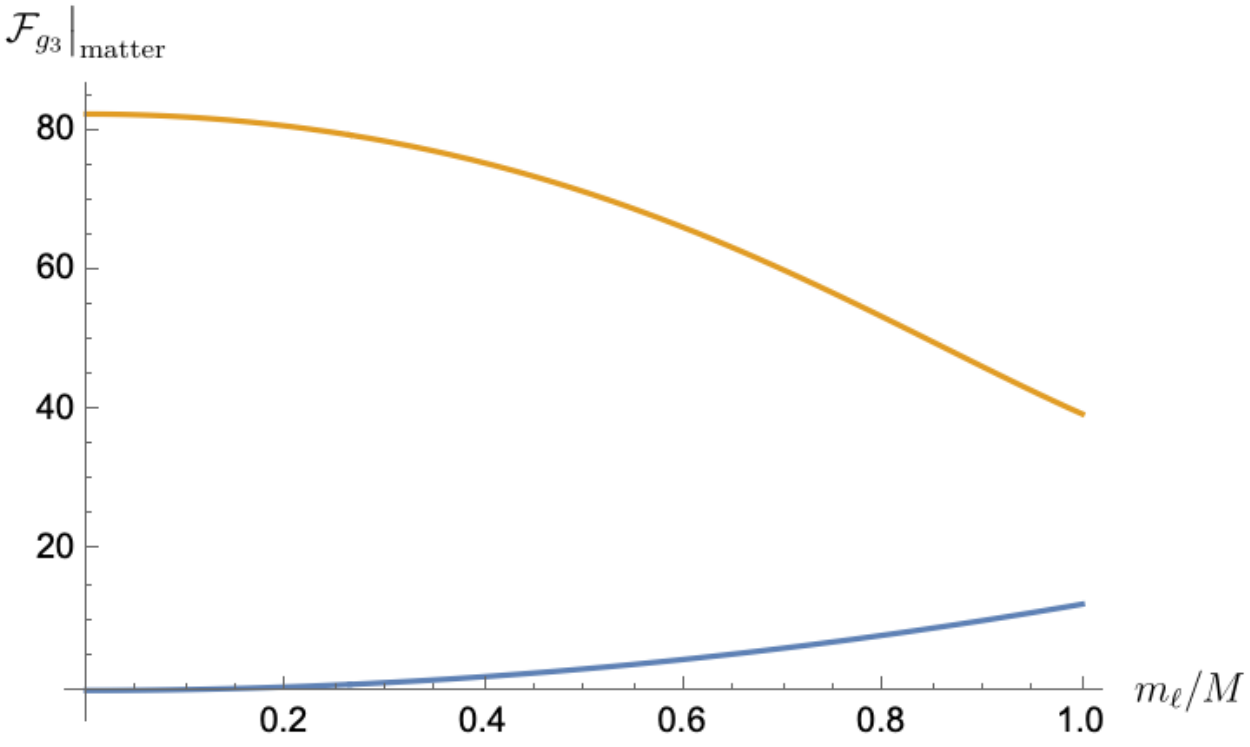} 
\caption{Positivity of functional $\mathcal{F}_{g_3}\big|_{\rm matter}$ for acting on light matter fields \eqref{g3 with KK}. We separate out the scalar and spin-$2$ contributions. The blue line is coefficient of $|g_0(m_\ell)|^2$ and the orange line is coefficient of $|g_2(m_\ell)|^2/m_\ell^4$ in $\mathcal{F}_{g_3}\big|_{\rm matter}$.}
\label{fig: positive KK}
\end{figure}

The key feature is that the unknown couplings $|g_J(\mlight)|^2$ are sign-definite.
Therefore, if we take combination of sum rules such that the coefficient of each unknown is positive for $\mlight\in (0,M)$, 
then the unknown light and heavy states will all contribute with the same sign leading again to a valid inequality
on EFT parameters. For the functionals in eq.~\eqref{functionals g3g4}, for example, we get:  
\begin{align} \label{F KK}
         \mathcal{F}_{g_3}\big|_{\rm matter} &= \sum_{\mlight<M} |g_0(\mlight)|^2\left(\fft{2591}{210}\mlight^2+\fft{1}{18}\right)  \\ 
                                    & + \sum_{\mlight<M} \frac{|g_2(\mlight)|^2}{\mlight^4} {}\left(\frac{2591}{210}\mlight^6-\frac{239}{18}\mlight^4-\frac{64573}{1540} \mlight^2+\frac{330151}{4004}\right) \,.   \nn
\end{align}
in units where $M=1$.
These polynomials are positive-definite as demonstrated in figure \ref{fig: positive KK}.

Alternatively, we could move the matter contribution in eq.~\eqref{g3 bound} to the left-hand-side,
giving:
\be
|\hatg_{3}|^2 + \mathcal{F}_{g_3}\big|_{\rm matter} &\leq \frac{1}{M^8}\left(37.8 \log(M/\pmin)- 45.4\right)\,. \label{g3 with KK}
\ee
We conclude that our bounds on $\hatg_3$ limit the \emph{sum} of squared cubic couplings to \emph{all}
light particles below the higher-spin scale $M$.

It is remarkable that light scalars or spin-2 particles cannot couple strongly to two gravitons.
This is very different than for the scalar EFT studied in \cite{Caron-Huot:2020cmc}, where the only limit
on the interaction strength of light scalars would be the unitarity bound ${\rm Im}\ a_J(s) \leq 2$.

This result can be interpreted as follows. In Einstein's gravity,
the decay rate of a Kaluza-Klein graviton to two massless gravitons
is proportional to an overlap integral $\int dy \sqrt{g} \chi(y)=0$, 
which vanishes by orthogonality of eigenfunctions (here $y$ is a coordinate on the internal manifold and
$\chi(y)$ is the eigenfunction corresponding to the mode in question).
Thus Kaluza-Klein modes only decay through higher-derivative corrections from the higher-dimensional
perspective. The $1/M^8$ suppression in \eqref{g3 with KK} confirms that such a suppression
exists for \emph{any} massive spin-two particle, irrespective of its microscopic origin.  The scale of suppression is controlled
the mass of \emph{higher-spin} particles. 

\subsection{Systematic strategy: improved sum rules}

\label{sec: improved sum rules}

In general, a low-energy gravitational EFT includes an infinite number of contact terms. When evaluated at low-energies, the spin-$2$ and spin-$3$  sum rules $B_2(t)$ and $B_3(t)$ are each sensitive to an infinite subset of these contact terms. It is useful to subtract from $B_2(t)$ a linear combination of forward-limits of higher spin sum rules to define ``improved" sum rules \cite{Caron-Huot:2021rmr}, which are sensitive only to a finite number of higher-derivative corrections.
 Not all sum rules need to be improved, for example, the $B_{2,3}^{(1)}$ and $B_2^{(3)}$ sum rules are automatically free of  higher-dimension contact terms.  For the $B_2$ sum rules, we only need to define
 \begin{subequations}
\begin{align}
B_2^{(2)\, {\rm imp}}(p^2) &= B_2^{(2)}(p)+ \oint_{\infty} \fft{ds}{4\pi i} \left[-\fft{\partial_tf(p^2-s,0)}{s(s+p^2)}+\fft{\partial_t f (0,s)}{s(s-p^2)}\right]\,,
\\
B_2^{(4)\, {\rm imp}}(p^2) &= B_2^{(4)}(p) + \oint_{\infty} \fft{ds}{4\pi i}\left[\fft{p^4(4s-3p^2)}{s^4(s-p^2)^2}h(s,p^2-s) + \fft{2p^6}{s^3 (s^2-p^4)}\partial_t h(s,-s)\right]\,.
\end{align}
\end{subequations}
The improved spin-$2$ sum rules have the low-energy contributions
\bea
-B_2^{(2)\,{\rm imp}}(p^2)|_{\rm low,\ grav} = \fft{2\pi G}{p^2}|\hatg_3|^2\,,\quad
-B_{2}^{(4)\,{\rm imp}}(p^2)|_{\rm low,\ grav} = -40 \pi \hatg_3 p^2\,,
\eea
where we have written only the contribution of graviton exchange for brevity.
Similarly we can also define improved versions of higher-spin sum rules. An example is:
\bea
B_4^{(1)\,{\rm imp}}(p^2) &&=B_4^{(1)}(p^2) + \oint \fft{ds}{4\pi i} \Big[\fft{p^2}{s(s-p^2)}f(0,-s) -\fft{2p^{12}}{s^5(s^2-p^4)}f(s,-s) 
\cr &&
\cr &&  +
\fft{p^2(s^4-2s^3 p^2-2s^2 p^4-2s p^6-2p^8)}{s^5(s+p^2)}f(0,s)+ \fft{p^4(s^2+2p^4)}{s^4}\partial_t f(0,s)  
\cr &&
\cr && -\fft{p^4(s^2-p^4)}{s^4}\partial_t f(0,-s) -\fft{p^8}{2s^3}\big(\partial_t^2 f(0,-s)+\partial_t^2 f(0,-s)\big)\Big]\,,
\eea
which nicely gives
\be
 -B_4^{(1)\,{\rm imp}}(p^2)|_{\rm low,\ grav} = 4\pi G |\hatg_3|^2 + 2 g_4\,.
\ee

In this section, we apply these improved sum rules (together with some forward-limit sum rules) to derive bounds involving higher-derivative Wilson coefficients and gravity, using the parameter choices listed in appendix \ref{app: numerics}. Following \cite{Caron-Huot:2021rmr}, we consider wavefunctions $\psi_i(p)$ that are polynomials in $p$
\be
\psi_i(p)=\sum_{n=n_{i\,{\rm min}}} c_{in} p^n\,.
\ee
Our wavefunctions start with specific exponents $n_{i{\rm min}}$ that depend on the sum rule. Specifically, we require that all sum rules possess the same large-$b$ behavior in the scaling limit $m\to \oo$ with $b=\frac{2J}{m}$ fixed and large. (One can prove that this is a necessary condition for obtaining a positive functional.) Precisely, the minimal values of $n$ for the improved spin-$2$ sum rules are
\be
(n_{2\,{\rm min}}^{(1)} , n_{2\,{\rm min}}^{(2)} , n_{2\,{\rm min}}^{(3)} , n_{2\,{\rm min}}^{(4)})=(1,10, 6, 1)\,.
\ee
Since higher-spin sum rules are subleading at large-$m$ (with fixed $b$), their wavefunctions can safely start with $n_{\rm min}=0$, as long as spin-2 sum rules are included in the functional.

It is worth noting that compact support in $p$ can lead to oscillations at large $b$, which can potentially hinder positivity. To suppress oscillations, it is useful to make the wavefunctions smoother near $p=1$ (in units where $M=1$). We do this by multiplying the wave functions by a power of $(1-p)$. For example, our example functionals in the previous section included factors of $(1-p)^3$ to suppress large-$b$ oscillations. In this section, we consider wavefunctions of the form
\be
\psi_i(p)= \sum_{n=n_{i\,{\rm min}}}^{n_{i\,{\rm min}}+n_{\rm max}} c_{in} (1-p)^2 p^n\,,\label{eq: condition for wave function}
\ee
where we truncate $n$ to $n_{\rm max}$. Larger values of $n_{\rm max}$ will correspond to more complicated functionals and stronger bounds.

As mentioned before, to allow for low-energy matter particles, we add additional constraints to the solver SDPB  \cite{Simmons-Duffin:2015qma,Landry:2019qug} that impose negativity of matter contributions to low-energy sum rules for $m_\ell<M$. In practice, this means including positive matrices that are negatives of the matrices paired with low-energy three-point couplings $g_s(\mlight), g_s^\ast(\mlight)$, and $g_2(\mlight), g_2^\ast(\mlight)$ for $\mlight <M$. In the MHV sum rules, matrices for matter particles are simply the coefficients of $|g_s(\mlight)|^2, |g_2(\mlight)|^2$. In numerics, we discretize $m_\ell$ to the following values: $\mlight \in (0, 0.01, \ldots, 1)M$.

\subsection{Relation with the CEMZ argument}\label{sec: CEMZ}

\begin{figure}[t]
  \centering
\includegraphics[width=.25\linewidth]{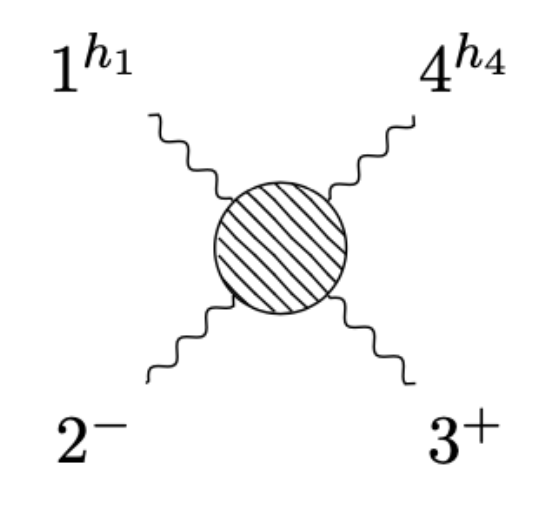} 
 \caption{Scattering of gravitons with arbitrary helicity $1^{h_1}\to 4^{-h_4}$ against a $2\to 3$ ``target'',
 as used in the CEMZ-like argument \eqref{eq: CEMZ scattering}. Time runs from left to right.}
 \label{fig: CEMZ scattering}
\end{figure}

In \cite{Camanho:2014apa}, CEMZ considered classical scattering of gravitons with various polarizations against a target
(for example, a black hole) and argued that a too-large value of $|\hatg_3|M^4/G$ would lead to a time advance for one of
the polarizations. 
It is instructive to see how the CEMZ argument is paralled in our formalism.

The basic idea is to consider the $B$ sum rules at large impact parameters $bM\gg 1$.
Momentarily ignoring the compact-support constraint $p<M$, we can define impact-parameter sum rules
by Fourier transforming in the transverse space
\be
\widehat{B}_2^{(1)}(b) = \int \frac{d^2p}{(2\pi)^2} e^{i p\cdot b} B_2^{(1)}(p) = \frac{1}{2\pi}\int_0^\infty dp\, p J_0(b p) B_2^{(1)}(p)\,.
\ee
At large $b$, the integral is highly oscillatory, which suppresses the
heavy action \eqref{B1 heavy} on states with small $J$.
Acting on states with $J\gg 1$, the $\tilde{d}$ function simplifies according to \eqref{Bessel limit},
and the integral localizes\footnote{A way to see this is using the identity $\int_0^\infty dp\,pJ_\nu(xp)J_\nu(yp) = \delta(x-y)/x$ for $\nu>-1/2$.} to
\be
 \widehat{B}_2^{(1)}(b)\big|_{\rm high} \to \frac{1}{\pi} \left\langle \left(|c^{++}_{J,m^2}|^2+|c^{+-}_{J,m^2}|^2\right)
 \frac{\delta\left(\frac{2J}{m}-b\right)}{bm^4} \right\rangle \,.\label{eq: transform of B21}
\ee
This is positive, yielding a constraint on low-energy coefficients $-\widehat{B}_2^{(1)}(b)\big|_{\rm low}$.

More generally, we can consider scattering of an arbitrary helicity graviton (1 and 4)
against a positive-helicity target (2 and 3), see fig.~\ref{fig: CEMZ scattering}, which motivates us to consider a matrix of sum rules:
\begin{align}
        \mathbb{B}(p^2) = \oint_{s=\infty} \frac{ds}{s^3}
\begin{pmatrix} {\cal M}(1^+2^-3^+4^-) &\ \  {\cal M}(1^+2^-3^+4^+)
\\ {\cal M}(1^-2^-3^+4^-)  &\ \  {\cal M}(1^-2^-3^+4^+)\end{pmatrix}(s,-p^2) = 0\,,\label{eq: CEMZ scattering}
\end{align}
evaluated in the forward limit.
In the center of mass frame, we have
\be
\begin{pmatrix} 
        {\cal M}(1^+2^-3^+4^-) &\ \  {\cal M}(1^+2^-3^+4^+) \\
        {\cal M}(1^-2^-3^+4^-) &\ \  {\cal M}(1^-2^-3^+4^+)
\end{pmatrix} 
\sim 
        s^4\begin{pmatrix}
          f(s,t) &   \hat{p}^4 g(s,t)        \\
          \hat{p}^{*4}  \bar{g}(s,t) & f(s,t)
        \end{pmatrix}\,,
\ee
where we have defined $\hat{p} = p_1 + i p_2$, where $p_i$ are the components of transverse momentum transfer. Thus, we recognize the matrix elements of Eq.~\eqref{eq: CEMZ scattering} as our forward $B_2^{(1)}$, $B_2^{(3)}$ sum rules
\be
\mathbb{B}(p^2) = \begin{pmatrix} \tfrac12 B_2^{(1)}(p^2) & \hat{p}^4 B_2^{(3)}(p^2) \\ \hat{p}^{*4} B_2^{(3)}(p^2)^* & \tfrac12 B_2^{(1)}(p^2) \end{pmatrix}\,.
\ee

Let us transform to to $b$ space using \eqref{eq: transform of B21} for the diagonal terms. For the off-diagonal terms, we use
\be
\widehat{B}_2^{(3)}(b) = \int \frac{d^2p}{(2\pi)^2} \hat{p}^4 e^{i p\cdot b} B_2^{(3)}(p) = \frac{\beta^2}{2\pi} \int_0^\infty  dp\, p^5 J_4(b p) B_2^{(3)}(p)\,,
\ee
where $\beta = \hat{b}/\hat{b}^\ast$ with $\hat{b}=b_1+b_2$. We find the following heavy contribution:
\begin{align}
\mathbb{B}(b)|_{\rm high}
 \propto&
\left\langle 
\left[
\begin{pmatrix} c^{++}_{J,m^2}\beta^* \\ c^{-+}_{J,m^2} \beta\end{pmatrix}
\begin{pmatrix} c^{++}_{J,m^2}\beta^* & c^{-+}_{J,m^2}\beta\end{pmatrix}^*
+
\begin{pmatrix} c^{+-}_{J,m^2}\beta^* \\ c^{--}_{J,m^2}\beta\end{pmatrix}
\begin{pmatrix} c^{+-}_{J,m^2}\beta^* & c^{--}_{J,m^2}\beta\end{pmatrix}^*
 \right]
 \frac{\delta\left(\frac{2J}{m}-b\right)}{bm^4}
\right\rangle,
\label{CEMZ 1}
\end{align}
which is positive-definite, since the bracket is (by construction) a positive definite-matrix. 
On the other hand, evaluating the low-energy contribution in eqs.~\eqref{eq:fghlow} we find
\begin{equation}
-\mathbb{B}(b)|_{\rm low}
 = 4G \begin{pmatrix} \log [1/(b m_{\rm IR})]  & \frac{24 \hatg_3}{\hat{b}^4} \\
 \frac{24 \hatg_3^*}{\hat{b}^{\ast 4}} & \log [1/(b m_{\rm IR})]\end{pmatrix}\,.
\label{CEMZ 2}
\end{equation}
The eigenvalues of this matrix precisely reproduce
the eikonal time delays given in eq.~(3.22) of \cite{Camanho:2014apa} (with $\gamma^{\rm there} =8\hatg_3^{\rm here}$).
It is instructive to see how these calculations relate.
In that reference the eikonal phase is computed from the Fourier transform of the non-analytic parts of the amplitude
\begin{equation}
        \delta(s,b) = \frac{1}{2s}   \int \frac{d^2p}{(2\pi)^2} e^{i p\cdot b}   \begin{pmatrix} 
         \frac{8\pi G s^2}{p^2} & \frac{4\pi G s^2 \hat p^4}{p^2} \\
        \frac{4\pi G s^2 \hat p^{*4}}{p^2} & \frac{8\pi G s^2}{p2}
\end{pmatrix}
\,.
\label{CEMZ delta}
\end{equation}
A classical time delay (matrix) is then extracted from the energy derivative:
$t = \frac{\partial}{\partial E} \delta(E^2,b)$.
We see that the energy-growing part of the time delay is precisely what our matrix of sum rules captures:
\be
 t = E\ \mathbb{B}(b)|_{\rm low}.
\ee
Now the dispersive sum rules state that \eqref{CEMZ 1} equals \eqref{CEMZ 2}.
In particular, they imply that the energy-growing part of time delays (in the linear regime) must be positive.

The argument of ref.~\cite{Camanho:2014apa} is concluded by noting that the low-energy
calculation (using graviton exchange) is only valid then $b\gg M^{-1}$ where $M$ is the mass of heavy states.
Thus we should only impose positivity of \eqref{CEMZ 2} in that range, which gives the parametric bound:
\begin{equation}
        |\hatg_3| \lsim \frac{1}{M^4} \log[1/(bm_{\rm IR})]\,.
\end{equation}
Ref.~\cite{Camanho:2014apa} further argued that an infinite tower of higher-spin states needed to appear.

This discussion highlights that CEMZ constraints are \emph{built into} 
dispersive sum rules, they are a subset of the functionals enumerated in the preceding subsection.
Namely, this subset consists of spin-2 functionals $B_2(p)$ integrated against wavepackets that are peaked at impact parameters $b$. The CEMZ requirement $b\gg 1/M$ then has a clear origin in our compact support property $p<M$ (see section \ref{sec2: Regge bound}).

Despite the fact that the same mathematics appear, the physical assumptions are quite distinct.
Ref.~\cite{Camanho:2014apa} considered very large center of mass energy, where the amplitude exponentiates, whereas we consider $Gs\ll 1$ where the tree-level approximation is sufficient. (Appendix D of \cite{Camanho:2014apa} also presented an argument valid in the linear regime and related to the ``chaos bound'' \cite{Maldacena:2015waa}.)
The upside of using large center of mass energies there was that the acausality becomes classical and macroscopic, which led to transparent ``grandparent paradoxes''.  The downside is that the cutoff is imprecise, $b\ll 1/M$: such a
method gives parametric bounds, in contrast with the precise cutoff $p<M$ which yields sharp bounds.

In addition, the fact that we scatter waves rather than particles enable the precise energy resolution that is
needed to probe the mass and couplings of states above the cutoff.  One might say that time advances constitute a classical statement of causality, while crossing symmetry and analyticity of the S-matrix provide a quantum version.


\section{Results}\label{sec:results}
In this section we will present our results for bounds on EFT modifications of Einstein gravity.

\subsection{Comparison with model amplitudes}
\label{sec:models}

\begin{table}[t]\centering
\begin{tabular}{c|ccccc}
spin & $8\pi G\hatg_3 M^2$ & $g_4M^4$ &  $g_5 M^6$ & $g_6M^8$ & $g_6^\prime M^8$ \\\hline
0 &$+\frac{1\times2}{315}$ & $\frac{16}{1575}$  & $\frac{64}{10395}$ & $\frac{192}{35035}$ & $\frac{64}{135135}$ \\[1mm]
$\tfrac12$ &$-\frac{2\times2}{315}$ & $\frac{58}{1575}$  & $\frac{944}{51975}$ & $\frac{32}{1911}$ & $\frac{1856}{675675}$\\[1mm]
1 &$+\frac{3\times2}{315}$ & $\frac{248}{1575}$ & $\frac{848}{17325}$ & $\frac{258304}{4729725}$ & $\frac{13856}{675675}$ \\[1mm]
$\tfrac32$ &$-\frac{4\times2}{315}$ & $\frac{1676}{1575}$ & $\frac{160}{2079}$ & $\frac{434272}{1576575}$ & $\frac{32192}{135135}$ \\[1mm]
2 &$+\frac{5\times2}{315}$ & $\frac{5368}{315}$ & $-\frac{31888}{10395}$ & $\frac{6024448}{1576575}$ & $\frac{800288}{135135}$
\end{tabular}
\caption{\label{tab:models} One-loop contributions to low-energy parameters from heavy particles of various spins,
divided by an overall factor $G^2N$ where $N$ is the number of particles of the given spin circulating in the loop.
Extracted from \cite{Bern:2021ppb}.}
\end{table}

Following ref.~\cite{Bern:2021ppb} it will be instructive to compare our bounds with explicit models, with higher-dimension operators arising from integrating out a loop of massive particles of mass $m$ and spin two or less. For this one loop process, the mass of the lightest higher-spin states, which are two-particle states, is $M=2m$. The resulting Wilson coefficients are shown in Table.~\ref{tab:models}.

Note that the complex coupling $\hatg_3$ (which enter non-MHV amplitudes) vanishes for supersymmetric spectra;
correspondingly, the first column in the table is proportional to the numbers of degrees of freedom and a fermionic sign.
Other complex couplings follow the same pattern and it thus suffices to record the contribution from a heavy spin-0 particle:
\be
\left\{ \hatg_4 M^2,\ \hatg_5 M^4,\ \hatg_6 M^8,\  \hatg_6^\prime M^8,\  \hatg_6^{\prime\prime}M^8\right\}_{\rm spin\ 0}
= G^2N\left\{ \frac{8}{945}, \frac{4}{495}, \frac{4}{3003}, \frac{3488}{225225}, \frac{512}{1576575}\right\}.
\ee

It is important to note that since we neglect light graviton loops, the effect of integrating out a particle
are suppressed by a power of $GM^2\ll 1$ compared with the effects we focus on, 
and are thus beyond the accuracy of our bounds.  Nonetheless, the comparison with these models becomes physically meaningful
(larger than neglected graviton loops) if a large number $N$ of heavy particles circulate.
In this sense, the upper bounds on couplings recorded below in fig.~\ref{fig: compare with and without KK for g3sq-g4-grav} can be interpreted as bounds on the number of species above mass $M$:
$N\lsim \#\frac{\Mpl^2}{M^2}\log(m/m_{\rm IR})$, as far as their effects on low energies are concerned.
This has the expected dependence on $\Mpl$ from ``species bounds,''
although our bounds have an extra factor of an infrared logarithmic due to the nature of the probes we are using.

In addition, following \cite{Bern:2021ppb}, we will consider three models of string theory:
\be \label{strings}
 f= \frac{8\pi G}{stu} \frac{\Gamma\big(1-\tfrac{s}{M^2}\big)\Gamma\big(1-\tfrac{t}{M^2}\big)\Gamma\big(1-\tfrac{u}{M^2}\big)}{\Gamma\big(1+\tfrac{s}{M^2}\big)\Gamma\big(1+\tfrac{t}{M^2}\big)\Gamma\big(1+\tfrac{u}{M^2}\big)} \times X^p\,,
\ee
where $X=1-\frac{su}{M^2(t+M^2)}$ and $p=0,1,2$ representing respectively the superstring theory,
the heterotic string, and the bosonic string.
In all cases $M$ coincides with the mass of the first spin-4 particle exchanged between gravitons.

For the string models, we extract low-energy parameters by matching with the low-energy expansion
\eqref{eq:fghlow} and subtracting the low-energy poles. For the bosonic string, we additionally subtract a tachyon pole $\frac{1}{t+M^2}$.
(Because of the tachyon, the bosonic string model is not fully ``physical'',
however after subtracting this pole it still has a positive heavy spectral density.)
There is an ambiguity of whether we include the spin-0 and spin-2 contributions to the $t=M^2$ pole
as part of $f_{\rm matter}$ or as part of the ``heavy'' contributions: our bounds are valid in either case. 
In the plots below the string models thus span extended regions,
depending on what fraction $0\leq \gamma_i\leq 1$ of each pole we choose to subtract. For example, for the heterotic string we find
\be
 g_4\big|_{\rm hs}= 8\pi G \left(1+2\zeta_3 - \frac{11}{12}\gamma_0 - \frac{1}{12}\gamma_2\right),\qquad
 g_5\big|_{\rm hs}= -8\pi G \left(1+\frac{11}{12}\gamma_0 + \frac{1}{12}\gamma_2\right).
\ee

\subsection{Bounds involving $R^3$, $R^4$ and gravity}

\begin{figure}[ht]
\centering
\includegraphics[width=.5\linewidth]{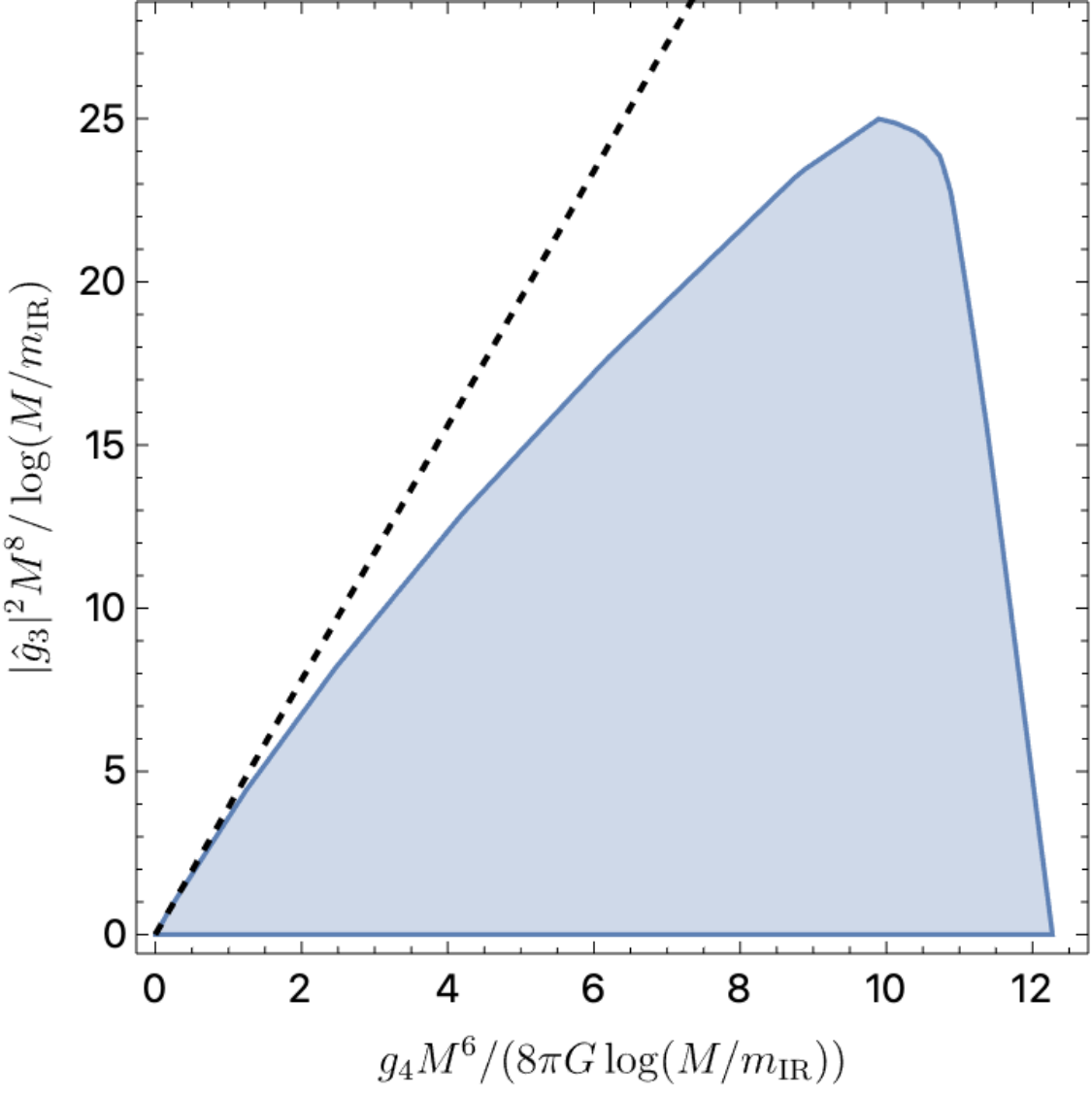} 
\caption{Allowed region for $|\hatg_3|^2$ and $g_4$ in terms of Newton's constant and the spin-$4$ mass gap $M$.
Note that both axes are rescaled by an infrared logarithm $\log(M/\pmin)$. 
Manifestly, both $|\hatg_3|^2$ and $g_4$ obey two-sided bounds;
a nonvanishing cubic coupling $\hatg_3$ requires a nonvanishing quartic $g_4$. The dashed line gives the bound eq.~\ref{bound g4 g3}.
}
\label{fig: exclusion plot for g3sq-g4-grav}
\end{figure}

In the preceding section, we provided example functionals giving upper bounds \eqref{eq: example bound} on $|\hatg_3|^2$ and $g_4$ in terms of gravity. As we emphasized, these bounds were not optimal, and we get stronger ones
by using the numerical parameter choices in appendix \ref{app: numerics}. Our optimal bounds are:
\begin{align} \label{optimal}
|\hatg_3|^2M^8 &\leq 24.9 \log(M/\pmin)-27.6\,,\\
\fft{g_4M^6}{8\pi G} &\leq 12.3 \log(M/\pmin)-13.5\,.
\end{align}
To obtain these, we included all improved sum rules $B_2^{\rm imp}$ and $B_3^{\rm imp}$ with $n_{\rm max}=6$,
and we included additional $\partial_{p^2}^q B_{4}^{(1)\,{\rm imp}}(0)$ up to $q=2$ to get the bound on $g_4$.

A finer way to present the constraint is to carve out the allowed space in the three EFT parameters
$|\hatg_3|^2$, $g_4$ and $G$, as shown in figure \ref{fig: exclusion plot for g3sq-g4-grav}.
These were are computed by using all improved $B_2$ and $B_3$ for $n_{\rm max}=5$ and additional forward-limit contributions from $\partial_{p^2}^q B_{4}^{(1)\,{\rm imp}}(0)$ up
to $q=2$.

A special limit of the bound is the dashed line in figure \ref{fig: exclusion plot for g3sq-g4-grav} which
is tangent to the allowed region near origin; from its slope we find numerically that
\be
\fft{g_4}{8\pi G}  \geq 0.26 |\hat{g}_3|^2 M^2\,.\label{bound g4 g3}
\ee
This is effectively equivalent to the bound $\fft{g_4}{8\pi G}  \geq \frac14 |\hat{g}_3|^2 M^2$
reported in (6.13) of \cite{Maldacena:2015waa} using forward-limit bounds of spin $k\geq 4$.
This bound indicates that it is not possible to turn on a cubic coupling without having a quartic coupling as well.

\begin{figure}[t]
\centering
\includegraphics[width=.6\linewidth]{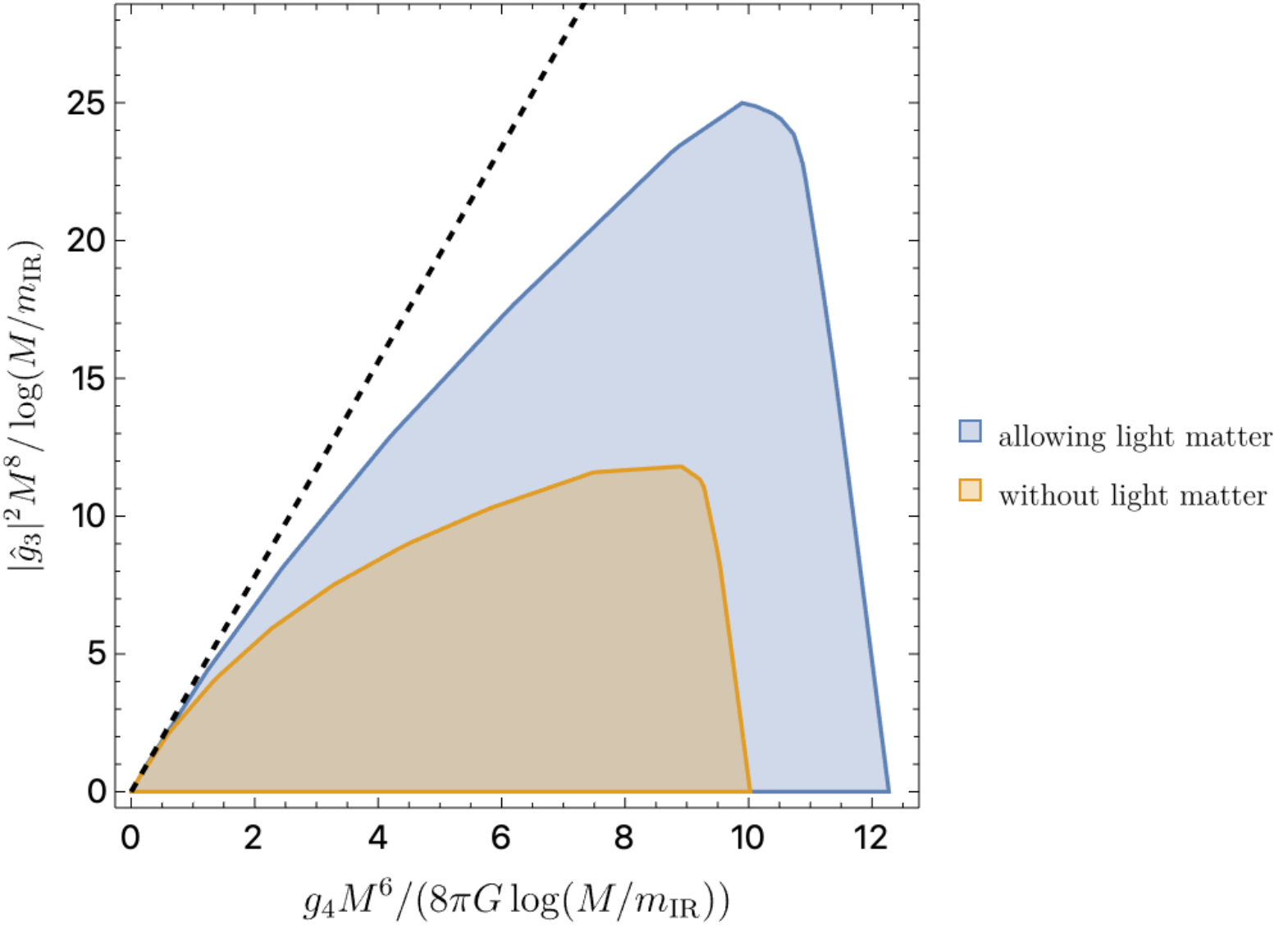} 
\caption{Similar to figure \ref{fig: exclusion plot for g3sq-g4-grav}:
the allowed region for $|\hatg_3|^2$ and $g_4$ in terms of Newton's constant.
For the blue region, light spin-0 and spin-2 massive states are allowed in the spectrum,
while the smaller orange region applies to theories that have no such light fields.
Both regions share the same tangent \eqref{bound g4 g3}.}
\label{fig: compare with and without KK for g3sq-g4-grav}
\end{figure}

It is instructive to see the impact of allowing light spin-0 and spin-2 matter fields on the bounds.
If we assume that such particles are absent, we of course obtain stronger bounds, as shown in 
figure \ref{fig: compare with and without KK for g3sq-g4-grav}, where we used the same space of functionals. Notice both regions with or without light matter share the same tangent line \eqref{bound g4 g3}.

\begin{figure}[t]
\centering
\includegraphics[width=.6\linewidth]{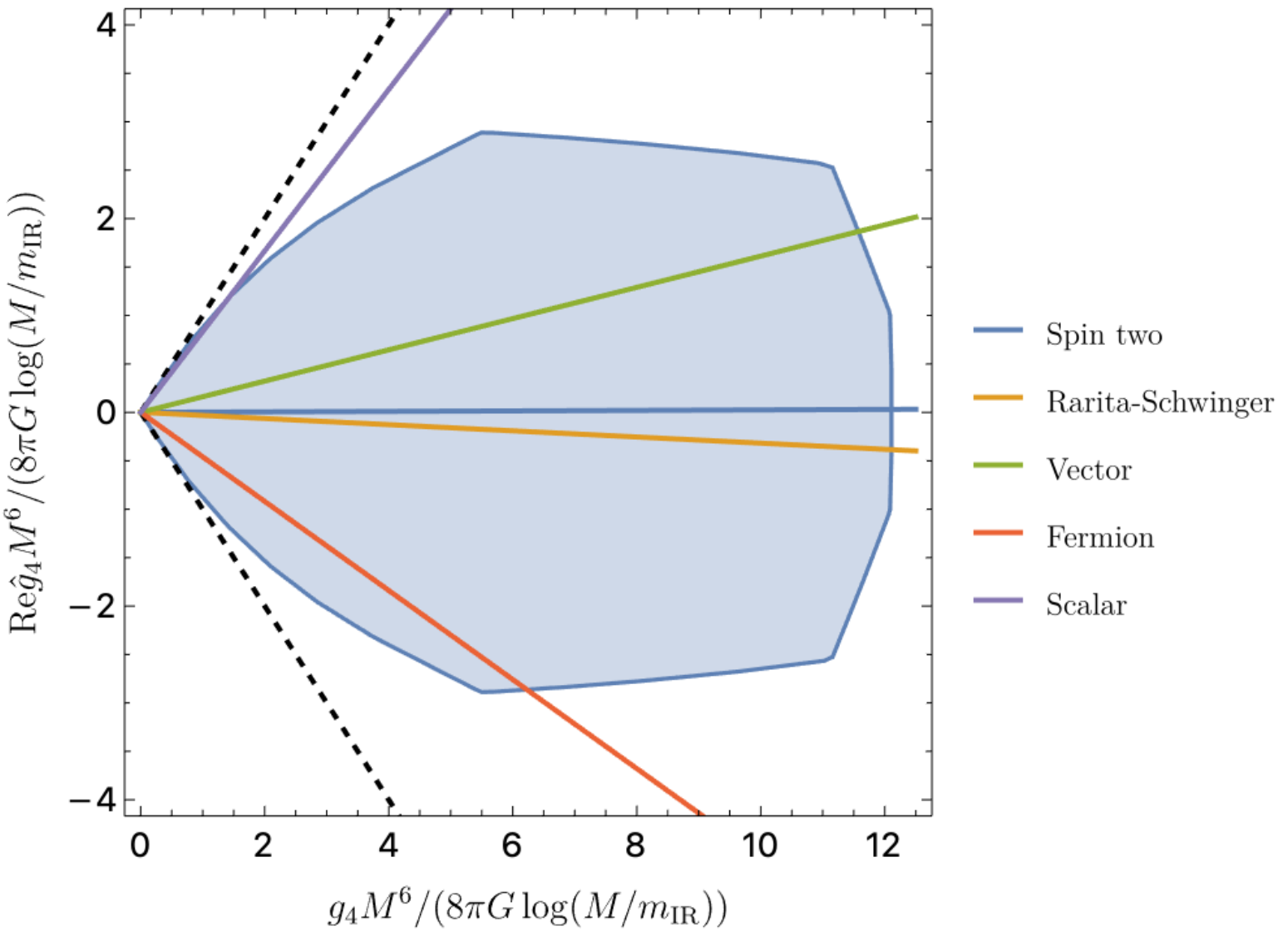} 
\caption{Allowed region for ${\rm Re}\ \hat{g}_4$ and $g_4$ in terms of Newton's constant, the spin-$4$ mass gap $M$,
and an infrared cutoff $\pmin$, with light spin-0 and spin-2 matter fields allowed.
The dashed lines show the positivity bounds \eqref{positivity 4 b}.
Bounds obtained from improved spin-$2,3$ sum rules and $B_4^{(2)\,{\rm imp}}$ at the second derivatives order.
The lines display the loop amplitudes of subsection \ref{sec:models}.}
\label{fig: exclusion plot for g4hat-g4-grav}
\end{figure}

Our approach can also compute bounds for mixed amplitudes, beyond MHV.
In figure \ref{fig: exclusion plot for g4hat-g4-grav}, we derive bounds for ${\rm Re}\ \hat{g}_4$ (i.e., in all positive helicity configuration) and $g_4$ in terms of gravity. In particular, we instead use $n_{\rm max}=6$ to help convergence of functionals built from combining $B_{2,3}^{\rm imp}$ and forward-limit of improved sum rules $\partial_{p^2}^q B_{4}^{(2)\,{\rm imp}}$ up to $q=2$. Although we consider only ${\rm Re}\ \hat{g}_4$, this bound really applies to the magnitude $|\hat{g}_4|$ since from the perspective of the graviton scattering
amplitude the overall phase of a complex coupling can be removed by a little-group rotation and thus cannot be constrained
(only relative phases between couplings can).

The dashed lines display the positivity constraints  \eqref{positivity 4}, which we reproduce here:
\be \label{positivity 4 b}
 g_4 \pm |\hatg_4|\geq 0\,.
\ee
We see that the allowed region is much smaller than the cone \eqref{positivity 4}, which appears
to be tangent to the allowed region.  In particular, a theory which would saturate one of these inequalities is ruled out;
this would correspond to a theory where either of $\alpha$ or $\alpha^\prime$ in \eqref{eq: full S to R4} is set to zero.
(A similar conclusion was reached recently using seemingly different arguments which used only causality within the EFT \cite{deRham:2021bll}; it would be interesting to understand the relationship.)

The solid lines in the figure \ref{fig: exclusion plot for g4hat-g4-grav} display the loop amplitudes of subsection \ref{sec:models} as a function of $N$, assuming a large number of species $N$.
As visible from the plot, theories with $N\gg \frac{M^2}{G}$ are excluded.

\subsection{Bounds involving $D^2R^4$}

\begin{figure}[thb!]
\begin{subfigure}{.42\textwidth}
\centering
\includegraphics[width=\linewidth]{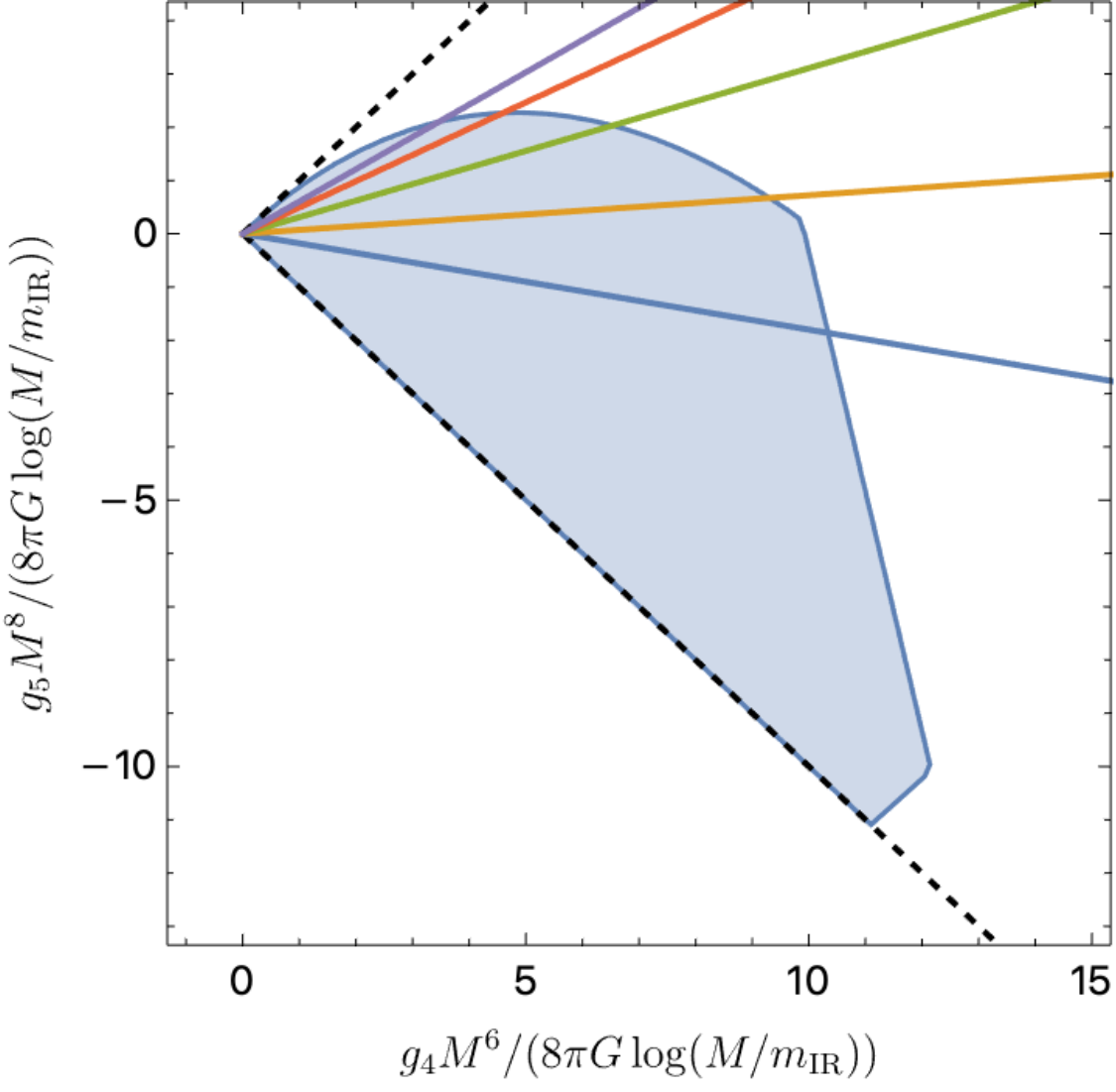} 
\caption{}
\label{fig: exclusion plot for g5-g4-grav}
\end{subfigure}
\hfill
\begin{subfigure}{.57\textwidth}
\centering
\includegraphics[width=\linewidth]{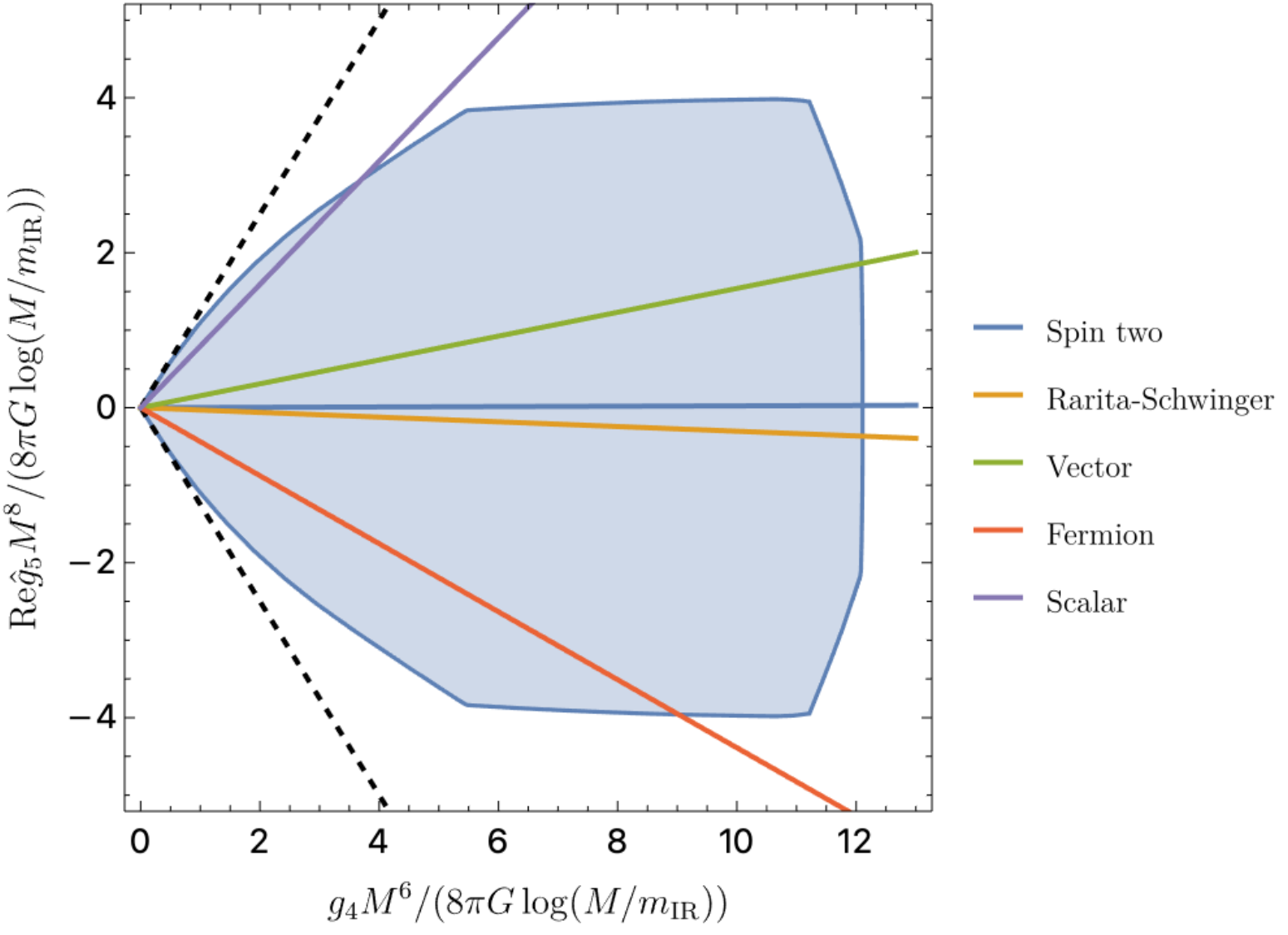} 
\caption{}
\label{fig: exclusion plot for g5hat-g4-grav}
\end{subfigure}
\caption{Allowed region for (a) $g_5$ and $g_4$ and (b) ${\rm Re}\ \hatg_5$ and $g_4$ in units of Newton's constant, the spin-$4$ mass gap $M$, and an infrared cutoff $\pmin$, with light spin-0 and spin-2 matter fields allowed.
The dashed lines show the positivity bounds \eqref{eq: simple bound g5/g4} and \eqref{eq: simple bound g5hat/g4} respectively.
Solid lines display the loop amplitudes of subsection \ref{sec:models}.}
\end{figure}

Moving to the next derivative order, 
we apply spin-$2$ sum rules to relate $g_5$ to $g_4$ and $G$, see figure \ref{fig: exclusion plot for g5-g4-grav}.
Of course, we already know from eq.~\eqref{eq: to do g5/g4 bound} that $g_5/g_4$ has two-sided bounds, and that $g_4/G\log$ is bounded,
and therefore we expected two-sided bounds on $g_5/G\log$ as well.
Similarly we can compute bounds on non-MHV couplings, i.e., for $\hatg_5$ in figure \ref{fig: exclusion plot for g5hat-g4-grav}. We follow the same strategy as for $\hatg_4$ above to project onto the parity eigenstates and thus present bounds for ${\rm Re}\ \hatg_5$, 
although the bounds really apply to $|\hatg_5|$.

An interesting aspect of these plots is that the allowed regions are rather smaller than a cone.
The dashed lines in figure \ref{fig: exclusion plot for g5-g4-grav} are tangent to the plot at origin, which reproduces
the positive bounds \eqref{eq: to do g5/g4 bound}:
\be
g_4 \pm g_5 M^2\geq 0\,.\label{eq: simple bound g5/g4}
\ee 
Similarly, the dashed lines that are tangent to figure \ref{fig: exclusion plot for g5hat-g4-grav} at the origin correspond to simple positive bounds
\be
g_4 \pm 0.8 |\hat{g}_5 M^2| \geq 0\,.\label{eq: simple bound g5hat/g4}
\ee
Curiously, this bound appears stronger than what we could derive from forward limit functionals.

To compute bounds plotted in figures \ref{fig: exclusion plot for g5-g4-grav} and \ref{fig: exclusion plot for g5hat-g4-grav}, we truncated the space of functionals to $n_{\rm max}=5$ built from improved sum rules to from spin-$2$ to spin-$6$, i.e., $B_i^{\rm imp}$ with $i=2\ldots 6$. In the former case we also include forward-limit of sum rules $\partial_t^k B_4^{(1)\,{\rm imp}}(0)$ and $\partial_t^k B_5^{(1)\,{\rm imp}}(0)$ and in the latter 
$\partial_{p^2}^q B_4^{(1)\,{\rm imp}}(0)$ and $\partial_{p^2}^q B_4^{(2)\,{\rm imp}}(0)$, with
up to $q=4$ to guarantee the large $J$ behaviour of functionals are positive. 
Other detailed parameter choices are listed in table \ref{table: numerics}.

\subsection{Bounds involving $D^4R^4$ and low spin dominance} \label{D4R4}

\begin{figure}[t]
\begin{subfigure}{.56\textwidth}
\centering
\includegraphics[width=1\linewidth]{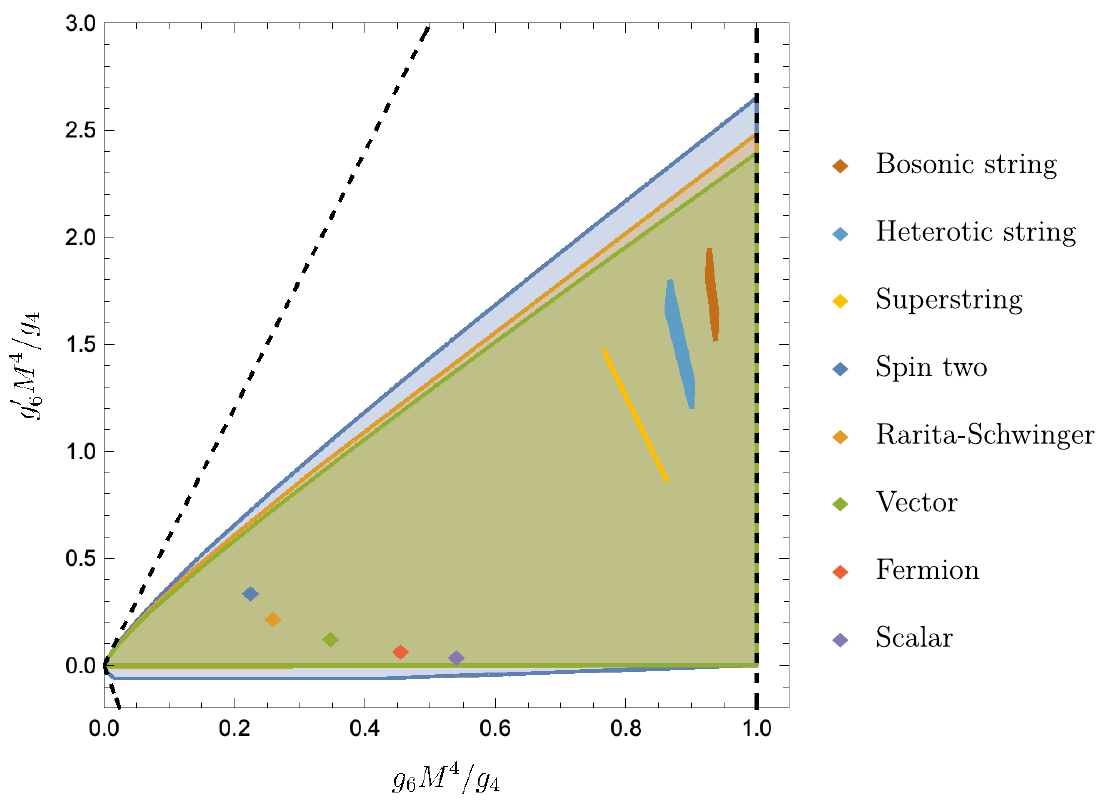}  
\caption{}
\label{subfig: g6p-g6-g4-imp}
\end{subfigure}
\begin{subfigure}{.42\textwidth}
\centering
\includegraphics[width=1\linewidth]{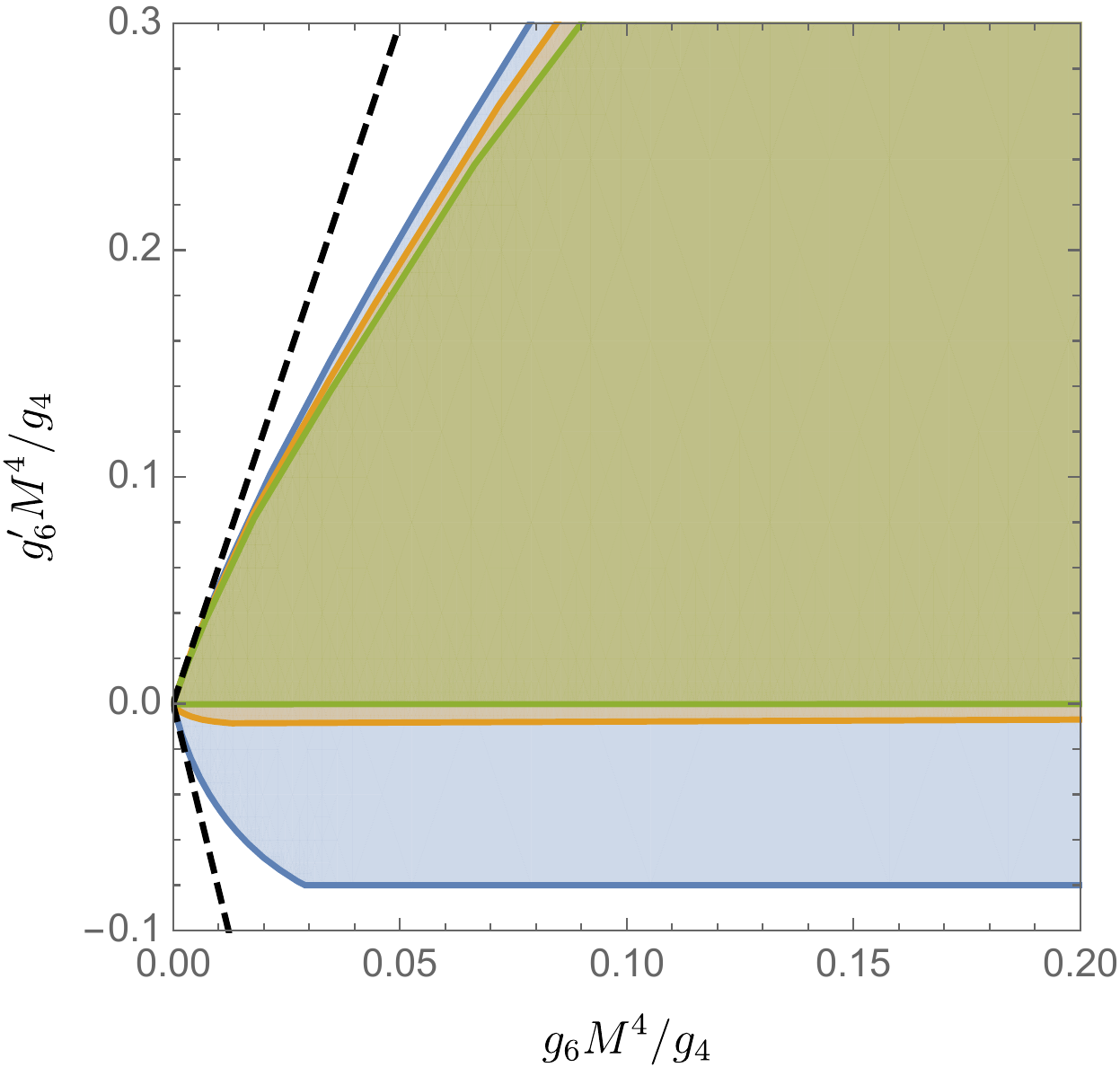} 
\caption{}
\label{fig: negative region for g6p}
\end{subfigure}
\caption{Allowed regions for $g_6$ and $g_6^{\prime}$, normalized by the quartic self-coupling, $g_4$.
The blue, orange and olive regions
show increasing derivative orders $6$, $7$ and $15$ respectively.
On the left we superimposed the values realized in the models of subsection \ref{sec:models}.
(b) is a zoom near the origin of the allowed region, showing the rapid convergence with derivative order. 
Although negative values of $g_6'$ are excluded asymptotically, at any finite derivative order the boundary appears to be tangent to lines of slope $6$ and $-\frac{90}{11}$.
}
\label{fig: bounds for g6p and g6}
\end{figure}

In section \ref{sec2: simple bounds} we reviewed how expanding higher-spin sum rules around forward-limit produces
homogeneous bounds involving e.g., $g_6^{\prime}/g_6$, which gives eq.~\eqref{eq: bound g6 and g6prime}, which we reproduce here:
\be
-\fft{90}{11} \leq \fft{g_6^\prime}{g_6} \leq 6 \qquad \text{(using forward limits and a single null constraint)}\,.
\label{eq:forwardg6}
\ee
An important observation made in \cite{Bern:2021ppb} was that the space of couplings spanned by the theories in Section~\ref{sec:models}, a.k.a ``the theory island'', is much smaller than that given by such homogeneous bounds. In \cite{Bern:2021ppb}, in order to approach the theory island, the authors propose an additional assumption called low-spin-dominance (LSD), which is a constraint on possible UV spectra stating that higher-spin states are suppressed compared to low-spin states.
Quantitively, for MHV amplitudes, LSD implies
\be
{\rm LSD}:\qquad |c^{+\pm}_{4,m^2}| \geq \alpha |c^{+\pm}_{J>4,m^2}|^2\,,
\ee 
where $\alpha\geq 1$ was used to parameterize the size of suppression of higher-spin states. By increasing $\alpha$, ref \cite{Bern:2021ppb} pushes the bounds asymptotically to
\be
0 \leq \frac{g_6'}{g_6} \leq 2 \qquad \text{(assuming LSD)}\,,
\ee 
thus narrowing down the space of couplings to that spanned by the aforementioned theories.

In this paper we do not assume LSD. However, by considering inhomogeneous bounds involving $g_6/g_4$ and $g_6^{\prime}/g_4$
of increasing derivative orders $n$ (meaning null constraints having up to the same scaling dimension as the coupling $g_n$),
we find that we can further narrow down the space of couplings as shown in figure \ref{fig: bounds for g6p and g6}.\footnote{This result was found concurrently in Ref.~\cite{yutin}, which also studies inhomogeneous bounds of the form $g_nM^{2(n-m)}/g_m$, and makes similar observations. We thank the authors of that paper for sharing their draft with us and coordinating submission.}  
As we increase the number of null constraints at higher derivative order, we observe that $g_6^{\prime}$ is approaching $g_6^{\prime}\geq 0$. We can thus claim that positivity of $g_6^{\prime}$ holds asymptotically, which agrees with the prediction of LSD \cite{Bern:2021ppb}.
On the rightmost edge $g_6M^4/g_4=1$ we find the absolute upper bound $g_6^{\prime}M^4/g_4\lesssim 2.38$,
which is significantly closer to the ratio predicted by LSD.

\begin{figure}[t]
\begin{subfigure}{.45\textwidth}
\centering
\includegraphics[height=0.32\textheight, trim= 25 0 0 0]{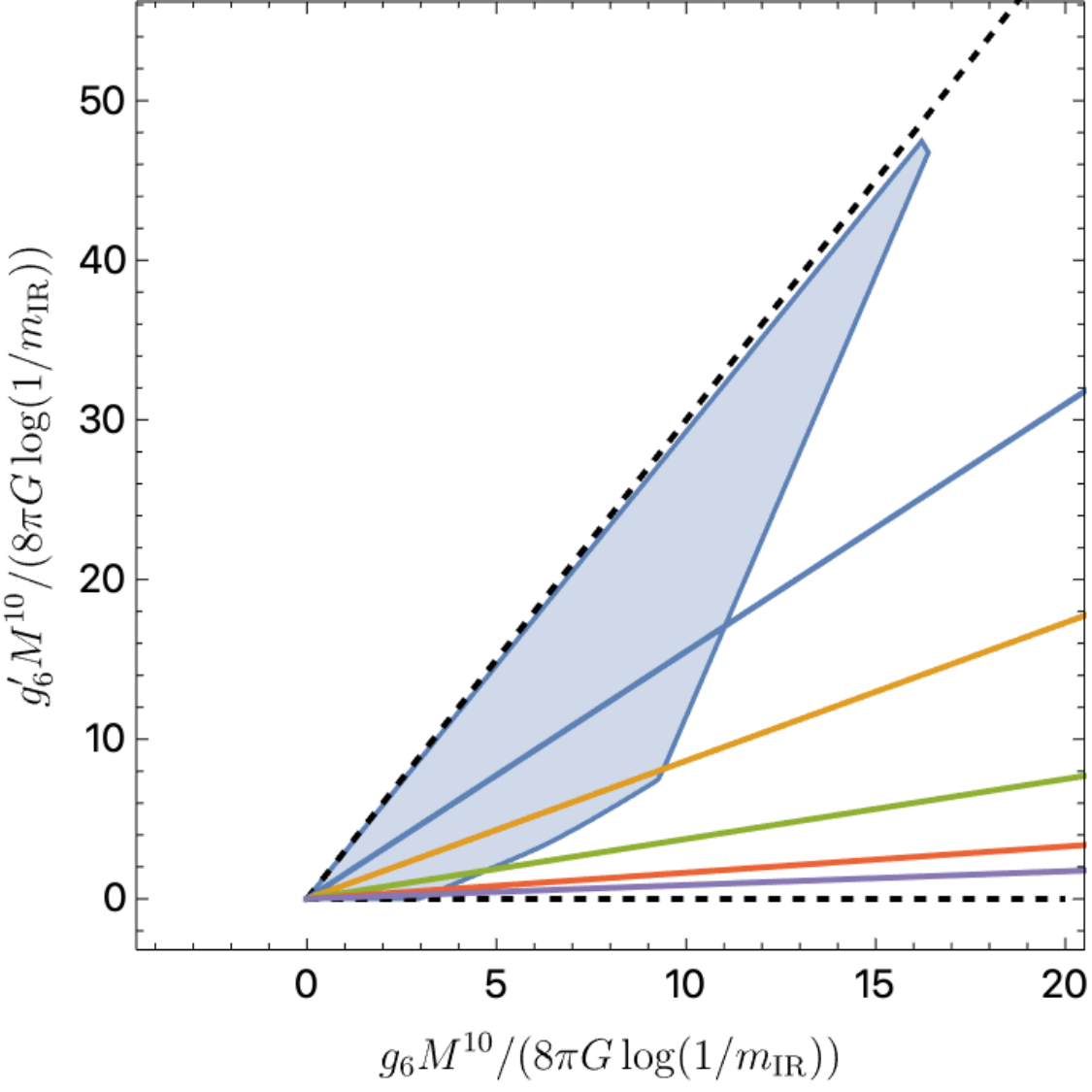}  
\caption{}
\label{subfig: g6p-g6-grav}
\end{subfigure}
\hspace{20pt}
\begin{subfigure}{.46\textwidth}
\centering
\includegraphics[height=0.32\textheight, trim= 20 0 0 0]{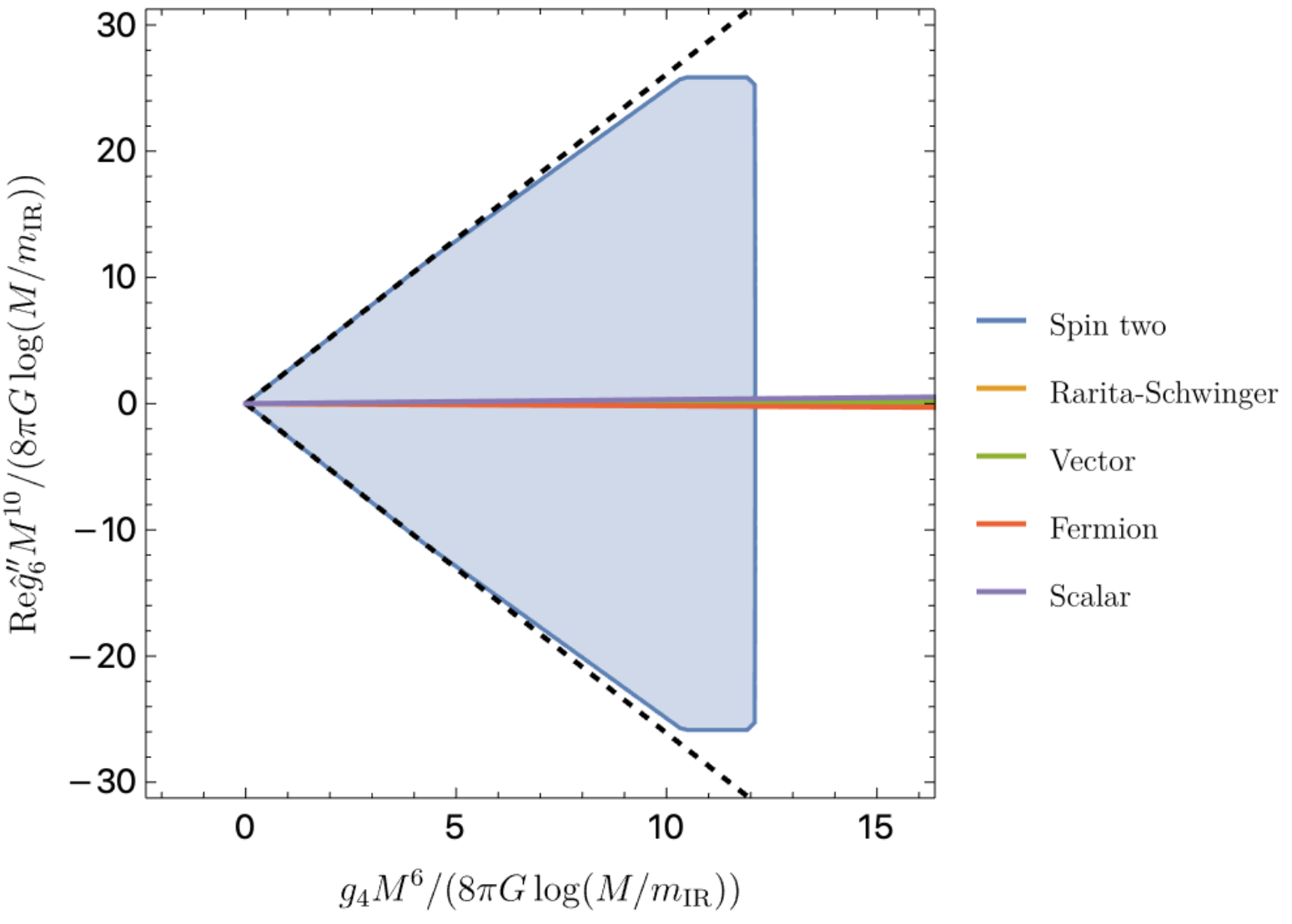} 
\caption{}
\label{fig: exclusion plot for g6bar-g4-grav}
\end{subfigure}
\caption{Allowed regions for (a) $g_6$ and $g_6^{\prime}$ and (b) ${\rm Re}\ \hatg_6^{\prime\prime}$ and $g_4$, normalized by  gravity, and allowing light matter.
The left panel rules out negative values of $g_6'$. 
Moving along the model lines corresponds to changing the number $N$ of particles in the loop.
Curiously, all considered models have very small values of $\hatg_6^{\prime\prime}$.
}
\label{fig: bounds g6 with grav}
\end{figure}

We can do even better by considering impact-parameter bounds on $g_6$ and $g_6^{\prime}$ normalized by gravity, which are the main novelty of this paper. 
This can be computed by using $B_{i}^{\rm imp}$ with $i=2\ldots 6$ with $n_{\rm max}=8$, with the result shown in Fig.~\ref{subfig: g6p-g6-grav}.
Surprisingly, we find, as we include spin-$2$ sum rules, we have the dashed line tangent to the plot at origin that gives bounds
\be
0\leq \frac{g_6'}{g_6} \leq 3\, \qquad\mbox{(our bounds, using spin $k\geq 2$ sum rules)}\,. \label{eq: LSD bound}
\ee
This lower bound is exactly the one predicted by LSD \cite{Bern:2021ppb}. Furthermore, although the absolute upper bound is weaker than that predicted by LSD, the whole region is narrower since the upper dashed line is mostly excluded.

As an additional example, we can also provide bounds involving couplings which contribute to the ${+++-}$ helicity configuration. More concretely, we compute bounds on ${\rm Re}\ \hatg_6^{\prime\prime}$ and $g_4$ in terms of gravity using $B_{i}^{\rm imp}$ with $i=2\ldots 7$ together with  $\partial_{p^2}^q B_4^{(1)\,{\rm imp}}(0)$ and $\partial_{p^2}^q B_4^{(3)\,{\rm imp}}(0)$ up to $q=4$,
with the result shown in  Fig \ref{fig: exclusion plot for g6bar-g4-grav} where the dashed lines correspond to
\be
g_4 \pm 2.61 |\hatg_6^{\prime\prime}| M^4 \geq 0\,.\label{eq: bounds for g6bar/g4}
\ee

We conclude that the discrepancy between dispersive bounds and the known ``theory space''
is very much reduced when more sum rules are used.
Compared with previous work, our constraints are tighter (without making additional assumptions such as ``low spin dominance'') mainly as a result of including more functionals and considering inhomogeneous bounds.
The later greatly helps
since tangent slopes near the origin in figure~\ref{fig: bounds for g6p and g6} converge more slowly than other constraints. 
Given the tendency of higher-derivative coefficients to grow geometrically, we expect even more dramatic reductions in
the allowed volume at higher derivative orders, although we have not studied those systematically.

\subsection{Can higher-spin states be hidden from the Standard Model?} \label{sec:SM}

\begin{figure}
\centering\includegraphics[width=\textwidth]{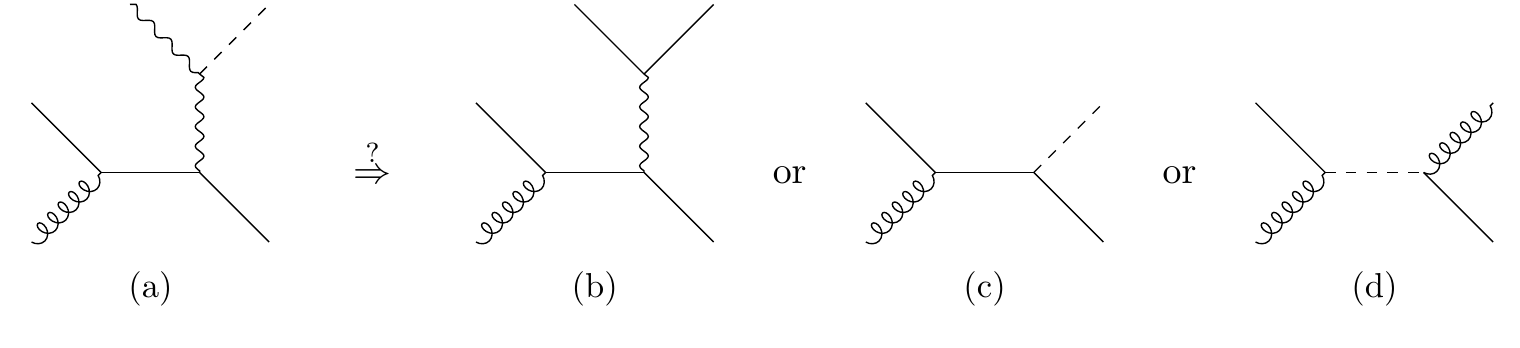}
\caption{Production of a quark jet plus missing energy from a collision of a quark with a gluon;
wavy lines denote graviton and dashed line higher-spin particles.
The rapid high-energy growth of (a) could be softened by modifying its various vertices or propagators,
however each option opens new processes: (b) if the higher-spin states are composite their components can be produced directly with ${\cal M}\sim (\Mpl)^{-2}$ (c) if other vertical exchanges cancel the graviton contribution
they can be produced directly with ${\cal M}\sim (\Mpl)^{-1}$ (d) new horizontal exchanges would lead to resonances with ${\cal M}\sim (\Mpl)^0$. \label{fig:missing E}
}
\end{figure}

This subsection is less rigorous than the rest of this paper; we limit ourselves to non-exhaustive
arguments and order-of-magnitude estimates.

What do collider searches tell us about higher spin particles, of the kind that can lead
to modifications of GR?
Heuristically, because gravity is universal and couples to all matter, one might expect that modifications to it also couple to everything.
Indeed, many specific scenarios of modified gravity, such as string theory models,
predict resonances that couple directly to Standard Model matter.
The non-observation of such resonances impose strong constraints on the string scale:
$M\gsim $7.7 TeV,
and on many other scenarios as well \cite{ATLAS:2017eqx,CMS:2018mgb}.
Can model-independent constraints be made on potential low-scale modifications to GR?

To orient the discussion,
let us imagine a (very hypothetical) scenario where the dynamics of a 10$M_\odot$ black hole,
of size $L\sim 30$km$\sim 1/(10^{-11}\rm eV)$,
were somehow observed to differ by more than $10^{-20}$ from GR's prediction.
(Such a signal strength is orders-of-magnitude weaker than considered in either LIGO or EHT contexts
\cite{Sennett:2019bpc,Carson:2020dez}, but has been chosen to illustrate collider constraints.)
An effective field theorist might try to attribute this to higher-derivative terms of size:\footnote{We recall from the introduction that ``fifth forces'' which do not grow with energy,
involving for example direct couplings of matter to new spin 0 and spin 1 particles,  are unconstrained by our arguments.
We thus omit this possibility here.}
\be
\hatg_3\sim \frac{10^{-20}}{(10^{-11}\rm eV)^4}
\quad\mbox{or}\quad
\frac{\hatg_4}{16\pi G}\sim \frac{10^{-20}}{(10^{-11}\rm eV)^6} \,,\label{param g3 g4}
\ee
the latter being only relevant if the former vanishes. 
Note that $g_5$ and higher-derivative couplings can never dominate over $g_4$, by \eqref{eq: to do g5/g4 bound}.
The bounds \eqref{optimal} indicate that light higher-spin states
with $M<10^{-6}$eV or $M<10^{-8}$eV would then have to exist,
depending on whether $\hatg_3$ or $\hatg_4$ dominates, respectively.\footnote{In this subsection we ignore infrared logarithms
since $(\log MR_{\rm Universe})^{1/4}\lsim 3$.}
Could this be compatible with null results from collider searches?


Temporarily treating higher-spin states as point-like particles (left-panel of figure \ref{fig:missing E}),
there are two possible index structures for their couplings to two gravitons.
These have respectively eight and four derivatives, corresponding to same- and opposite-helicity of the two gravitons. 
The production cross-section of the higher-spin particle at a hadron collider, in association with a graviton and a jet
(to make missing momentum to trigger on) are then
\be
 |{\cal M}|^2_{pp\to {\rm jet}+\slash\!\!\!\! E}|_{(a)} \sim \left\{\begin{array}{l}
 \alpha_s \frac{\hat{s}^{8}}{\Mpl^4M^{12}}, \quad \mbox{if $|c_{J=4}^{++}|^2\neq 0$,} \\
 \alpha_s \frac{\hat{s}^{4}}{\Mpl^4M^{4}}, \quad \mbox{if $|c_{J=4}^{++}|^2= 0$.}\end{array}\right.
\ee
Here $\hat{s}$ is the partonic center of mass energy, and we have included a factor of $\alpha_s$ to create the jet.
For crude estimates, we will consider that missing-energy searches \cite{ATLAS:2017bfj,CMS:2017zts}
exclude amplitudes of size $|{\cal M}|\gsim 1$ when $\sqrt{\hat{s}}\sim {\rm TeV}$.
Thus, despite the strong suppression by $\Mpl\sim 10^{15}$TeV,
we see that in the first scenario $M\lsim {\rm MeV}$ is clearly ruled out.
However, the analysis in this paper does not rule out the option that $|c^{++}|^2=0$, which is the boundary of slope $-1$ in fig.~\ref{fig: exclusion plot for g5-g4-grav}; but in this case there is still a tension with collider searches if $M\lsim 10^{-3}$eV.
We conclude that if higher-spin particles are point-like, colliders and our bounds
easily exclude effects of the size \eqref{param g3 g4}; effects on larger distances would be even smaller.

Now, the new states don't have to be point-like: the above amplitudes could be softer.
However, softening mechanisms generically open up other production mechanisms that are also constrained,
as in the stringy example discussed above.  These are depicted in figure \ref{fig:missing E}.

For example, the higher-spin states could be two-particle states of some light fields that couple to us
only through gravity, which would soften the top vertex.
However since two-graviton couplings in this scenario are small $\sim 1/\Mpl^2$,
long-distance effects are negligible unless there are a very large number of such light fields (for the same reason
that quantum effects from Standard Model loops are extremely small for simple observables around
macroscopic black holes \cite{Bjerrum-Bohr:2014zsa}).
Specifically, in order to have a $10^{-20}$ effect at $L$, at a minimum $N\gsim 10^{60}$ of them would be needed
(such that $N/(\Mpl^2L^2)\sim10^{-20}$; fields with mass $m<L^{-1}$ can be treated as effectively massless in this estimate).
This would lead to a very low ``cutoff'' $\Mpl/\sqrt{N}$ above which gravity becomes strongly coupled,
and values $N>10^{32}$ are typically not considered for this reason \cite{Dvali:2007hz}.
Besides possible cosmological constraints, let us simply mention that even with the minimal coupling (b),
since $({\rm TeV}/\Mpl)^4\sim 10^{-60}$, for such $N$ even current colliders could have detected a production cross-section 
\be
 |{\cal M}|^2_{pp\to {\rm jet}+\slash\!\!\!\! E}\big|_{(b)} \sim \alpha_s\frac{\hat{s}^2}{\Mpl^4} \times N.
\ee
Other softening mechanisms may modify other ingredients in figure (a).
For example new higher-spin states could be exchanged between Standard Model fields and the higher-spin particle.
But these could then be produced directly with a four-derivative graviton-strength coupling as in (c):
\be
 |{\cal M}|^2_{pp\to {\rm jet}+\slash\!\!\!\! E}\big|_{(c)} \sim \alpha_s\frac{\hat{s}^6}{\Mpl^2M^4}, 
\ee
detectable if $M\lsim 10{\rm keV}$.  This amplitude could again be softened,
but again the cure is worse than the disease since
this almost certainly requires new states exchanged on the horizontal propagator, leading to string-like resonances
(d) as discussed above.

These estimates, if correct, challenge the notion that modifications of gravity at large scales could be hidden from colliders.
We hope that more robust and model-independent statements will be obtained in the future.

\section{Conclusions}\label{sec:conclusion}

In this paper, we analyzed constraints on low-energy graviton-graviton scattering, assuming that causality
and other basic principles apply at all energies.
Theoretically, graviton scattering is an ideal way to probe potential modifications of Einstein's theory of gravity.
At the same time, causality of a scattering process translates into well-understand mathematical properties like analyticity.
This implies Kramers-Kronig-type dispersion relations, which express
low-energy observables in terms of unknown but positive absorption probabilities at high energies,
and which we have systematically analyzed.


Our key result \eqref{optimal} is simple to state:
if a higher-derivative correction were measured, corresponding to ${\cal L}= \frac{1}{16\pi G}(R +r_0^4 {\rm Riem}^3+\ldots)$,
and if Nature respects causality as we understand it,
then a spin-4 particle must exist whose Compton wavelength is at least as long as the length $r_0$: $M^{-1}> r_0$.

The notion that higher-derivative corrections can
come from heavy particles may not surprise an effective field theorist.  The scaling with $\Mpl$ of our bounds is however interesting.  An effective field theorist might argue that the apparent breakdown of the derivative expansion at the length $r_0$ suggests the existence of new states with mass $\sim r_0^{-1}$ or lighter, but their friend could have objected that the states could be much heavier: the required three-point couplings respect unitarity bounds as long as $M\lsim (\frac{\Mpl}{r_0})^{\frac13}$.
We excluded the second option: new higher-spin states \emph{must} exist with mass $M\lsim r_0^{-1}$ or lighter.
The $\Mpl$ scaling of our bounds means that heavy states
couple to two gravitons with a strength that never exceeds the three-graviton coupling of Einstein's theory.

This means that the effective field theory approach to gravity can never describe modifications to
Einstein's theory that are not parametrically small:
in any situation where an EFT is justified at the distance $L$ ($ML\gg 1$), corrections are small ($r_0/L\ll 1$).

Our bounds on cubic couplings are qualitatively similar to the parametric bounds obtained in \cite{Camanho:2014apa},
with the important novelty that we find sharp bounds with precise numerical coefficients.
As discussed in section \ref{sec: CEMZ},
this is done by imposing a quantum notion of causality based on commutators and crossing symmetry,
rather than a classical one based on time advances.
Technically, we apply dispersion relations to a physical region where transverse momenta are below the cutoff
$M$ but energies are above $M$ (but still much less than $\Mpl$, so that tree-level diagrams give a good approximation).
The finite momentum transfer also allows us to constrain local (quartic) and non-local (cubic)
self-interactions on the same footing.

Our bounds assume that new states satisfy $M\ll \Mpl$. They are expected to be valid modulo
relative corrections $\sim \frac{M^2}{\Mpl^2}$ from loop effects within the EFT. It would be interesting to explicitly compute  some of these corrections.

We treated the graviton as exactly massless.
By continuity, we expect our bounds to also control the scattering of transversely polarized gravitons with $m\ll M$.
For massive gravitons, one however anticipates much stronger constraints
coming from the study of longitudinal polarizations, which do not have a smooth $m\to 0$ limit
(see for example \cite{deRham:2017zjm,Cheung:2016yqr,Afkhami-Jeddi:2018apj}). 

Our results significantly reduce the gap observed in \cite{Bern:2021ppb}
between dispersive bounds and ``theory space'', as further discussed in section \ref{D4R4}.
Yet, it is noteworthy that our bounds are not saturated by known theories.
This could indicate that stronger bounds are still possible; this is of course also suggested by the fact that all our bounds involving $G$ contain infrared logarithms.
It will be interesting to find a way to obtain infrared-safe bounds, and to further investigate the gap between theory space and dispersive bounds.

As explained in \cite{Caron-Huot:2021enk}, dispersive bounds for scalar amplitudes in flat space can be lifted to AdS and
imply corresponding bounds on OPE coefficients of holographic CFTs.
Physically, the same must be true for graviton scattering as well. 
For example, (\ref{optimal}) translates to a bound on three-point OPE coefficients of stress tensors
in holographic 3d CFTs. In AdS${}_4$, infrared logarithms are simply replaced by $\log (MR_\mathrm{AdS})$.

The notion of causality that we used in the paper applies to scattering processes in an otherwise empty and flat region of space.
Physically, since we mainly probe energies of order $M$ and impact parameters $b\sim M^{-1}$,
we expect our conclusions to remain valid in any spacetime in which a flat patch of radius $\gg M^{-1}$ exists,
including our own. (This is the reason for the agreement between bounds in flat space and AdS.)
Evidently, studying energies above the EFT cutoff was essential to probe the mass and couplings of higher-spin states.
An important question is whether some of our bounds can be understood from thought experiments within the EFT itself.

From the generic collider estimates given in section \ref{sec:SM}, assuming causality as we understand it,
we find it difficult to imagine that higher-derivative terms could be visible at macroscopic distances.
A better understanding of interactions between gravitational higher-spin states and matter
should lead to more precise constraints.

In summary, experimental verifications of Einstein's theory test
a basic question: can signals travel faster than the speed of light?

\section*{Acknowledgements}	

We thank Robert Brandenberger, Clifford Cheung, Cyuan-Han Chang, Jim Cline, Claudia de Rham,
Yanky Landau, Dalimil Mazac, Leonardo Rastelli, Katelin Schutz, Andrew Tolley and Sasha Zhiboedov for discussions.
DSD, SCH and YZL are supported by the Simons Foundation through the Simons Collaboration on the Nonperturbative
Bootstrap.
DSD is also supported by a DOE Early Career Award under grant no.\ DE-SC0019085.
SCH is also supported by the Canada Research Chair program and the Sloan Foundation.
JPM is supported by the DOE under grant no.\ DE-SC0011632.
The computations presented here were conducted in the Resnick High Performance Computing Center, a facility supported by Resnick Sustainability Institute at the California Institute of Technology.
Some figures were made using the TikZ-Feynman package \cite{Ellis:2016jkw}.

\pagebreak
	
\appendix



\section{Spinning partial waves and Wigner-d functions}\label{app:Wigner}

The Wigner-d functions are defined as
\begin{equation} \label{Wigner d}
        d_{h,h'}^J(x) = {\cal N}_{h,h'}^J \left(\frac{1+x}{2}\right)^{\frac{h+ h'}{2}}  \left(\frac{1-x}{2}\right)^{\frac{h-h'}{2}} {}_2F_1\left(h - J, J + h +1; h - h' + 1;\tfrac{1-x}{2}\right)\,,
\end{equation}
where we defined the normalization
\begin{equation}
 {\cal N}_{h,h'}^J =  \frac{1}{\Gamma(h-h'+1)} \sqrt{\frac{\Gamma(J + h + 1) \Gamma(J - h' + 1)}{\Gamma(J - h + 1)\Gamma(J + h' + 1) }}\,,
\end{equation}
so that $d_{h,h'}^J(1) = \delta_{h,h'}$.
In the main text we will use instead the helicity-stripped d-functions 
\begin{equation}
        \tilde d_{h',h}^J(x) = {\cal N}_{h,h'}^J \,  {}_2F_1\left(h - J, J + h +1; h - h' + 1;\tfrac{1-x}{2}\right)\,,
\end{equation}
Note that $\tilde d_{0,0}^J(x)$ is the familiar Legendre polynomial.
Some useful properties of these functions are the symmetries
\begin{equation}
        d_{-h - h'}^J(x) =  d_{h' h}^J(x)\,, \qquad  d_{h', h}^J(x) = (-1)^{h-h'}  d_{h,h'}^J(x)\,,
\end{equation}
and their transformation under parity
\begin{equation}
        d_{h, h'}^J(-x) = (-1)^{J + h}  d_{h, -h'}^J(x)\,, 
\end{equation}
as well as orthogonality
\begin{equation}
        \int_{-1}^1 dx \, d_{h,h'}^J(x) \, d_{h,  h'}^{J'}(x) =  \frac{2}{2J+1} \delta_{J J'} \,\,.
\end{equation}

Physically, the decomposition \eqref{Im a from c} corresponds to exchanging heavy states $X$
with the following three-point amplitudes, which are uniquely fixed by Poincar\'{e} symmetry \cite{Arkani-Hamed:2017jhn}:
\begin{align} \label{three points}
{\cal M}(1^+,2^-,3^X) &= c_{J,m^2}^{+-} \, \sqrt{16\pi m^2} \, \langle 1 \mathbf{3} \rangle^{J-4}  \langle 2 \mathbf{3} \rangle^{J+4}  \langle 1 2 \rangle^{-J} \\
{\cal M}(1^+,2^+,3^X) &= c_{J,m^2}^{++}\, \sqrt{16\pi m^2} \langle 1 \mathbf{3} \rangle^{J}  \langle 2 \mathbf{3} \rangle^{J}  \langle 1 2 \rangle^{-2-J} [12]^2 \,.
\end{align}
We can then glue two three-point amplitudes by summing over immediate polarizations to conclude the partial wave expansion eq.~\eqref{eq: partial wave expansion} (where $a_J^{h}(s)$ is replaced by eq.~\eqref{Im a from c}). For example, for $\mathcal{M}^{+--+}$, we have \cite{Arkani-Hamed:2017jhn}
\be
 &\sum_{X}\mathcal{M}(1^+,2^-,5^X)\mathcal{M}(3^-,4^+,5^{X^\ast})
 \nn\\
 &=|c_{J,m^2}^{+-}|^2 16\pi m^2 \langle 12\rangle^{-J}\langle 34\rangle^{-J} \times
\sum_{j} \fft{(J+4)!^2 (J-4)!^2}{j! (J+4-j)!^2 (j-8)!} \langle 13\rangle^j \langle 14\rangle^{J+4-j}\langle
23\rangle^{J+4-j}\langle 24\rangle^{j-8}\,.
\label{eq: glue three-point heavy}
\ee
We can then convert to the center-of-mass frame where
\begin{align}
& \lambda_1 =\sqrt{m} \left(
\begin{array}{c}
 1 \\
 0 \\
\end{array}
\right)\,,\quad \lambda_2 =\sqrt{m} \left(
\begin{array}{c}
 0 \\
 1 \\
\end{array}
\right)\,,\quad \lambda_3 =i\sqrt{m} \left(
\begin{array}{c}
  \sqrt{\fft{1+x}{2}}\\
 \sqrt{\fft{1-x}{2}} \\
\end{array}
\right)\,,\quad
\lambda_4 =\sqrt{m} \left(
\begin{array}{c}
 \sqrt{\fft{1-x}{2}}  \\
 -\sqrt{\fft{1+x}{2}}  \\
\end{array}
\right)\,.
\end{align}
Using this parameterization, eq.~\eqref{eq: glue three-point heavy} can then be analytically summed to Wigner-d functions.

We will also be interested in the behavior at large $J$, taken with fixed impact parameter $b = 2J/m$, which is given by
\begin{equation} \label{Bessel limit}
        \lim_{ J,m \to \infty}  \tilde d_{h,h'}^{J} \left(1 - \tfrac{2p^2}{m^2}\right) = \frac{J_{h-h'}( bp )}{(p/m)^{h-h'}} \,,
\end{equation}
where $J_{h-h'}( bp )$ on the right-hand side is a Bessel function of the first kind.

The following are also useful when expanding around the forward limit, which corresponds to expanding $x$ around 1
\begin{align}
&\frac{d^n}{dx^n} \tilde d_{0,0}^J(x)\Big|_{x=1} \ge 0\,, \qquad   \frac{d^n}{dx^n} (\tilde d_{4,4}^J(x) - (-1)^J \tilde d_{4,-4}^J(x))\Big|_{x=1} \ge 0  \,.
\end{align}

\section{Amplitudes from matter and Kaluza-Klein exchanges}
\label{app:kk}

In this appendix, we record the contribution from light ``matter'' fields to the low-energy amplitudes \eqref{eq:fghlow}.
If the light matter fields have discrete masses $\mlight$, these
amplitudes are rational functions whose poles match the spin-0 and spin-2 contributions to the above partial waves:
\begin{align}
        f_{\rm matter}(s,u) =
                \sum_{\mlight < M}\frac{|g_0^{++}(\mlight)|^2}{\mlight^2-t}
                +\sum_{\mlight < M}\frac{|g_2^{++}(\mlight)|^2(1-6su/\mlight^4)}{\mlight^2 -t}\,.
\label{eq: KK MHV amp}
\end{align}
To get the last factor, for example, we used the $J=2$ partial wave \eqref{f: tpole} on the $t=\mlight^2$ pole:
\begin{align}
\tilde{d}_{0,0}^2\left(1+\tfrac{2s}{\mlight^2}\right) = 1+ \frac{6s(s+\mlight^2)}{\mlight^4} \simeq 1-\frac{6su}{\mlight^4} \qquad\mbox{when $t=\mlight^2$}\,.
\end{align}
Away from the pole, the spin-2 exchange \eqref{eq: KK MHV amp}
is unique up to contact interactions, which are, to the same derivative order, either linear in $t$ or constant.
(That is, different choices of writing \eqref{eq: KK MHV amp} could shift the $g_4$ and $g_5$ coefficients in
\eqref{eq:fghlow} by multiples of $|g_2^{++}(\mlight)|^2$.)  The above choice will lead to the amplitude with the best possible Regge growth (spin-4 at fixed-$s$, and spin-2 at fixed-$t$), and to the simplest sum rules,
but other choices would not significantly change the analysis.

In reality, since they can decay (in particular, to two gravitons),
the light fields cannot give sharp poles but at best resonances;
this simply replaces the sum by an integral:
\be
\sum_{\mlight < M} \frac{|g_J^{++}(\mlight)|^2}{\mlight^2-t}(\cdots) \mapsto 16\pi(2J+1)
\int_0^{M^2} \frac{d\mlight^2}{2\pi \mlight^8} \frac{|c^{++}_{J,\mlight^2}|^2}{\mlight^2-t-i0}(\cdots)\,,
\ee
where the partial waves $c^{++}_{J,\mlight^2}$ are normalized as in \eqref{eq:Imf}.

Only same-helicity couplings $g_2^{++}$ appear in the above due to angular momentum
rules, which forbid a particle of spin $<4$ from decaying to a $({+}{-})$ two-graviton state. 
For the same reason, the single minus amplitudes do not receive contributions 
\begin{align}
g_{\rm matter}(s,u) = 0.
\end{align}
Finally, the all-plus amplitudes are
\begin{subequations}
\begin{align}
h_{\rm matter}(s,u) & = \sum_{\mlight< M}g_0^{++}(\mlight)^2 \Big(\fft{s^4}{\mlight^2-s}+\fft{t^4}{\mlight^2-t}+\fft{u^4}{\mlight^2-u}\Big) \\  & +  \sum_{\mlight<M} g_2^{++}(\mlight)^2 \Big(\fft{s^2(s^2-6 t u)}{\mlight^2-s}+
\fft{t^2(t^2-6 s u)}{\mlight^2-t}+\fft{u^2(u^2-6 s t)}{\mlight^2-u}\Big)\,.
\end{align}
\label{eq: KK 2 amp}
\end{subequations}
It is important here that each numerator is quartic in Mandelstam invariants,
so as to cancel the denominator in \eqref{eq:Mdef}.
Contact ambiguities have been fixed to avoid high-energy growth with Regge spin greater than 2.

\section{Details on numerics}\label{app: numerics}

In this appendix, we give details on our numerical implementation of dispersive bounds. We mostly follow \cite{Caron-Huot:2021rmr}, though some new complications arise when considering spinning external particles (as opposed to scalars). Most of the issues are not specific to 4d, so we phrase our discussion in general spacetime dimensions.

A generic dispersive optimization problem for takes the following form: 
\be
\label{eq:continuousoptimizationproblem}
\textrm{maximize}&\ \  \sum_i \int_0^M dp\, \psi_i(p) \left.B_{i}(p^2)\right|_\mathrm{light} \nn\\
\textrm{such that}&\ \ \sum_i \int_0^M dp\, \psi_i(p) B_{i}(p^2)[m^2,J,\l] \succeq 0
\quad
\forall\textrm{$m>M$, reps $\rho=[J,\l]$}\,.
\ee
The $\cA_{i,-p^2}$ are dispersive sum rules, labeled by an index $i$, evaluated at $t=-p^2$. Here, $\rho$ runs over little-group representations of massive particles that can appear in intermediate channels. We can associate to each $\rho$ a Young diagram $[J,m_2,\dots,m_n]$ where $n=\lfloor\frac{d-1}{2}\rfloor$. We denote $\l=[m_2,\dots,m_n]$ and write $\rho=[J,\l]$. Aside from a few special cases at small $J$, the $\rho$'s fall into families where $J$ can be arbitrarily large, while $\l$ remains fixed.
For each $\rho$-family, the heavy density $B_{i}[p^2,m^2,J,\l]$ is an $n\x n$ matrix, where $n$ (which depends on $\l$) is the number of three-point structures for two external particles producing an internal $\rho$-particle.

Following \cite{Caron-Huot:2021rmr}, we choose a polynomial basis for the wavefunctions $f_i(p)$:
\be
\psi_i(p) &= \sum_{n=0}^N a_{i,n} p^n.
\ee
Then (\ref{eq:continuousoptimizationproblem}) becomes a semidefinite program with decision variables $a_{i,n}$ and an infinite number of positivity constraints labeled by $m,\rho$. We truncate to a finite number of constraints using a combination of discretization and polynomial approximations. Specifically, we split $(m,\rho)$-space into the following four regimes:

\paragraph{Fixed $J$, fixed $m$.} Choosing some $J_\mathrm{max}$, we impose positivity at the finite set of representations $\rho=[J,\l]$ with $J\leq J_\mathrm{max}$ and masses $m^2=1/(1-x)$, where $x \in \{0,\de x,2\de x,\dots,\lfloor \frac{1}{\de x}\rfloor \de x\}$.  Our detailed  parameter choices are listed in table~\ref{table: numerics}.

\paragraph{Large $J$, fixed $m$.} Unlike in the case of external scalar particles \cite{Caron-Huot:2021rmr}, we found that it is important to explicitly control the large-$J$ limit of our functionals at fixed $m$. To do so, we compute a series expansion of the integrated heavy density at large $J$
\be
\int_0^1 dp\, p^n B_{i}[p^2,m^2,J,\l] &\sim \# J^r + \# J^{r-1} + \dots\,.
\ee
Truncating this expansion to a finite number of terms, and multiplying by an appropriate power of $J$, we obtain a polynomial approximation for the heavy density as a function of $J$. We discretize $m$ as before and, for each $m$, impose positivity of the resulting matrix polynomial of $J$ on the interval $[J_\mathrm{max},\oo)$.

\paragraph{Large $m$, fixed $b=\frac{2J}{m}$.} An important set of positivity constraints come from the impact parameter scaling limit $m\to \oo$ with fixed $b=\frac{2J}{m}$. In this regime, the heavy densities behave as
\be
\label{eq:scalinglimitapproxes}
\lim_{m\to \oo} B_{i}[p^2,m^2,\frac{bm}{2},\l] &\sim \frac{\mathcal{J}_{i,\l}(pb)}{m^{2k_i}}\,,
\ee
where $k_i$ is the Regge spin of the $i$-th sum rule. Consequently, only the lowest-spin sum rules contribute in this limit. Thus, in the scaling limit, we set all higher-spin sum rules to zero and replace the lowest-spin sum rules with their approximations (\ref{eq:scalinglimitapproxes}). The matrices $\cJ_{i,\l}(pb)$ have entries that involve Bessel functions and their derivatives. Following \cite{Caron-Huot:2021rmr}, we choose a cutoff $B$ and impose positivity at a discrete set of $b\leq B$, $b\in \{\e_b,\e_b+\de_b,\dots,\e_b+\lceil\frac{B-\e_b}{\de_b}-1\rceil \de_b\}$.

\paragraph{Large $m$, large $b$.} As in \cite{Caron-Huot:2021rmr}, we must take care to impose positivity in the impact parameter scaling limit for $b>B$. Let us review the trick used there. For scalar scattering, the heavy density $\mathcal{C}^\mathrm{imp}_{2,-p^2}[m^2,J]$ is simply a $1\x1$ matrix, with scaling limit
\be
\lim_{m\to \oo} \mathcal{C}^\mathrm{imp}_{2,-p^2}[m^2,\tfrac{mb}{2}] &= \frac{2\Gamma(\tfrac{d-2}{2})}{m^4} \frac{J_{\frac{d-4}{2}}(bp)}{(bp/2)^{\frac{d-4}{2}}}\,.
\ee
Integrating against a wavefunction, we have
\be
\int_0^1 dp \psi(p) \frac{J_{\frac{d-4}{2}}(bp)}{(bp/2)^{\frac{d-4}{2}}}
&= A(b) +B(b) \cos\p{b-\frac{\pi(d-1)}{4}} + C(b) \sin\p{b-\frac{\pi (d-1)}{4}}\,,
\label{eq:integralinscalinglimit}
\ee
where $A(b),B(b),C(b)$ have well-behaved asymptotic expansions in inverse powers of $b$. The large-$b$ behavior of $A(b)$ can be determined from the small-$p$ behavior of the integral on the left-hand side. The oscillating terms $B(b)$ and $C(b)$ come from the other endpoint of the integral $p=1$.

The key idea is to replace (\ref{eq:integralinscalinglimit}) with 
\be
\label{eq:thingwithphi}
A(b)+B(b)\cos\f+C(b)\sin\f
\ee
and impose positivity for $b>B$ and {\it arbitrary\/} $\f$ --- not just for $\f=b-\frac{\pi(d-1)}{4}$. Although this is a stronger condition than positivity of (\ref{eq:integralinscalinglimit}) (and thus, could lead to weaker bounds), we expect it to be a good approximation at large $b$, where $b-\frac{\pi(d-1)}{4}$ is a rapidly-varying phase relative to $A(b), B(b), C(b)$.

To see how to impose positivity of (\ref{eq:thingwithphi}), we write
\be
\label{eq:keytrick}
A(b)+B(b)\cos\f + C(b) \sin\f &= \left(\cos\tfrac \f 2\ \sin\tfrac\f 2\right)
M(b)
\begin{pmatrix}
\cos\tfrac \f 2\\
\sin\tfrac\f 2
\end{pmatrix}\,,
\ee
where
\be
M(b)=\begin{pmatrix}
A(b)+B(b) & C(b) \\
C(b) & A(b)-B(b)
\end{pmatrix}\,.
\label{eq:matrixpositivitycondition}
\ee
Positivity of (\ref{eq:keytrick}) for arbitrary $\f$ is equivalent to the statement that $M(b)$ is positive-semidefinite: 
\be
M(b)\succeq 0\,.
\ee
We can now expand $M(b)$ at large $b$, and approximate it as a power of $b$ times a $2\x2$ matrix polynomial of $b$. We then impose positivity of this matrix polynomial for $b\geq B$.

Let us now consider the case of interest for this work, where the heavy density is a matrix. In the impact parameter scaling limit, the integral against wavefunctions $f_i(p)$ has the same form as before:
\be
\sum_i\int_0^1 dp \psi_i(p) \cJ_{i,\l}(pb) &= A_\l(b) +B_\l(b) \cos\p{b-\frac{\pi(d-1)}{4}} + C_\l(b) \sin\p{b-\frac{\pi (d-1)}{4}}\,,
\ee
where now $A_\l(b),B_\l(b),C_\l(b)$ are $n\x n$ matrices ($n$ depends on $\l$). As before, we replace $b-\frac{\pi(d-1)}{4}$ with $\f$ and seek to impose positivity for arbitrary $\f$. In other words, we would like to impose
\be
\left(\vec v \cos\tfrac \f 2\ \vec v \sin\tfrac\f 2\right)
M_\l(b)
\begin{pmatrix}
\vec v\cos\tfrac \f 2\\
\vec v \sin\tfrac\f 2
\end{pmatrix} \geq 0 \qquad \textrm{ for all } \f\in \R,\ \vec v\in \R^n\,,
\label{eq:positivityconditionwewouldlike}
\ee
where $M_\l(b)$ is a $2n\x2n$ matrix of the form (\ref{eq:matrixpositivitycondition}) with $n\x n$ block entries.

Note that we can freely rescale the vectors $\vec v$ and $(\cos \frac\f 2,\sin\frac\f2)$ without changing the positivity conditions (\ref{eq:positivityconditionwewouldlike}). Thus, we can think of $\vec v$ as parametrizing a point $[v_1:\cdots:v_n]\in \mathbb{RP}^{n-1}$, and $(\cos \frac\f 2,\sin\frac\f2)$ as parametrizing a point $[w_1:w_2]=[\cos\frac\f 2:\sin\frac \f 2]\in\mathbb{RP}^1$.
In this language,~(\ref{eq:positivityconditionwewouldlike}) is equivalent to imposing that $M_\l(b)$ is positive on the image of the Segre embedding
\be
\s&:\mathbb{RP}^1 \x \mathbb{RP}^{n-1} \to \mathbb{RP}^{2n-1} \nn\\
\s&:([w_1:w_2],[v_1:\cdots:v_n]) \mto [w_1v_1:\cdots:w_1v_n:w_2v_1:\cdots:w_2v_n]\,.
\label{eq:segreembedding}
\ee
A theorem of \cite{2015quadratic} states that if a quadratic form $Q$ is nonnegative on a variety $X\subseteq \mathbb{RP}^{k-1}$ of minimal degree, then $Q$ is a sum of squares of linear forms (and hence represented by a positive semidefinite matrix on $\R^k$). $X$ has minimal degree if it is nondegenerate (not contained in a hyperplane) and $\mathrm{deg}(X)=1+\mathrm{codim}(X)$. Fortunately, the image of the Segre embedding $\s$ has precisely these properties: it is nondegenerate, has degree $n$, and has codimension $n-1$.\footnote{In fact, the image of $\s$ is an example of a {\it rational normal scroll}, which is one of the three families of minimal degree varieties, according to the classification \cite{EisenbudHarris}.} Hence, we conclude that (\ref{eq:positivityconditionwewouldlike}) is equivalent to the statement that $M_\l(b)$ is a positive-semidefinite $2n\x2n$ matrix:
\be
M_\l(b) \succeq 0\,.
\ee
It is remarkable that such a simple condition captures the necessary positivity conditions, even for $n\neq 1$. We can now proceed as in the scalar case: we expand $M_\l(b)$ at large $b$ to $r_\mathrm{max}$ subleading orders and approximate it in terms of a matrix polynomial of $b$. We then impose positivity of this matrix polynomial for $b\geq B$.

Having imposed positivity of the heavy density in these four regimes (fixed $m$ and $J$, large $J$ and fixed $m$, large $m$ and fixed $b$, and large $m$ and $b$), we find that the resulting functionals are positive in practice for almost all $m,J$. Violations of positivity come from the functional becoming slightly negative between discretized values of $m$ or $b$. Such violations can usually be fixed by perturbing the functional slightly (for example by including a small admixture of another nearly positive functional).

Our parameter choices are listed in table~\ref{table: numerics}.

\begin{table}
\centering
\begin{tabular}{c|c}
\hline
$J_\mathrm{max}$ & 42  \\
$\de_x$ & 1/400   \\
$\e_b$ & 1/250  \\
$\de_b$ & 1/32  \\
$B$ & 40  \\
$r_\mathrm{max}$ & 4  \\
\hline
 non-default \\
{\tt SDPB} parameters
 & {\tt ----precision=768} \\
 \end{tabular}
\caption{Table of parameters used for discretization and polynomial approximations.}
\label{table: numerics}
\end{table}

\bibliographystyle{JHEP}
\bibliography{refs}

\providecommand{\href}[2]{#2}\begingroup\raggedright\begin{thebibliography}{10}

\bibitem{Clifton:2011jh}
T.~Clifton, P.~G. Ferreira, A.~Padilla and C.~Skordis, \emph{{Modified Gravity
  and Cosmology}},
  \href{http://dx.doi.org/10.1016/j.physrep.2012.01.001}{\emph{Phys. Rept.}
  {\bf 513} (2012) 1--189}, [\href{https://arxiv.org/abs/1106.2476}{{\tt
  1106.2476}}].

\bibitem{Bull:2015stt}
P.~Bull et~al., \emph{{Beyond $\Lambda$CDM: Problems, solutions, and the road
  ahead}}, \href{http://dx.doi.org/10.1016/j.dark.2016.02.001}{\emph{Phys. Dark
  Univ.} {\bf 12} (2016) 56--99}, [\href{https://arxiv.org/abs/1512.05356}{{\tt
  1512.05356}}].

\bibitem{Ishak:2018his}
M.~Ishak, \emph{{Testing General Relativity in Cosmology}},
  \href{http://dx.doi.org/10.1007/s41114-018-0017-4}{\emph{Living Rev. Rel.}
  {\bf 22} (2019) 1}, [\href{https://arxiv.org/abs/1806.10122}{{\tt
  1806.10122}}].

\bibitem{Bueno:2016ypa}
P.~Bueno, P.~A. Cano, V.~S. Min and M.~R. Visser, \emph{{Aspects of general
  higher-order gravities}},
  \href{http://dx.doi.org/10.1103/PhysRevD.95.044010}{\emph{Phys. Rev. D} {\bf
  95} (2017) 044010}, [\href{https://arxiv.org/abs/1610.08519}{{\tt
  1610.08519}}].

\bibitem{Weinberg:1964ew}
S.~Weinberg, \emph{{Photons and Gravitons in $S$-Matrix Theory: Derivation of
  Charge Conservation and Equality of Gravitational and Inertial Mass}},
  \href{http://dx.doi.org/10.1103/PhysRev.135.B1049}{\emph{Phys. Rev.} {\bf
  135} (1964) B1049--B1056}.

\bibitem{Weinberg:1965rz}
S.~Weinberg, \emph{{Photons and gravitons in perturbation theory: Derivation of
  Maxwell's and Einstein's equations}},
  \href{http://dx.doi.org/10.1103/PhysRev.138.B988}{\emph{Phys. Rev.} {\bf 138}
  (1965) B988--B1002}.

\bibitem{Camanho:2014apa}
X.~O. Camanho, J.~D. Edelstein, J.~Maldacena and A.~Zhiboedov, \emph{{Causality
  Constraints on Corrections to the Graviton Three-Point Coupling}},
  \href{http://dx.doi.org/10.1007/JHEP02(2016)020}{\emph{JHEP} {\bf 02} (2016)
  020}, [\href{https://arxiv.org/abs/1407.5597}{{\tt 1407.5597}}].

\bibitem{Gao:2000ga}
S.~Gao and R.~M. Wald, \emph{{Theorems on gravitational time delay and related
  issues}}, \href{http://dx.doi.org/10.1088/0264-9381/17/24/305}{\emph{Class.
  Quant. Grav.} {\bf 17} (2000) 4999--5008},
  [\href{https://arxiv.org/abs/gr-qc/0007021}{{\tt gr-qc/0007021}}].

\bibitem{Chen:2021bvg}
C.~Y.~R. Chen, C.~de~Rham, A.~Margalit and A.~J. Tolley, \emph{{A cautionary
  case of casual causality}},  \href{https://arxiv.org/abs/2112.05031}{{\tt
  2112.05031}}.

\bibitem{Adams:2006sv}
A.~Adams, N.~Arkani-Hamed, S.~Dubovsky, A.~Nicolis and R.~Rattazzi,
  \emph{{Causality, analyticity and an IR obstruction to UV completion}},
  \href{http://dx.doi.org/10.1088/1126-6708/2006/10/014}{\emph{JHEP} {\bf 10}
  (2006) 014}, [\href{https://arxiv.org/abs/hep-th/0602178}{{\tt
  hep-th/0602178}}].

\bibitem{deRham:2017avq}
C.~de~Rham, S.~Melville, A.~J. Tolley and S.-Y. Zhou, \emph{{Positivity bounds
  for scalar field theories}},
  \href{http://dx.doi.org/10.1103/PhysRevD.96.081702}{\emph{Phys. Rev. D} {\bf
  96} (2017) 081702}, [\href{https://arxiv.org/abs/1702.06134}{{\tt
  1702.06134}}].

\bibitem{Caron-Huot:2020cmc}
S.~Caron-Huot and V.~Van~Duong, \emph{{Extremal Effective Field Theories}},
  \href{https://arxiv.org/abs/2011.02957}{{\tt 2011.02957}}.

\bibitem{Caron-Huot:2021rmr}
S.~Caron-Huot, D.~Mazac, L.~Rastelli and D.~Simmons-Duffin, \emph{{Sharp
  Boundaries for the Swampland}},
  \href{http://dx.doi.org/10.1007/jhep07(2021)110}{\emph{JHEP} {\bf 07} (2021)
  110}, [\href{https://arxiv.org/abs/2102.08951}{{\tt 2102.08951}}].

\bibitem{Tolley:2020gtv}
A.~J. Tolley, Z.-Y. Wang and S.-Y. Zhou, \emph{{New positivity bounds from full
  crossing symmetry}},  \href{https://arxiv.org/abs/2011.02400}{{\tt
  2011.02400}}.

\bibitem{Arkani-Hamed:2020blm}
N.~Arkani-Hamed, T.-C. Huang and Y.-T. Huang, \emph{{The EFT-Hedron}},
  \href{https://arxiv.org/abs/2012.15849}{{\tt 2012.15849}}.

\bibitem{Chiang:2021ziz}
L.-Y. Chiang, Y.-t. Huang, W.~Li, L.~Rodina and H.-C. Weng, \emph{{Into the
  EFThedron and UV constraints from IR consistency}},
  \href{https://arxiv.org/abs/2105.02862}{{\tt 2105.02862}}.

\bibitem{Bellazzini:2020cot}
B.~Bellazzini, J.~Elias~Mir\'o, R.~Rattazzi, M.~Riembau and F.~Riva,
  \emph{{Positive Moments for Scattering Amplitudes}},
  \href{https://arxiv.org/abs/2011.00037}{{\tt 2011.00037}}.

\bibitem{Sinha:2020win}
A.~Sinha and A.~Zahed, \emph{{Crossing Symmetric Dispersion Relations in
  QFTs}}, \href{http://dx.doi.org/10.1103/PhysRevLett.126.181601}{\emph{Phys.
  Rev. Lett.} {\bf 126} (2021) 181601},
  [\href{https://arxiv.org/abs/2012.04877}{{\tt 2012.04877}}].

\bibitem{Alberte:2020bdz}
L.~Alberte, C.~de~Rham, S.~Jaitly and A.~J. Tolley, \emph{{QED positivity
  bounds}},  \href{https://arxiv.org/abs/2012.05798}{{\tt 2012.05798}}.

\bibitem{Li:2021lpe}
X.~Li, H.~Xu, C.~Yang, C.~Zhang and S.-Y. Zhou, \emph{{Positivity in Multifield
  Effective Field Theories}},
  \href{http://dx.doi.org/10.1103/PhysRevLett.127.121601}{\emph{Phys. Rev.
  Lett.} {\bf 127} (2021) 121601},
  [\href{https://arxiv.org/abs/2101.01191}{{\tt 2101.01191}}].

\bibitem{Haldar:2021rri}
P.~Haldar, A.~Sinha and A.~Zahed, \emph{{Quantum field theory and the
  Bieberbach conjecture}},
  \href{http://dx.doi.org/10.21468/SciPostPhys.11.1.002}{\emph{SciPost Phys.}
  {\bf 11} (2021) 002}, [\href{https://arxiv.org/abs/2103.12108}{{\tt
  2103.12108}}].

\bibitem{Raman:2021pkf}
P.~Raman and A.~Sinha, \emph{{QFT, EFT and GFT}},
  \href{https://arxiv.org/abs/2107.06559}{{\tt 2107.06559}}.

\bibitem{Henriksson:2021ymi}
J.~Henriksson, B.~McPeak, F.~Russo and A.~Vichi, \emph{{Rigorous Bounds on
  Light-by-Light Scattering}},  \href{https://arxiv.org/abs/2107.13009}{{\tt
  2107.13009}}.

\bibitem{Zahed:2021fkp}
A.~Zahed, \emph{{Positivity and geometric function theory constraints on pion
  scattering}}, \href{http://dx.doi.org/10.1007/JHEP12(2021)036}{\emph{JHEP}
  {\bf 12} (2021) 036}, [\href{https://arxiv.org/abs/2108.10355}{{\tt
  2108.10355}}].

\bibitem{Alberte:2019xfh}
L.~Alberte, C.~de~Rham, A.~Momeni, J.~Rumbutis and A.~J. Tolley,
  \emph{{Positivity Constraints on Interacting Spin-2 Fields}},
  \href{http://dx.doi.org/10.1007/JHEP03(2020)097}{\emph{JHEP} {\bf 03} (2020)
  097}, [\href{https://arxiv.org/abs/1910.11799}{{\tt 1910.11799}}].

\bibitem{Alberte:2021dnj}
L.~Alberte, C.~de~Rham, S.~Jaitly and A.~J. Tolley, \emph{{Reverse
  Bootstrapping: IR lessons for UV physics}},
  \href{https://arxiv.org/abs/2111.09226}{{\tt 2111.09226}}.

\bibitem{Chowdhury:2021ynh}
S.~D. Chowdhury, K.~Ghosh, P.~Haldar, P.~Raman and A.~Sinha, \emph{{Crossing
  Symmetric Spinning S-matrix Bootstrap: EFT bounds}},
  \href{https://arxiv.org/abs/2112.11755}{{\tt 2112.11755}}.

\bibitem{Bellazzini:2021oaj}
B.~Bellazzini, M.~Riembau and F.~Riva, \emph{{The IR-Side of Positivity
  Bounds}},  \href{https://arxiv.org/abs/2112.12561}{{\tt 2112.12561}}.

\bibitem{Zhang:2021eeo}
C.~Zhang, \emph{{SMEFTs living on the edge: determining the UV theories from
  positivity and extremality}},  \href{https://arxiv.org/abs/2112.11665}{{\tt
  2112.11665}}.

\bibitem{Wang:2020jxr}
Y.-J. Wang, F.-K. Guo, C.~Zhang and S.-Y. Zhou, \emph{{Generalized positivity
  bounds on chiral perturbation theory}},
  \href{http://dx.doi.org/10.1007/JHEP07(2020)214}{\emph{JHEP} {\bf 07} (2020)
  214}, [\href{https://arxiv.org/abs/2004.03992}{{\tt 2004.03992}}].

\bibitem{Trott:2020ebl}
T.~Trott, \emph{{Causality, Unitarity and Symmetry in Effective Field Theory}},
   \href{https://arxiv.org/abs/2011.10058}{{\tt 2011.10058}}.

\bibitem{deRham:2017zjm}
C.~de~Rham, S.~Melville, A.~J. Tolley and S.-Y. Zhou, \emph{{UV complete me:
  Positivity Bounds for Particles with Spin}},
  \href{http://dx.doi.org/10.1007/JHEP03(2018)011}{\emph{JHEP} {\bf 03} (2018)
  011}, [\href{https://arxiv.org/abs/1706.02712}{{\tt 1706.02712}}].

\bibitem{deRham:2018qqo}
C.~de~Rham, S.~Melville, A.~J. Tolley and S.-Y. Zhou, \emph{{Positivity Bounds
  for Massive Spin-1 and Spin-2 Fields}},
  \href{http://dx.doi.org/10.1007/JHEP03(2019)182}{\emph{JHEP} {\bf 03} (2019)
  182}, [\href{https://arxiv.org/abs/1804.10624}{{\tt 1804.10624}}].

\bibitem{Wang:2020xlt}
Z.-Y. Wang, C.~Zhang and S.-Y. Zhou, \emph{{Generalized elastic positivity
  bounds on interacting massive spin-2 theories}},
  \href{http://dx.doi.org/10.1007/JHEP04(2021)217}{\emph{JHEP} {\bf 04} (2021)
  217}, [\href{https://arxiv.org/abs/2011.05190}{{\tt 2011.05190}}].

\bibitem{Kologlu:2019bco}
M.~Kologlu, P.~Kravchuk, D.~Simmons-Duffin and A.~Zhiboedov, \emph{{Shocks,
  Superconvergence, and a Stringy Equivalence Principle}},
  \href{https://arxiv.org/abs/1904.05905}{{\tt 1904.05905}}.

\bibitem{Caron-Huot:2021enk}
S.~Caron-Huot, D.~Mazac, L.~Rastelli and D.~Simmons-Duffin, \emph{{AdS bulk
  locality from sharp CFT bounds}},
  \href{http://dx.doi.org/10.1007/JHEP11(2021)164}{\emph{JHEP} {\bf 11} (2021)
  164}, [\href{https://arxiv.org/abs/2106.10274}{{\tt 2106.10274}}].

\bibitem{Weinberg:1995mt}
S.~Weinberg, \emph{{The Quantum theory of fields. Vol. 1: Foundations}}.
\newblock Cambridge University Press, 6, 2005.

\bibitem{Lee:1969fy}
T.~D. Lee and G.~C. Wick, \emph{{Negative Metric and the Unitarity of the S
  Matrix}}, \href{http://dx.doi.org/10.1016/0550-3213(69)90098-4}{\emph{Nucl.
  Phys. B} {\bf 9} (1969) 209--243}.

\bibitem{Grinstein:2008bg}
B.~Grinstein, D.~O'Connell and M.~B. Wise, \emph{{Causality as an emergent
  macroscopic phenomenon: The Lee-Wick O(N) model}},
  \href{http://dx.doi.org/10.1103/PhysRevD.79.105019}{\emph{Phys. Rev. D} {\bf
  79} (2009) 105019}, [\href{https://arxiv.org/abs/0805.2156}{{\tt
  0805.2156}}].

\bibitem{Donoghue:2019fcb}
J.~F. Donoghue and G.~Menezes, \emph{{Unitarity, stability and loops of
  unstable ghosts}},
  \href{http://dx.doi.org/10.1103/PhysRevD.100.105006}{\emph{Phys. Rev. D} {\bf
  100} (2019) 105006}, [\href{https://arxiv.org/abs/1908.02416}{{\tt
  1908.02416}}].

\bibitem{Woodard:2015zca}
R.~P. Woodard, \emph{{Ostrogradsky's theorem on Hamiltonian instability}},
  \href{http://dx.doi.org/10.4249/scholarpedia.32243}{\emph{Scholarpedia} {\bf
  10} (2015) 32243}, [\href{https://arxiv.org/abs/1506.02210}{{\tt
  1506.02210}}].

\bibitem{Cline:2003gs}
J.~M. Cline, S.~Jeon and G.~D. Moore, \emph{{The Phantom menaced: Constraints
  on low-energy effective ghosts}},
  \href{http://dx.doi.org/10.1103/PhysRevD.70.043543}{\emph{Phys. Rev. D} {\bf
  70} (2004) 043543}, [\href{https://arxiv.org/abs/hep-ph/0311312}{{\tt
  hep-ph/0311312}}].

\bibitem{Cutkosky:1969fq}
R.~E. Cutkosky, P.~V. Landshoff, D.~I. Olive and J.~C. Polkinghorne, \emph{{A
  non-analytic S matrix}},
  \href{http://dx.doi.org/10.1016/0550-3213(69)90169-2}{\emph{Nucl. Phys. B}
  {\bf 12} (1969) 281--300}.

\bibitem{Elvang:2013cua}
H.~Elvang and Y.-t. Huang, \emph{{Scattering Amplitudes}},
  \href{https://arxiv.org/abs/1308.1697}{{\tt 1308.1697}}.

\bibitem{Bern:2021ppb}
Z.~Bern, D.~Kosmopoulos and A.~Zhiboedov, \emph{{Gravitational effective field
  theory islands, low-spin dominance, and the four-graviton amplitude}},
  \href{http://dx.doi.org/10.1088/1751-8121/ac0e51}{\emph{J. Phys. A} {\bf 54}
  (2021) 344002}, [\href{https://arxiv.org/abs/2103.12728}{{\tt 2103.12728}}].

\bibitem{Bueno:2019ltp}
P.~Bueno, P.~A. Cano, J.~Moreno and A.~Murcia, \emph{{All higher-curvature
  gravities as Generalized quasi-topological gravities}},
  \href{http://dx.doi.org/10.1007/JHEP11(2019)062}{\emph{JHEP} {\bf 11} (2019)
  062}, [\href{https://arxiv.org/abs/1906.00987}{{\tt 1906.00987}}].

\bibitem{Hebbar:2020ukp}
A.~Hebbar, D.~Karateev and J.~Penedones, \emph{{Spinning S-matrix Bootstrap in
  4d}},  \href{https://arxiv.org/abs/2011.11708}{{\tt 2011.11708}}.

\bibitem{Pajer:2020wnj}
E.~Pajer, D.~Stefanyszyn and J.~Supel, \emph{{The Boostless Bootstrap:
  Amplitudes without Lorentz boosts}},
  \href{http://dx.doi.org/10.1007/JHEP12(2020)198}{\emph{JHEP} {\bf 12} (2020)
  198}, [\href{https://arxiv.org/abs/2007.00027}{{\tt 2007.00027}}].

\bibitem{Grall:2021xxm}
T.~Grall and S.~Melville, \emph{{Positivity Bounds without Boosts}},
  \href{https://arxiv.org/abs/2102.05683}{{\tt 2102.05683}}.

\bibitem{Du:2021byy}
Z.-Z. Du, C.~Zhang and S.-Y. Zhou, \emph{{Triple crossing positivity bounds for
  multi-field theories}},  \href{https://arxiv.org/abs/2111.01169}{{\tt
  2111.01169}}.

\bibitem{Martin:1962rt}
A.~Martin, \emph{{Unitarity and high-energy behavior of scattering
  amplitudes}}, \href{http://dx.doi.org/10.1103/PhysRev.129.1432}{\emph{Phys.
  Rev.} {\bf 129} (1963) 1432--1436}.

\bibitem{Maldacena:2015waa}
J.~Maldacena, S.~H. Shenker and D.~Stanford, \emph{{A bound on chaos}},
  \href{http://dx.doi.org/10.1007/JHEP08(2016)106}{\emph{JHEP} {\bf 08} (2016)
  106}, [\href{https://arxiv.org/abs/1503.01409}{{\tt 1503.01409}}].

\bibitem{Chandorkar:2021viw}
D.~Chandorkar, S.~D. Chowdhury, S.~Kundu and S.~Minwalla, \emph{{Bounds on
  Regge growth of flat space scattering from bounds on chaos}},
  \href{https://arxiv.org/abs/2102.03122}{{\tt 2102.03122}}.

\bibitem{Gell-Mann:1954ttj}
M.~Gell-Mann, M.~L. Goldberger and W.~E. Thirring, \emph{{Use of causality
  conditions in quantum theory}},
  \href{http://dx.doi.org/10.1103/PhysRev.95.1612}{\emph{Phys. Rev.} {\bf 95}
  (1954) 1612--1627}.

\bibitem{DeLacroix:2018arq}
C.~De~Lacroix, H.~Erbin and A.~Sen, \emph{{Analyticity and Crossing Symmetry of
  Superstring Loop Amplitudes}},
  \href{http://dx.doi.org/10.1007/JHEP05(2019)139}{\emph{JHEP} {\bf 05} (2019)
  139}, [\href{https://arxiv.org/abs/1810.07197}{{\tt 1810.07197}}].

\bibitem{Mizera:2021fap}
S.~Mizera, \emph{{Crossing symmetry in the planar limit}},
  \href{http://dx.doi.org/10.1103/PhysRevD.104.045003}{\emph{Phys. Rev. D} {\bf
  104} (2021) 045003}, [\href{https://arxiv.org/abs/2104.12776}{{\tt
  2104.12776}}].

\bibitem{Fu:2013cza}
C.-H. Fu, J.-C. Lee, C.-I. Tan and Y.~Yang, \emph{{BCFW Deformation and Regge
  Limit}},  \href{https://arxiv.org/abs/1305.7442}{{\tt 1305.7442}}.

\bibitem{Chowdhury:2019kaq}
S.~D. Chowdhury, A.~Gadde, T.~Gopalka, I.~Halder, L.~Janagal and S.~Minwalla,
  \emph{{Classifying and constraining local four photon and four graviton
  S-matrices}}, \href{http://dx.doi.org/10.1007/JHEP02(2020)114}{\emph{JHEP}
  {\bf 02} (2020) 114}, [\href{https://arxiv.org/abs/1910.14392}{{\tt
  1910.14392}}].

\bibitem{Caron-Huot:2017vep}
S.~Caron-Huot, \emph{{Analyticity in Spin in Conformal Theories}},
  \href{http://dx.doi.org/10.1007/JHEP09(2017)078}{\emph{JHEP} {\bf 09} (2017)
  078}, [\href{https://arxiv.org/abs/1703.00278}{{\tt 1703.00278}}].

\bibitem{Simmons-Duffin:2017nub}
D.~Simmons-Duffin, D.~Stanford and E.~Witten, \emph{{A spacetime derivation of
  the Lorentzian OPE inversion formula}},
  \href{http://dx.doi.org/10.1007/JHEP07(2018)085}{\emph{JHEP} {\bf 07} (2018)
  085}, [\href{https://arxiv.org/abs/1711.03816}{{\tt 1711.03816}}].

\bibitem{Kravchuk:2018htv}
P.~Kravchuk and D.~Simmons-Duffin, \emph{{Light-ray operators in conformal
  field theory}}, \href{http://dx.doi.org/10.1007/JHEP11(2018)102}{\emph{JHEP}
  {\bf 11} (2018) 102}, [\href{https://arxiv.org/abs/1805.00098}{{\tt
  1805.00098}}].

\bibitem{Bellazzini:2015cra}
B.~Bellazzini, C.~Cheung and G.~N. Remmen, \emph{{Quantum Gravity Constraints
  from Unitarity and Analyticity}},
  \href{http://dx.doi.org/10.1103/PhysRevD.93.064076}{\emph{Phys. Rev. D} {\bf
  93} (2016) 064076}, [\href{https://arxiv.org/abs/1509.00851}{{\tt
  1509.00851}}].

\bibitem{Arkani-Hamed:2021ajd}
N.~Arkani-Hamed, Y.-t. Huang, J.-Y. Liu and G.~N. Remmen, \emph{{Causality,
  Unitarity, and the Weak Gravity Conjecture}},
  \href{https://arxiv.org/abs/2109.13937}{{\tt 2109.13937}}.

\bibitem{Simmons-Duffin:2015qma}
D.~Simmons-Duffin, \emph{{A Semidefinite Program Solver for the Conformal
  Bootstrap}}, \href{http://dx.doi.org/10.1007/JHEP06(2015)174}{\emph{JHEP}
  {\bf 06} (2015) 174}, [\href{https://arxiv.org/abs/1502.02033}{{\tt
  1502.02033}}].

\bibitem{Landry:2019qug}
W.~Landry and D.~Simmons-Duffin, \emph{{Scaling the semidefinite program solver
  SDPB}},  \href{https://arxiv.org/abs/1909.09745}{{\tt 1909.09745}}.

\bibitem{deRham:2021bll}
C.~de~Rham, A.~J. Tolley and J.~Zhang, \emph{{Causality Constraints on
  Gravitational Effective Field Theories}},
  \href{https://arxiv.org/abs/2112.05054}{{\tt 2112.05054}}.

\bibitem{yutin}
L.-Y. Chiang, Y.-t. Huang, W.~Li, L.~Rodina and H.-C. Weng, \emph{{Unitarity
  constraints on helicity amplitudes in Quantum Gravity}}, {\emph{to appear
  concurrently} (2022) }.

\bibitem{ATLAS:2017eqx}
{\scshape ATLAS} collaboration, M.~Aaboud et~al., \emph{{Search for new
  phenomena in dijet events using 37 fb$^{-1}$ of $pp$ collision data collected
  at $\sqrt{s}=$13 TeV with the ATLAS detector}},
  \href{http://dx.doi.org/10.1103/PhysRevD.96.052004}{\emph{Phys. Rev. D} {\bf
  96} (2017) 052004}, [\href{https://arxiv.org/abs/1703.09127}{{\tt
  1703.09127}}].

\bibitem{CMS:2018mgb}
{\scshape CMS} collaboration, A.~M. Sirunyan et~al., \emph{{Search for narrow
  and broad dijet resonances in proton-proton collisions at $ \sqrt{s}=13 $ TeV
  and constraints on dark matter mediators and other new particles}},
  \href{http://dx.doi.org/10.1007/JHEP08(2018)130}{\emph{JHEP} {\bf 08} (2018)
  130}, [\href{https://arxiv.org/abs/1806.00843}{{\tt 1806.00843}}].

\bibitem{Sennett:2019bpc}
N.~Sennett, R.~Brito, A.~Buonanno, V.~Gorbenko and L.~Senatore,
  \emph{{Gravitational-Wave Constraints on an Effective Field-Theory Extension
  of General Relativity}},
  \href{http://dx.doi.org/10.1103/PhysRevD.102.044056}{\emph{Phys. Rev. D} {\bf
  102} (2020) 044056}, [\href{https://arxiv.org/abs/1912.09917}{{\tt
  1912.09917}}].

\bibitem{Carson:2020dez}
Z.~Carson and K.~Yagi, \emph{{Asymptotically flat, parameterized black hole
  metric preserving Kerr symmetries}},
  \href{http://dx.doi.org/10.1103/PhysRevD.101.084030}{\emph{Phys. Rev. D} {\bf
  101} (2020) 084030}, [\href{https://arxiv.org/abs/2002.01028}{{\tt
  2002.01028}}].

\bibitem{ATLAS:2017bfj}
{\scshape ATLAS} collaboration, M.~Aaboud et~al., \emph{{Search for dark matter
  and other new phenomena in events with an energetic jet and large missing
  transverse momentum using the ATLAS detector}},
  \href{http://dx.doi.org/10.1007/JHEP01(2018)126}{\emph{JHEP} {\bf 01} (2018)
  126}, [\href{https://arxiv.org/abs/1711.03301}{{\tt 1711.03301}}].

\bibitem{CMS:2017zts}
{\scshape CMS} collaboration, A.~M. Sirunyan et~al., \emph{{Search for new
  physics in final states with an energetic jet or a hadronically decaying $W$
  or $Z$ boson and transverse momentum imbalance at $\sqrt{s}=13\text{ }\text{
  }\mathrm{TeV}$}},
  \href{http://dx.doi.org/10.1103/PhysRevD.97.092005}{\emph{Phys. Rev. D} {\bf
  97} (2018) 092005}, [\href{https://arxiv.org/abs/1712.02345}{{\tt
  1712.02345}}].

\bibitem{Bjerrum-Bohr:2014zsa}
N.~E.~J. Bjerrum-Bohr, J.~F. Donoghue, B.~R. Holstein, L.~Plant\'e and
  P.~Vanhove, \emph{{Bending of Light in Quantum Gravity}},
  \href{http://dx.doi.org/10.1103/PhysRevLett.114.061301}{\emph{Phys. Rev.
  Lett.} {\bf 114} (2015) 061301}, [\href{https://arxiv.org/abs/1410.7590}{{\tt
  1410.7590}}].

\bibitem{Dvali:2007hz}
G.~Dvali, \emph{{Black Holes and Large N Species Solution to the Hierarchy
  Problem}}, \href{http://dx.doi.org/10.1002/prop.201000009}{\emph{Fortsch.
  Phys.} {\bf 58} (2010) 528--536},
  [\href{https://arxiv.org/abs/0706.2050}{{\tt 0706.2050}}].

\bibitem{Cheung:2016yqr}
C.~Cheung and G.~N. Remmen, \emph{{Positive Signs in Massive Gravity}},
  \href{http://dx.doi.org/10.1007/JHEP04(2016)002}{\emph{JHEP} {\bf 04} (2016)
  002}, [\href{https://arxiv.org/abs/1601.04068}{{\tt 1601.04068}}].

\bibitem{Afkhami-Jeddi:2018apj}
N.~Afkhami-Jeddi, S.~Kundu and A.~Tajdini, \emph{{A Bound on Massive Higher
  Spin Particles}},
  \href{http://dx.doi.org/10.1007/JHEP04(2019)056}{\emph{JHEP} {\bf 04} (2019)
  056}, [\href{https://arxiv.org/abs/1811.01952}{{\tt 1811.01952}}].

\bibitem{Ellis:2016jkw}
J.~Ellis, \emph{{TikZ-Feynman: Feynman diagrams with TikZ}},
  \href{http://dx.doi.org/10.1016/j.cpc.2016.08.019}{\emph{Comput. Phys.
  Commun.} {\bf 210} (2017) 103--123},
  [\href{https://arxiv.org/abs/1601.05437}{{\tt 1601.05437}}].

\bibitem{Arkani-Hamed:2017jhn}
N.~Arkani-Hamed, T.-C. Huang and Y.-t. Huang, \emph{{Scattering amplitudes for
  all masses and spins}},
  \href{http://dx.doi.org/10.1007/JHEP11(2021)070}{\emph{JHEP} {\bf 11} (2021)
  070}, [\href{https://arxiv.org/abs/1709.04891}{{\tt 1709.04891}}].

\bibitem{2015quadratic}
G.~Blekherman, G.~Smith and M.~Velasco, \emph{Sums of squares and varieties of
  minimal degree}, \href{http://dx.doi.org/10.1090/jams/847}{\emph{Journal of
  the American Mathematical Society} {\bf 29} (Sep, 2015) 893–913}.

\bibitem{EisenbudHarris}
D.~Eisenbud and J.~Harris, \emph{On varieties of minimal degree (a centennial
  account)},  in \emph{Algebraic geometry, Bowdoin, 1985 (Brunswick, Maine,
  1985), 3 to 13, Proc. Sympos. Pure Math., 46, Part 1}.
\newblock American Math. Soc., Providence, RI, 1987.

\end{thebibliography}\endgroup

\end{document}